\documentclass[pdftex,twocolumn,epjc3,runningheads]{svjour3}
\usepackage[T1]{fontenc}
\usepackage{lmodern}
\usepackage{calc}
\usepackage{graphicx}
\usepackage{booktabs}
\usepackage{textcomp}
\usepackage{xspace}
\usepackage{relsize}
\usepackage{amssymb}
\usepackage{amsmath}
\usepackage{listings}
\usepackage{microtype}
\usepackage{multirow}
\usepackage{tabularx}
\usepackage{array}
\usepackage{placeins}
\usepackage{cuted}
\usepackage{soul} 
\usepackage{fixltx2e}
\usepackage[numbers,sort&compress]{natbib}
\usepackage[labelfont=bf,font=small]{caption}
\usepackage[skip=-2pt]{subcaption}
\usepackage{hyperref}
\usepackage[dvipsnames]{xcolor}
\usepackage[clockwise,figuresright]{rotating}
\usepackage{siunitx}

\allowdisplaybreaks

\newcolumntype{L}{>{\raggedright\let\newline\\\arraybackslash\hspace{0pt}}X}
\newcolumntype{R}{>{\raggedleft\let\newline\\\arraybackslash\hspace{0pt}}X}
\newcolumntype{C}{>{\centering\let\newline\\\arraybackslash\hspace{0pt}}X}

\newcommand{\imperial}{Department of Physics, Imperial College London, Blackett Laboratory, Prince Consort Road, London SW7 2AZ, UK}
\newcommand{\nordita}{NORDITA, Roslagstullsbacken 23, SE-10691 Stockholm, Sweden}
\newcommand{\oslo}{Department of Physics, University of Oslo, N-0316 Oslo, Norway}
\newcommand{\adelaide}{Department of Physics, University of Adelaide, Adelaide, SA 5005, Australia}
\newcommand{\glasgow}{SUPA, School of Physics and Astronomy, University of Glasgow, Glasgow, G12 8QQ, UK}
\newcommand{\monash}{School of Physics and Astronomy, Monash University, Melbourne, VIC 3800, Australia}
\newcommand{\coepp}{Australian Research Council Centre of Excellence for Particle Physics at the Tera-scale}
\newcommand{\okc}{Oskar Klein Centre for Cosmoparticle Physics, AlbaNova University Centre, SE-10691 Stockholm, Sweden}

\newcommand{\annecy}{LAPTh, Universit\'e de Savoie, CNRS, 9 chemin de Bellevue B.P.110, F-74941 Annecy-le-Vieux, France}
\newcommand{\harvard}{Department of Physics, Harvard University, Cambridge, MA 02138, USA}
\newcommand{\grappa}{GRAPPA, Institute of Physics, University of Amsterdam, Science Park 904, 1098 XH Amsterdam, Netherlands}
\newcommand{\sydney}{Centre for Translational Data Science, Faculty of Engineering and Information Technologies, School of Physics, The University of Sydney, NSW 2006, Australia}

\newcommand{\gambitacknos    }{We warmly thank the Casa Matem\'aticas Oaxaca, affiliated with the Banff International Research Station, for hospitality whilst part of this work was completed, and the staff at Cyfronet, for their always helpful supercomputing support.  \GB has been supported by STFC (UK; ST/K00414X/1, ST/P000762/1), the Royal Society (UK; UF110191), Glasgow University (UK; Leadership Fellowship), the Research Council of Norway (FRIPRO 230546/F20), NOTUR (Norway; NN9284K), the Knut and Alice Wallenberg Foundation (Sweden; Wallenberg Academy Fellowship), the Swedish Research Council (621-2014-5772), the Australian Research Council (CE110001004, FT130100018, FT140100244, FT160100274), The University of Sydney (Australia; IRCA-G162448), PLGrid Infrastructure (Poland), Polish National Science Center (Sonata UMO-2015/17/D/ST2/03532), the Swiss National Science Foundation (PP00P2-144674), the European Commission Horizon 2020 Marie Sk\l{}odowska-Curie actions (H2020-MSCA-RISE-2015-691164), the ERA-CAN+ Twinning Program (EU \& Canada), the Netherlands Organisation for Scientific Research (NWO-Vidi 680-47-532), the National Science Foundation (USA; DGE-1339067), the FRQNT (Qu\'ebec) and NSERC/The Canadian Tri-Agencies Research Councils (BPDF-424460-2012).}

\makeatletter
\g@addto@macro\bfseries{\boldmath}
\makeatother

\newcommand{\subparagraph}{} 
\usepackage{titlesec}
\titleformat*{\paragraph}{\bfseries}

\journalname{Eur. Phys. J. C}
\smartqed
\let\underscore\_
\renewcommand{\_}{\discretionary{\underscore}{}{\underscore}}

\makeatletter
\let\orgdescriptionlabel\descriptionlabel
\renewcommand*{\descriptionlabel}[1]{%
  \let\orglabel\label
  \let\label\@gobble
  \phantomsection
  \protected@edef\@currentlabel{#1}%
  \let\label\orglabel
  \orgdescriptionlabel{#1}%
}
\makeatother

\newcommand\postnewlinemarker{\hbox{\ensuremath{\hookrightarrow}}}
\newcommand\cpp[1]{{\lstinline!#1!}}  
\newcommand\cpppragma[1]{{\CPPcommentstyle#1}}
\newcommand\yaml[1]{{\lstset{style=yaml}\lstinline!#1!\lstset{style=cpp}}}

\newcommand\term[1]{{\lstset{style=terminal}\lstinline!#1!\lstset{style=cpp}}}
\newcommand\fortran[1]{{\lstset{style=fortran}\lstinline!#1!\lstset{style=cpp}}}
\newcommand\py[1]{{\lstset{style=python}\lstinline!#1!\lstset{style=cpp}}}
\newcommand\customtilde{{\raisebox{0.2ex}{\scalebox{0.6}{\boldmath$\sim$}}}}

\lstnewenvironment{lstlistingyaml}{\lstset{style=yaml}}{\lstset{style=cpp}}
\lstnewenvironment{lstlistingterm}{\lstset{style=terminal}}{\lstset{style=cpp}}
\lstnewenvironment{lstlistingfortran}{\lstset{style=fortran}}{\lstset{style=cpp}}
\lstnewenvironment{lstcpp}{\lstset{style=cpp}}{\lstset{style=cpp}}
\lstnewenvironment{lstcppalt}{\lstset{style=cppalt}}{\lstset{style=cpp}}
\lstnewenvironment{lstcppnum}{\lstset{style=cppnum}}{\lstset{style=cpp}}
\lstnewenvironment{lstyaml}{\lstset{style=yaml}}{\lstset{style=cpp}}
\lstnewenvironment{lstterm}{\lstset{style=terminal}}{\lstset{style=cpp}}
\lstnewenvironment{lsttermalt}{\lstset{style=terminalalt}}{\lstset{style=cpp}}
\lstnewenvironment{lsttext}{\lstset{style=text}}{\lstset{style=cpp}}
\lstnewenvironment{lstfortran}{\lstset{style=fortran}}{\lstset{style=cpp}}
\lstnewenvironment{lstpy}{\lstset{style=python}}{\lstset{style=cpp}}

\newcommand{\tmpname}{}
\newcommand{\tmplistingname}{}
\makeatletter
\newif\ifATOlabelname
\lst@Key{labelname}{Listing}{\def\ATOlabelname{#1}\global\ATOlabelnametrue}
\makeatother
\lstnewenvironment{lstcpplabel}[1][]{
  \lstset{style=cpp,#1} 
  \ifATOlabelname
    \renewcommand{\tmpname}{\lstlistingname}
    \renewcommand{\tmplistingname}{\lstlistlistingname}
    \renewcommand{\lstlistingname}{\ATOlabelname}
    \renewcommand{\lstlistlistingname}{List of \lstlistingname s}
  \fi
}{
  \renewcommand{\lstlistingname}{\tmpname}
  \renewcommand{\lstlistlistingname}{\tmplistingname}
  \lstset{style=cpp}
}
\definecolor{solarized@base03}{HTML}{002B36}
\definecolor{solarized@base02}{HTML}{073642}
\definecolor{solarized@base01}{HTML}{586e75}
\definecolor{solarized@base00}{HTML}{657b83}
\definecolor{solarized@base0}{HTML}{839496}
\definecolor{solarized@base1}{HTML}{93a1a1}
\definecolor{solarized@base2}{HTML}{EEE8D5}
\definecolor{solarized@base3}{HTML}{FDF6E3}
\definecolor{solarized@yellow}{HTML}{B58900}
\definecolor{solarized@orange}{HTML}{CB4B16}
\definecolor{solarized@red}{HTML}{DC322F}
\definecolor{solarized@magenta}{HTML}{D33682}
\definecolor{solarized@violet}{HTML}{6C71C4}
\definecolor{solarized@blue}{HTML}{268BD2}
\definecolor{solarized@cyan}{HTML}{2AA198}
\definecolor{solarized@green}{HTML}{859900}
\definecolor{darkred}{HTML}{550003}
\definecolor{darkgreen}{HTML}{00AA00}

\newcommand\YAMLstringstyle{\footnotesize\color{solarized@green}\mdseries}
\newcommand\YAMLkeystyle{\footnotesize\color{solarized@blue}\ttfamily}
\newcommand\YAMLvaluestyle{\footnotesize\color{blue}\mdseries}
\newcommand\ProcessThreeDashes{\llap{\color{cyan}\mdseries-{-}-}}

\newcommand\CPPcommentstyle{\color{solarized@violet}\footnotesize\ttfamily}
\newcommand\CPPdirectivestyle{\color{solarized@magenta}\footnotesize\ttfamily}
\newcommand\termplainstyle{\footnotesize\ttfamily}

\newcommand\processLongMacroDelimiter
{%
\CPPdirectivestyle%
\#define%
}

\lstdefinestyle{cpp}
{
  language=C++,
  basicstyle=\footnotesize\ttfamily,
  basewidth={0.53em,0.44em}, 
  numbers=none,
  tabsize=2,
  breaklines=true,
  escapeinside={@}{@},
  showstringspaces=false,
  numberstyle=\tiny\color{solarized@base01},
  keywordstyle=\color{solarized@orange},
  stringstyle=\color{solarized@red}\ttfamily,
  identifierstyle=\color{solarized@blue},
  commentstyle=\CPPcommentstyle,
  directivestyle=\CPPdirectivestyle,
  emphstyle=\color{solarized@green},
  frame=single,
  rulecolor=\color{solarized@base2},
  rulesepcolor=\color{solarized@base2},
  literate={~} {\customtilde}1,
  moredelim=*[directive]\ \ \#,
  moredelim=*[directive]\ \ \ \ \#
}

\lstdefinestyle{cppalt}
{
  language=C++,
  basicstyle=\footnotesize\ttfamily,
  basewidth={0.53em,0.44em}, 
  numbers=none,
  tabsize=2,
  breaklines=true,
  escapeinside={*@}{@*},
  showstringspaces=false,
  numberstyle=\tiny\color{solarized@base01},
  keywordstyle=\color{solarized@orange},
  stringstyle=\color{solarized@red}\ttfamily,
  identifierstyle=\color{solarized@blue},
  commentstyle=\CPPcommentstyle,
  directivestyle=\CPPdirectivestyle,
  emphstyle=\color{solarized@green},
  frame=single,
  rulecolor=\color{solarized@base2},
  rulesepcolor=\color{solarized@base2},
  literate={~}{\customtilde}1,
  moredelim=**[is][\processLongMacroDelimiter]{BeginLongMacro}{EndLongMacro} 
}

\lstdefinestyle{cppnum}
{
  language=C++,
  basicstyle=\footnotesize\ttfamily,
  basewidth={0.53em,0.44em}, 
  numbers=none,
  tabsize=2,
  breaklines=true,
  escapeinside={@}{@},
  numberstyle=\tiny\color{solarized@base01},
  showstringspaces=false,
  numberstyle=\tiny\color{solarized@base01},
  keywordstyle=\color{solarized@orange},
  stringstyle=\color{solarized@red}\ttfamily,
  identifierstyle=\color{solarized@blue},
  commentstyle=\CPPcommentstyle,
  directivestyle=\CPPdirectivestyle,
  emphstyle=\color{solarized@green},
  frame=single,
  rulecolor=\color{solarized@base2},
  rulesepcolor=\color{solarized@base2},
  literate={~} {\customtilde}1,
  moredelim=*[directive]\ \ \#,
  moredelim=*[directive]\ \ \ \ \#
}

\lstdefinestyle{python}
{
  language=Python,
  basicstyle=\footnotesize\ttfamily,
  basewidth={0.53em,0.44em},
  numbers=none,
  tabsize=2,
  breaklines=true,
  escapeinside={@}{@},
  showstringspaces=false,
  numberstyle=\tiny\color{solarized@base01},
  keywordstyle=\color{blue},
  stringstyle=\color{orange}\ttfamily,
  identifierstyle=\color{darkred},
  commentstyle=\color{purple},
  emphstyle=\color{green},
  frame=single,
  rulecolor=\color{solarized@base2},
  rulesepcolor=\color{solarized@base2},
  literate = {~}{\customtilde}1
             {\ as\ }{{\color{blue}\ as\ \color{black}}}3
}

\lstdefinestyle{fortran}
{
  language=Fortran,
  basicstyle=\footnotesize\ttfamily,
  basewidth={0.53em,0.44em},
  numbers=none,
  tabsize=2,
  breaklines=true,
  escapeinside={@}{@},
  showstringspaces=false,
  numberstyle=\tiny\color{solarized@base01},
  keywordstyle=\color{blue},
  stringstyle=\color{orange}\ttfamily,
  identifierstyle=\color{Periwinkle},
  commentstyle=\color{purple},
  emphstyle=\color{green},
  morekeywords={and, or, true, false},
  frame=single,
  rulecolor=\color{solarized@base2},
  rulesepcolor=\color{solarized@base2},
  literate={~}{\customtilde}1
}

\lstdefinestyle{terminal}
{
  language=bash,
  basicstyle=\termplainstyle,
  numbers=none,
  tabsize=2,
  breaklines=true,
  escapeinside={@}{@},
  frame=single,
  showstringspaces=false,
  numberstyle=\tiny\color{solarized@base01},
  keywordstyle=\color{solarized@orange},
  stringstyle=\color{solarized@red}\ttfamily,
  identifierstyle=\color{black},
  commentstyle=\color{solarized@violet},
  emphstyle=\color{solarized@green},
  frame=single,
  rulecolor=\color{solarized@base2},
  rulesepcolor=\color{solarized@base2},
  morekeywords={gambit, cmake, make, mkdir},
  deletekeywords={test},
  literate = {\ gambit}{{\ }{\color{black}}gambit}7
             {/gambit}{{/}{\color{black}}gambit}6
             {gambit/}{{\color{black}}gambit{/}}6
             {/include}{{/}{\color{black}}include}8
             {cmake/}{{\color{black}}cmake/}6
             {.cmake}{{.}{\color{black}}cmake}6
             {~}{\customtilde}1
}

\lstdefinestyle{terminalalt}
{
  language=bash,
  basicstyle=\footnotesize\ttfamily,
  numbers=none,
  tabsize=2,
  breaklines=true,
  escapeinside={*@}{@*},
  frame=single,
  showstringspaces=false,
  numberstyle=\tiny\color{solarized@base01},
  keywordstyle=\color{solarized@orange},
  stringstyle=\color{solarized@red}\ttfamily,
  identifierstyle=\color{black},
  commentstyle=\color{solarized@violet},
  emphstyle=\color{solarized@green},
  frame=single,
  rulecolor=\color{solarized@base2},
  rulesepcolor=\color{solarized@base2},
  morekeywords={gambit, cmake, make, mkdir},
  deletekeywords={test},
  literate = {\ gambit}{{\ }{\color{black}}gambit}7
             {/gambit}{{/}{\color{black}}gambit}6
             {gambit/}{{\color{black}}gambit{/}}6
             {/include}{{/}{\color{black}}include}8
             {cmake/}{{\color{black}}cmake/}6
             {.cmake}{{.}{\color{black}}cmake}6
             {~}{\customtilde}1
}

\lstdefinestyle{text}
{
  language={},
  basicstyle=\footnotesize\ttfamily,
  identifierstyle=\color{black},
  numbers=none,
  tabsize=2,
  breaklines=true,
  escapeinside={*@}{@*},
  showstringspaces=false,
  frame=single,
  rulecolor=\color{solarized@base2},
  rulesepcolor=\color{solarized@base2},
  literate={~}{\customtilde}1
}

\lstdefinestyle{yaml}
{
  language=bash,
  escapeinside={@}{@},
  keywords={true,false,null},
  otherkeywords={},
  keywordstyle=\color{solarized@base0}\bfseries,
  basicstyle=\footnotesize\color{black}\ttfamily,
  identifierstyle=\YAMLkeystyle,
  sensitive=false,
  commentstyle=\color{solarized@orange}\ttfamily,
  morecomment=[l]{\#},
  morecomment=[s]{/*}{*/},
  stringstyle=\YAMLstringstyle\ttfamily,
  moredelim=**[s][\YAMLkeystyle]{,}{:},   
  moredelim=**[l][\YAMLvaluestyle]{:},    
  morestring=[b]',
  morestring=[b]",
  literate =    {---}{{\ProcessThreeDashes}}3
                {>}{{\textcolor{solarized@red}\textgreater}}1
                {|}{{\textcolor{solarized@red}\textbar}}1
                {\ -\ }{{\mdseries\color{black}\ -\ \negmedspace}}3
                {\}}{{{\color{black} \}}}}1
                {\{}{{{\color{black} \{}}}1
                {[}{{{\color{black} [}}}1
                {]}{{{\color{black} ]}}}1
                {~}{\customtilde}1,
  breakindent=0pt,
  breakatwhitespace,
  columns=fullflexible
}

\newcommand{\cross}[1]{\ref{#1}}
\newcommand{\doublecross}[2]{\hyperref[#2]{\textbf{#1}}}
\newcommand{\doublecrosssf}[2]{\hyperref[#2]{\textbf{\textsf{#1}}}}
\newcommand{\gitem}[1]{\item[\textbf{#1}\label{#1}]}

\newcommand{\startglossary}{\section{Glossary}\label{glossary}Here we explain some terms that have specific technical definitions in \GB.\begin{description}}
\newcommand{\finishglossary}{\end{description}}

\newcommand{\bcode}{\begin{lstlisting}}
\newcommand{\ecode}{\end{lstlisting}}
\newcommand{\metavarf}[1]{\textit{\color{darkgreen}\footnotesize\textrm{#1}}}

\newcommand{\metavar}{\metavarf}

\newcommand{\eV}{\ensuremath{\text{e}\mspace{-0.8mu}\text{V}}\xspace}

\newcommand{\GeV}{\text{G\eV}\xspace}
\newcommand{\TeV}{\text{T\eV}\xspace}
\newcommand{\pb}{\text{pb}\xspace}
\newcommand{\fb}{\text{fb}\xspace}

\newcommand{\invfb}{\ensuremath{\fb^{-1}}\xspace}

\newcommand{\pt}{\ensuremath{p_\mathrm{T}}\xspace}
\newcommand{\et}{\ensuremath{E_\mathrm{T}}\xspace}
\newcommand{\etmiss}{\ensuremath{E_\mathrm{T}^\mathrm{\mspace{1.5mu}miss}}\xspace}

\newcommand{\hT}{\ensuremath{H_\mathrm{T}}\xspace}
\newcommand{\dphi}{\ensuremath{\Delta\phi}\xspace}

\newcommand{\CL}{\text{CL}\xspace}

\newcommand{\CLsb}{\ensuremath{\CL_{s+b}}\xspace}
\newcommand{\BR}{\ensuremath{\mathrm{BR}}\xspace}

\newcommand{\ie}{i.e.\ }

\newcommand{\atlas}{ATLAS\xspace}

\newcommand{\gambit}{\textsf{GAMBIT}\xspace}

\newcommand{\colliderbit}{\textsf{ColliderBit}\xspace}

\newcommand{\specbit}{\textsf{SpecBit}\xspace}
\newcommand{\decaybit}{\textsf{DecayBit}\xspace}

\newcommand{\scannerbit}{\textsf{ScannerBit}\xspace}

\newcommand{\BOSS}{\textsf{BOSS}\xspace}
\newcommand{\GB}{\gambit}

\newcommand{\omp}{\textsf{OpenMP}\xspace}
\newcommand{\mpi}{\textsf{MPI}\xspace}

\newcommand{\buckfast}{\textsf{BuckFast}\xspace}
\newcommand{\delphes}{\textsf{Delphes}\xspace}

\newcommand{\MGaMCNLO}{\textsf{MadGraph5\_aMC@NLO}\xspace}
\newcommand{\pythia}{\textsf{Pythia}\xspace}
\newcommand{\pythiaeight}{\textsf{Pythia\,8}\xspace}

\newcommand{\prospino}{\textsf{Prospino}\xspace}
\newcommand{\nllfast}{\textsf{NLL-fast}\xspace}

\newcommand{\fastjet}{\textsf{FastJet}\xspace}
\newcommand{\smodels}{\textsf{SModelS}\xspace}
\newcommand{\fastlim}{\textsf{FastLim}\xspace}
\newcommand{\checkmate}{\textsf{CheckMATE}\xspace}
\newcommand{\higgsbounds}{\textsf{HiggsBounds}\xspace}

\newcommand{\higgssignals}{\textsf{HiggsSignals}\xspace}
\newcommand{\ds}{\textsf{DarkSUSY}\xspace}

\newcommand{\micromegas}{\textsf{micrOMEGAs}\xspace}
\newcommand{\rivet}{\textsf{Rivet}\xspace}
\newcommand{\feynrules}{\textsf{Feynrules}\xspace}
\newcommand{\feynhiggs}{\textsf{FeynHiggs}\xspace}

\newcommand\flexiblesusy{\FlexibleSUSY}
\newcommand\FlexibleSUSY{\textsf{FlexibleSUSY}\xspace}
\newcommand\SOFTSUSY{\textsf{SOFTSUSY}\xspace}

\newcommand\SUSYHIT{\textsf{SUSY-HIT}\xspace}
\newcommand\susyhit{\SUSYHIT}

\newcommand\Mathematica{\textsf{Mathematica}\xspace}
\newcommand\lilith{\textsf{Lilith}\xspace}
\newcommand\nulike{\textsf{nulike}\xspace}

\newcommand\diver{\textsf{Diver}\xspace}

\newcommand\xx{\raisebox{0.2ex}{\smaller ++}\xspace}
\newcommand\Cpp{\textsf{C\xx}\xspace}

\newcommand\YAML{\textsf{YAML}\xspace}

\newcommand\beq{\begin{equation}}
\newcommand\eeq{\end{equation}}

\usepackage{nicefrac}

\title{ColliderBit: a GAMBIT module for the calculation of high-energy collider observables and likelihoods}

\author{The GAMBIT Collider Workgroup:
Csaba~Bal\'azs\thanksref{inst:a,inst:k} \and
Andy~Buckley\thanksref{inst:b,e1} \and
Lars~A.~Dal\thanksref{inst:n} \and
Ben~Farmer\thanksref{inst:c} \and
Paul~Jackson\thanksref{inst:d,inst:k} \and
Abram~Krislock\thanksref{inst:n} \and
Anders~Kvellestad\thanksref{inst:f,e2} \and
Daniel~Murnane\thanksref{inst:d,inst:k} \and
Antje~Putze\thanksref{inst:j} \and
Are~Raklev\thanksref{inst:n,e3} \and
Christopher~Rogan\thanksref{inst:g} \and
Aldo~Saavedra\thanksref{inst:l,inst:k} \and
Pat~Scott\thanksref{inst:h,e4} \and
Christoph~Weniger\thanksref{inst:i} \and
Martin~White\thanksref{inst:d,inst:k,e5}}

\institute{%
  \monash\label{inst:a} \and
  \coepp\label{inst:k}\and
  \glasgow\label{inst:b} \and
  \oslo\label{inst:n} \and
  \okc\label{inst:c} \and
  \adelaide\label{inst:d} \and
  \nordita\label{inst:f} \and
  \annecy\label{inst:j} \and
  \harvard\label{inst:g} \and
  \sydney\label{inst:l} \and
  \imperial\label{inst:h} \and
  \grappa\label{inst:i}
}

\thankstext{e1}{andy.buckley@glasgow.ac.uk}
\thankstext{e2}{anders.kvellestad@nordita.org}
\thankstext{e3}{ahye@fys.uio.no}
\thankstext{e4}{p.scott@imperial.ac.uk}
\thankstext{e5}{martin.white@adelaide.edu.au}

\titlerunning{ColliderBit}
\authorrunning{GAMBIT Collider Workgroup}

\date{Received: date / Accepted: date}

\begin{document}

\maketitle

\begin{abstract}
We describe \colliderbit, a new code for the calculation of high energy collider observables in theories of physics beyond the Standard Model (BSM). \colliderbit features a generic interface to BSM models, a unique parallelised Monte Carlo event generation scheme suitable for large-scale supercomputer applications, and a number of LHC analyses, covering a reasonable range of the BSM signatures currently sought by ATLAS and CMS. \colliderbit also calculates likelihoods for Higgs sector observables, and LEP searches for BSM particles. These features are provided by a combination of new code unique to \colliderbit, and interfaces to existing state-of-the-art public codes. \colliderbit is both an important part of the \gambit framework for BSM inference, and a standalone tool for efficiently applying collider constraints to theories of new physics.
\end{abstract}

\tableofcontents

\section{Introduction}
Despite decades of searches for physics beyond the Standard Model
(BSM), we still lack an unambiguous discovery of such
physics. The many null results from the Large Hadron Collider (LHC) and other
experiments allow us to constrain, to various degrees, the parameter spaces of
many extensions of the Standard Model (SM).  These include
effective theories and simplified models of dark matter, supersymmetric
theories, theories with extra space dimensions and composite Higgs
models. Because even the most minimal realistic theories of BSM physics
have observable consequences in multiple experiments, it is particularly
important to combine collider exclusions with other experiments in a
statistically rigorous way if one is to draw sound conclusions on the viability
of a theory.

Rigorously taking into account the sum of data relevant to a given model from
the many disparate experimental sources has become a challenging task. This
problem is addressed in \gambit (the Global And Modular
Beyond-the-Standard-Model Inference Tool) \cite{gambit}, which combines calculations of observables and
likelihoods in collider, flavour, dark matter and precision physics with a model
database, a flexible system for interfacing to external codes, and a wide
selection of different statistical methods and parameter scanning algorithms
that can be applied to the models~\cite{gambit}. In this paper, we introduce \colliderbit, a
\gambit module for the application of high-energy collider constraints to BSM
physics theories.

The ATLAS and CMS experiments~\cite{Aad:2008zzm,Chatrchyan:2008aa} have made
great progress in the search for evidence of BSM physics at high energies, but applying these constraints to a generic theory of such physics
remains challenging. Searches for new particles at the LHC are typically
presented either in specific planes of a restrictive high scale physics
hypothesis, {\it e.g.}\ the constrained minimal supersymmetric model (CMSSM), or
in simplified models that strictly apply only to a very small volume of the
total allowed space of particle masses and branching ratios. The computational
expense of simulating signal processes for hundreds of thousands of points in a
candidate model prevents an extended treatment by the experiments. In addition,
some LEP results remain useful~\cite{L3:gauginos,L3:gauginos_smallDeltaM,L3:sleptons_squarks,ALEPH:gauginos,Heister:2001nk,Heister:2002jca,ALEPH:sleptons_gauginos,ALEPH:squarks,OPAL:gauginos,OPAL:sleptons_gauginos,OPAL:squarks,DELPHI:SUSY}, and are not always rigorously applied
in the literature. Finally, the discovery of an
SM-like Higgs boson \cite{Aad:2012tfa,Chatrchyan2012} by the ATLAS and CMS experiments in 2012 --- and the subsequent measurement of its properties --- provides tight
constraints on variations in the Higgs branching ratios, which must be included in
any thorough exploration of a BSM physics
model. Given the ever-growing list of
constraints on BSM physics from experiments at the LHC, the need to
rigorously test those limits against various models is ever more pressing.

Partial solutions to each of these issues exist, but there is as yet no
comprehensive tool that tackles all of them. The package \smodels applies
constraints to supersymmetric (SUSY) models based on a combination of simplified
model results~\cite{Kraml:2013mwa}. \fastlim provides similar functionality for
SUSY models, but is extendible (in principle) to non-SUSY models through the use
of user-supplied efficiency tables~\cite{Papucci:2014rja}. Both of these tools
will provide limits that are much more conservative than a more rigorous
calculation, due to the limitations of simplified models.
\textsf{SUSY-AI}~\cite{SUSYAI} provides a random forest classifier for SUSY
models based on LHC exclusions, but as seen in earlier applications of machine
learning to this problem~\cite{2010JHEP...04..109D, SBCoverage,
  2012CoPhC.183..960B, LHC-FASER, Farmer13}, accuracy concerns exist when
applying the method to large-volume parameter spaces, due to the relative
sparsity of the training data~\cite{ATLAS15} in the model parameter space. Other approaches to SUSY model exclusion based on machine learning can be found in~\cite{Bornhauser:2013aya,Bechtle:2017vyu}. \checkmate provides a customised version of the \delphes
detector simulation, an event analysis framework and a list of ATLAS and CMS
analyses that can be used to apply LHC limits, and includes an interface to
\MGaMCNLO for event
generation~\cite{Drees:2013wra,deFavereau:2013fsa,Ovyn:2009tx,Alwall:2014hca}. However,
the time required to run a single BSM parameter combination through \checkmate
makes large-scale parameter scans a difficult prospect, and integration with a
global fitting framework is not within the scope of the package. To the best of our
knowledge, no general purpose tool exists to apply LEP BSM search limits,
although many theorists have implemented their own local codes over the
years. Packages such as \higgsbounds~\cite{Bechtle:2008jh,Bechtle:2011sb,Bechtle:2013wla,2015EPJC...75..421B},
\higgssignals~\cite{HiggsSignals} and \lilith~\cite{Lilith} allow the user to apply constraints on Higgs physics.

As \colliderbit is designed within the \gambit framework~\cite{gambit}, it
offers seamless integration with modules that provide statistical
fitting~\cite{gambit,ScannerBit}, the ability to impose constraints from
electroweak precision data~\cite{SDPBit}, flavour physics~\cite{FlavBit} and a
large range of astrophysical observations~\cite{DarkBit}. For LHC physics, we
use a combination of parallelised Monte Carlo (MC) simulation and fast detector
simulation to recast LHC limits without the approximations of the simplified
model approach. The first release of the code comes with a list of ATLAS and CMS
analyses that collectively present strong constraints on supersymmetry and
dark matter scenarios~\cite{ATLAS:2bStop_20invfb,ATLAS:2LEPStop_20invfb,ATLAS:0LEPStop_20invfb,ATLAS:2LEPEW_20invfb,ATLAS:3LEPEW_20invfb,ATLAS:1LEPStop_20invfb,ATLAS:0LEP_20invfb,CMS:3LEPEW_20invfb,CMS:1LEPDM_20invfb,CMS:2LEPDM_20invfb,CMS:MONOJET_20invfb}. It
contains interfaces to the \pythiaeight MC event
generator~\cite{Sjostrand:2006za,Sjostrand:2014zea}, to the \delphes detector
simulation~\cite{Ovyn:2009tx,deFavereau:2013fsa}, and a customised detector simulation based on four-vector smearing
(\buckfast).  In this paper we show that that \buckfast gives comparable results to \delphes, but
at a dramatically lower CPU cost. We also supply custom routines for
re-evaluating LEP limits on supersymmetric particle production, and include
interfaces to \higgsbounds and \higgssignals for calculating Higgs
observables. \colliderbit follows the modular design of \gambit, thus enabling
the user to easily swap components (e.g. choose a different detector simulation
without affecting the LHC analysis framework), add new collider analyses, or provide interfaces to standard particle physics tools.

This paper serves as both a description of the physics and design strategy of
\colliderbit, and a user manual for the first code release. In
Appendix~\ref{sec:quickstart}, we provide a quick start guide for users keen to
compile and use the software out of the box. Sec.~\ref{sec:code} describes
the physics and implementation of the \colliderbit software. The \colliderbit
user interface is outlined in
Sec.~\ref{sec:interface}. 
In Sec.~\ref{sec:examples} we cover two use cases: First, we point to an annotated \gambit input file that details the application of collider constraints in a scan of the constrained minimal supersymmetric standard model (CMSSM). 
Second, we provide a detailed example of how the user can add their own model to the \colliderbit code. 
The second of these examples shows the flexibility of \colliderbit in tackling generic theories supplied by the user, using existing codes for automatic generation of matrix elements.  
After summarising in Sec.\ \ref{sec:conclusions}, we also provide Appendices \ref{sec:colliderbitclasses} and \ref{glossary}, where we detail the \Cpp classes defined by \colliderbit, and a glossary of common \GB terms, respectively.

\colliderbit is released under the terms of the 3-clause BSD license\footnote{\href{http://opensource.org/licenses/BSD-3-Clause}{http://opensource.org/licenses/BSD-3-Clause}.  Note that \textsf{fjcore} \cite{Cacciari:2011ma} and some outputs of \flexiblesusy \cite{Athron:2014yba} (incorporating routines from \SOFTSUSY \cite{Allanach:2001kg}) are also shipped with \GB \textsf{1.0}.  These code snippets are distributed under the GNU General Public License (GPL; \href{http://opensource.org/licenses/GPL-3.0}{http://opensource.org/licenses/GPL-3.0}), with the special exception, granted to \GB by the authors, that they do not require the rest of \GB to inherit the GPL.}, and can be obtained from \href{http://gambit.hepforge.org}{gambit.hepforge.org}.

\section{Physics and implementation}
\label{sec:code}
To perform any calculations, \colliderbit requires numerical values for the free
parameters of a theory for new physics. If \colliderbit is run with other
\gambit modules, these will come from a scanning algorithm implemented in the
\scannerbit~\cite{ScannerBit} module, and other \gambit modules will then
perform the necessary spectrum generation and decay rate calculations. The user
may also run \colliderbit as a standalone code, in which case the parameters can
be supplied via a model description, such as an SLHA file for supersymmetric
models~\cite{Skands:2003cj,Allanach:2008qq}. In this case, the user must supply
spectrum and/or decay calculations as appropriate. The \colliderbit output is a
series of signal event rate predictions and likelihood terms derived from BSM searches at the LHC, as well as likelihood terms from SUSY searches at LEP and Higgs searches at LEP, the Tevatron and the LHC.
The terms may then be combined according to the user's request, to form a composite
likelihood. Here we describe the strategy for calculating each individual likelihood term,
along with the code implementation.

\subsection{LHC likelihood calculation}
\subsubsection{Overview of LHC constraints included in \colliderbit}
As the flagship collider at the energy frontier, the LHC provides the most
stringent constraints on BSM physics models in the majority of cases. The
search groups of the ATLAS and CMS experiments provide long
lists of results using data from LHC proton--proton collisions taken at
$\sqrt{s}=7, 8$ and $13$\,\TeV, including searches for specific particles
encountered in BSM physics models, and generic resonances in a multitude of
final
states~\cite{ATLAS-SUSY-pub,ATLAS-Exotics-pub,CMS-SUSY-pub,CMS-Exotics-pub}.

Implementing the full list of LHC constraints is a daunting task. The initial approach taken in \colliderbit is to provide a
representative set of searches that run out-of-the-box, supplemented by a
framework that makes it easy to add new LHC analyses. \colliderbit includes a selection of Run I and Run II LHC analyses, chosen for
their relevance to supersymmetry and dark matter simplified model
applications.

The \textbf{Run I} analyses included are:
\begin{itemize}
\item \emph{ATLAS 0-lepton supersymmetry search: }This targets squark and gluino
  production, and is the most constraining single ATLAS SUSY analysis in cases
  where the gluino, some or all squarks are expected to be light. The
  analysis looks for an excess of events in various signal regions defined by
  the jet multiplicity, the missing energy and other kinematic
  variables~\cite{ATLAS:0LEP_20invfb}.
\item \emph{ATLAS and CMS third generation squark searches: }It is possible for
  supersymmetry to remain a natural theory with only the third generation
  squarks accessible at LHC energies. In the limit of large stop mixing,
  only one squark may be light enough to be observed. Increasing theoretical
  interest in naturalness has prompted a series of optimised searches for top
  squarks in recent years, focussing primarily on stop decays to a top quark and
  the lightest neutralino, or to $b$ quarks and charginos with subsequent chargino
  decay via an on- or off-shell gauge boson. \colliderbit includes ATLAS
  searches for top squarks in 0-lepton, 1-lepton and 2-lepton final
  states~\cite{ATLAS:0LEPStop_20invfb,ATLAS:1LEPStop_20invfb,ATLAS:2LEPStop_20invfb},
  and the CMS 1- and 2-lepton
  searches~\cite{CMS:1LEPDM_20invfb,CMS:2LEPDM_20invfb}. We also include the
  ATLAS $b$-jets plus MET search~\cite{ATLAS:2bStop_20invfb}, which targets direct
  sbottom production. All of these searches are also expected to strongly
  constrain simplified dark matter models with a mediator that couples
  preferentially to third generation fermions. These models, which have gained
  popularity as explanations of the \textit{Fermi}-LAT Galactic Centre excess \cite{Berlin:2014tja}, give rise
  to similar final states and, indeed, the CMS searches that we implement have
  already been used in a non-supersymmetric dark matter context \cite{Buckley:2014fba}.
\item \emph{ATLAS and CMS multilepton supersymmetry searches: } In the case that
  all coloured superpartners are too heavy to observe at the LHC, electroweak
  gaugino searches are the only hope of finding evidence for supersymmetry. Even
  if coloured superpartners are accessible, direct searches for the electroweak
  gauginos would provide extra information on the parameters of the neutralino
  and chargino mixing matrices, in addition to telling us whether the weak
  gaugino sector is that of the MSSM, or an expanded sector from an exotic
  supersymmetric scenario. \colliderbit includes the 2- and 3-lepton ATLAS
  electroweak gaugino searches~\cite{ATLAS:2LEPEW_20invfb,ATLAS:3LEPEW_20invfb}
  and the CMS 3-lepton electroweak gaugino search~\cite{CMS:3LEPEW_20invfb},
  which should provide the dominant constraints on the electroweak sector of the
  MSSM.
\item \emph{Dark matter searches: }The classic technique for searching for dark
  matter at colliders is to look for events with a monojet plus missing energy.
  This signature results from pair production of a dark matter candidate,
  with the jet arising from QCD radiation. \colliderbit includes the CMS monojet
  search \cite{CMS:MONOJET_20invfb}, which provides a constraint on various dark matter scenarios, in
  addition to supersymmetric scenarios with compressed spectra. Some caution
  must be taken when applying this to e.g.~dark matter effective field
  theories. In cases where NLO QCD effects are significant, the user will need
  to interface GAMBIT to a suitable Monte Carlo generator capable of modelling
  these effects.
\end{itemize}

We also provide the ATLAS and CMS \textbf{Run II} (13 TeV) 0-lepton supersymmetry searches, based on 13 fb$^{-1}$ of analysed data~\cite{ATLAS-CONF-2016-078,CMS-PAS-SUS-16-014}. More \textbf{Run II} analyses will be added to \colliderbit in the near future, including searches sensitive to R-parity violating supersymmetry such as Ref.~\cite{Aaboud:2017faq}.

We consider this a reasonable minimum of LHC searches for covering a wide range of LHC phenomenology, but the average user will no doubt be keen to expand the collection. New analyses will be continuously added to the code repository, and information on how the user can add a new LHC analysis to \colliderbit is given in Sec.~\ref{sec:LHC_simulation_cap}.  It is worth noting, however, that the general treatment of the LHC analyses in \colliderbit means that even the LHC Run I results can provide previously unavailable insights when used to constrain models with large parameter spaces. We also emphasise that the above list does not contain searches for SUSY scenarios with compressed sparticle spectra. Since we use the LO Pythia generator in the current \colliderbit release, we would obtain less precise results than the ATLAS and CMS publications that use \MGaMCNLO to explicitly model initial state jet radiation through the addition of the relevant diagrams to the tree-level sparticle production process.

In the rest of this section, we describe the process by which LHC analysis
constraints are derived without employing any model-dependent assumptions, following a full simulation of
proton--proton collisions, including detector effects and an approximation of
the ATLAS and CMS statistical procedures.

\subsubsection{Strategy for applying LHC constraints without model-dependent assumptions}

A parameter point of a specified BSM model can in principle be expected to show
up in a variety of LHC BSM searches. For counting analyses, the relevant data to
model are the number of events that pass kinematic selection criteria (for
brevity referred to in what follows as `cuts') imposed in each
analysis. If a model predicts that $s$ signal events will pass the cuts for a
given signal region, and $b$ background events are expected from known SM
processes, the likelihood of observing $n$ events is given by the standard
Poisson formula,
\begin{equation}
\label{poisson_likelihood}
\mathcal{L}=\frac{e^{-(s+b)}(s+b)^n}{n!}.
\end{equation}
For now, we neglect effects of systematic uncertainties in the signal and background yields---but we return to this point in Sec.~\ref{sec:stats}. LHC BSM search papers provide details of $b$ and $n$ for each signal region, along with the background uncertainty, and some estimate of the signal uncertainty for representative models.

Calculating the likelihood for a given model thus requires an accurate estimate
of $s$.  This is given by
\begin{equation}
\label{eq:s}
s=\sigma \epsilon A L,
\end{equation}
where $\sigma$, $\epsilon$ and $A$ are the process-specific production
cross-section, detector efficiency and acceptance, respectively. $L$ is the
integrated luminosity of data used in the search.

The rigorous way to calculate $s$ is to perform a cross-section calculation at
the highest practically achievable level of accuracy in perturbation theory,
before evaluating the acceptance and efficiency via a Monte Carlo simulation of
the LHC collisions. This is usually augmented by simulating the reconstructed
signatures of the Monte Carlo events in the relevant detector---ATLAS or CMS in
the case of direct BSM searches at the LHC. One can then apply the analysis cuts
for a given LHC search to the results of the detector simulation. An approach
using look-up tables for efficiencies and extrapolations from simplified models
removes the need for time consuming simulation, but tends to give very
conservative results as a consequence. This is because the approach misses
models that do not resemble the simplified models under consideration, but still
have some acceptance to the analysis cuts that are used to generate the
simplified model results. Furthermore, generating look-up tables must be
repeated for the parameter space of every physics model of interest, making it
hard to produce a generic code for the application of LHC constraints.

\begin{figure}[tp]
\centering
\includegraphics[width=\columnwidth]{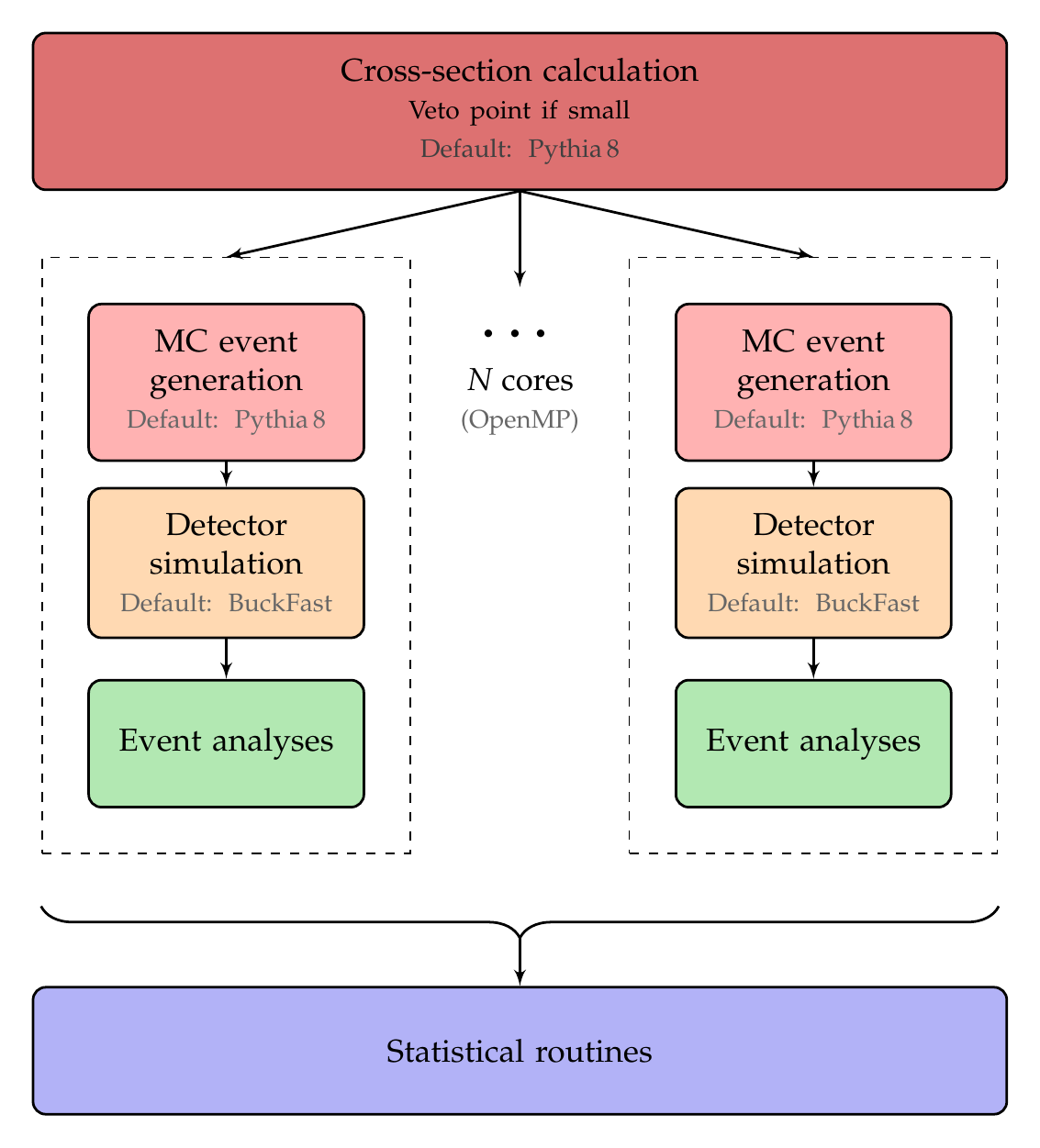}
\caption{Schematic diagram of the \colliderbit processing chain for LHC likelihoods.}
\label{fig:lhcchain}
\end{figure}

The core strategy of the \colliderbit LHC module is instead to make each step of
the simulation chain faster, using a combination of custom speed increases and
parallel computing. The package thus performs a cross-section calculation,
generates Monte Carlo events, performs an LHC detector simulation and then
applies the analysis cuts for a range of LHC analyses, using a custom event
analysis framework. The user can then utilise the \gambit statistical routines
to return LHC likelihoods. The basic processing chain is illustrated in
Figure~\ref{fig:lhcchain}.  The code is designed so that the user can choose,
among available options, which software performs each step of this process, or,
as an alternative, add an interface to an external code in place of an
implemented option. Nevertheless, \colliderbit has a default chain implemented,
and the current version contains the elements summarised in the following
subsections.

\subsubsection{Cross-section calculations}
\label{xsecs}

\colliderbit uses the LO+LL cross-sections calculated numerically by the
\pythiaeight event generator~\cite{Sjostrand:2006za,Sjostrand:2014zea}. For many
models, these are the state-of-the-art. For models where an NLO (or better)
calculation exists, e.g~the MSSM, this is a conservative approximation,
as the $K$-factors are predominantly greater than one.  The LO+LL MSSM
cross-sections are considerably quicker to evaluate than the full NLO results
obtained using e.g. \prospino
\cite{Beenakker:1996ch,Beenakker:1996ed,Beenakker:1997ut}. A single evaluation
of just the strong production cross-sections for a CMSSM benchmark point, with
all relevant processes kinematically available, takes around 15 minutes of CPU
time on a modern processor using \prospino \textsf{2.1} (Intel Core~i5 at
2.6\,GHz). This is clearly unusable in a scan where the evaluation of a single
parameter point must be done in times on the order of a few seconds.
For strong production there exist pre-computed grids of NLO cross-sections with added (N)NLL corrections, which in combination with fast interpolation routines allow accurate cross-sections to be obtained within fractions of a second \cite{Beneke:2016kvz,Beenakker:2016lwe,Kulesza:2008jb,Kulesza:2009kq,Beenakker:2009ha,Beenakker:2010nq,Beenakker:2011fu}. However, these interpolations are limited to models where all squarks except the stops are mass degenerate. While this approximation is suitable for many lower-dimensional parametrisations of the MSSM, it is not sufficiently general to serve as a default MSSM cross-section calculator for \colliderbit.

With the improvement to NLO+NLL, the error from the factorisation and
renormalisation scales has been shown to be as low as 10\%~\cite{Kulesza:2009kq}
for a wide range of processes and masses; however, PDF and $\alpha_s$
uncertainties must be included in the total error budget. These increase with
the sparticle masses because the PDFs are most poorly constrained at large
scales and at large parton $x$. As an example, at 8~TeV \nllfast~\textsf{2.1}~\cite{Kulesza:2008jb,Kulesza:2009kq,Beenakker:2009ha,Beenakker:2010nq,Beenakker:2011fu}
gives errors of $(+24.3\%, -22.2\%)$ and $(+8.3\%, -7.3\%)$, for the PDF and
$\alpha_s$, respectively, using the MSTW2008\,NLO PDF set~\cite{Martin:2009iq},
with gluino and squark masses set to 1.5~TeV. Because 1.5 TeV is at the edge of
the LHC reach at that energy, the total error budget here will not drop
much below 25\% even with NLO+NLL cross-sections.\footnote{With the CTEQ\,6.6M
PDF set~\cite{Nadolsky:2008zw}, the errors increase to $(+63.1\%, -38.5\%)$
and $(+15.6\%, -10.3\%)$; \textit{these} uncertainties will reduce somewhat as PDF fits
including higher-$x$ LHC data become available.}

In light of the above, we take the conservative path of calculating likelihoods
with the LO 
\pythiaeight cross-sections for the LHC. Assigning errors to these
cross-sections is rather meaningless, considering the monotonic nature of LO
scale-dependence, and the fact that the LO cross-sections in BSM models are
known to almost always lie significantly below the NLO and higher order cross-section, sometimes
by as much as a factor of two.\footnote{For a recent thorough exploration of $K$-factors in the MSSM up to approximate NNLO+NNLL order see \cite{Beenakker:2016lwe} and Fig.~2 within.} The LO cross-sections are hence nearly always more conservative than the lower edge of the most pessimistic NLO uncertainty
band due to renormalisation scale systematics.  Accordingly, we do not apply any systematic theory error
to our cross-sections, as any error due to finite statistics in the event generation is dwarfed by the systematic
underestimation of the cross-sections due to the LO approximation.  We have verified that these choices, combined with the approximations used in the event and detector
simulation, result in limits equal to or more conservative than those in the
included ATLAS and CMS analyses (see Sec.~\ref{sec:event_analysis}).

In future releases, we will allow the user to supply cross-sections and associated uncertainties as input to the LHC likelihood calculation, making it possible to calculate them using any preferred choice of external code (known in \gambit as a \cross{backend}).

\subsubsection{Monte Carlo event generation}
\label{MC}

For the \colliderbit event generation, we supply an interface to the
\pythiaeight~\cite{Sjostrand:2006za,Sjostrand:2014zea} event generator,
alongside custom code that parallelises the main event loop of \pythia using
\omp.\footnote{For an earlier similar approach, see Ref.~\cite{Lester:2005je}.}
This substantially reduces the runtime, as seen in Table~\ref{tab:cpucore}.
In a parameter scan with \gambit the parameter sampling is parallelised using \mpi. The additional \omp parallelisation of the LHC likelihood calculation in \colliderbit, along with similarly parallelised calculations in other \gambit modules, helps \gambit's overall scan performance to scale beyond the number of cores that the sampling algorithm alone can make efficient use of.

\begin{table}[t]
  \centering
  \begin{tabular}{lrr}
   \toprule
    \multicolumn{2}{l}{Num. cores \quad\hfill $t$ ($10^5$ events)} & Speed-up\\
    \midrule
    ~1                 & 421~sec & 1\\
    ~4                 & 128~sec & 3.3\\
    ~8                 &  67~sec & 6.3\\
    ~16                &  38~sec & 11.1\\
    ~20                &  33~sec & 12.8\\
    \bottomrule
  \end{tabular}
  \caption{Time taken for the \colliderbit LHC likelihood calculation as a function of the number of cores,
    for 100,000 SUSY events at the SPS1a parameter point~\cite{Allanach:2002nj,AguilarSaavedra:2005pw}, including all sub-processes.
    The processes were run on a single computer node, with ISR, FSR, and
    full hadronisation enabled, but multiple parton interactions and tau decay spin
    correlations disabled.  \gambit was compiled with full optimisation settings (cf.\ Sec.~11 of Ref.~\cite{gambit}).}
  \label{tab:cpucore}
\end{table}

For the purposes of BSM searches, many time-consuming generator components also
add little to the quality of relevant physics modelling, and can therefore be
safely disabled. The single-threaded timing effects of sequentially disabling
``soft physics'' modelling such as multi-parton interactions (MPI), $\tau$ polarisation, QCD
final-state radiation (FSR), and hadronisation are shown for a typical SUSY
model point in Table~\ref{tab:evgencpu}. Of the model components shown, removal
of MPI and tau correlations give the clearest gains. The detailed tau decay
correlation mechanism is not generally relevant for BSM hard processes. LHC
jet reconstruction includes a jet area correction~\cite{Cacciari:2007fd} that
removes the effects of pile-up and MPI on average, so disabling MPI is actually
a \emph{more} appropriate physical configuration than enabling it---and delivers
a 60\% CPU cost saving to boot.

The choices for FSR and hadronisation are less clear: these are responsible for
production of realistic track and cluster multiplicities and energies on which
detector simulation can be run.  Completely disabling FSR---which mainly
produces internal jet structure, not relevant to most BSM analyses---and all
hadron-level processes including both hadronisation and decays, are both rather
drastic options.  In practice there are intermediate alternatives, such as
raising the low-\pt cutoff of FSR evolution to balance CPU cost against physical
accuracy, or to produce physical primary hadrons but elide simulation of their
decays.

By default \colliderbit runs in the mode with MPI and ``sophisticated'' tau
decays disabled; there is potential for further significant speed-up if the
hadron-level or FSR simulation steps can be reduced, perhaps by use of
specialised detector smearing to compensate for the biased final state particle
distributions.

\begin{table}[t]
  \centering
  \begin{tabular}{lrr}
    \toprule
    \multicolumn{2}{l}{Configuration \quad\hfill $t$ ($10^5$ events)} & Speed-up\\
    \midrule
    All                                     & 1,529~sec & 1\\
    ~$\hookrightarrow-$MPI                  &   516~sec & 3.0\\
    ~~~$\hookrightarrow-\tau$ correlations  &   434~sec & 3.5\\
    ~~~~~$\hookrightarrow-$FSR              &   195~sec & 7.8\\
    ~~~~~~~$\hookrightarrow-$hadrons        &   102~sec & 15.0\\
    \bottomrule
  \end{tabular}
  \caption{Single-thread CPU effects of sequentially disabling event simulation components,
    for 100,000 SUSY events at the SPS1a parameter point~\cite{Allanach:2002nj,AguilarSaavedra:2005pw}, including all sub-processes. The
    disabled components have a major effect on CPU, and a minor (sometimes even positive)
    effect on physics performance. The third row corresponds to the first row in Table~\ref{tab:cpucore}. Note that the few percent difference is typical of the variation with local CPU load on the cluster on which this was tested.}
  \label{tab:evgencpu}
\end{table}


This combination of multi-threading and reduced generator functionality allows
generation of 20,000 all-subprocess SUSY events in about 7\,s on an Intel~Core~i7 processor using 8 cores, provided that the compilation makes use of the \textsf{gcc}
option \term{--ffast-math}, or a suitable equivalent. Generating 100,000 events with the same settings and number of cores takes 19\,s. When including FSR and hadronisation, as per the \colliderbit default, the time required to generate 20,000 and 100,000 events increase to 17\,s and 67\,s, respectively.

In the above examples a factor 5 increase in the number of generated events only lead to a factor 2.5--4 increase in the evaluation time. This illustrates that when the number of generated events is fairly low, other parts of the calculation besides the event loop itself account for a noticable fraction of the total evaluation time. The most important contribution comes from the initialisation of \pythiaeight. While this step has not been parallelised, we have optimised the \colliderbit \pythiaeight initialisation so that per-thread copies of the \pythia objects are only constructed at the beginning of a \gambit
sampling run, only requiring re-initialisation of process-specific physics
components for each new model point. This produces a further speed increase in
realistic applications.


In addition, in a \gambit-driven global fit, the event generation for a point can be skipped on the basis of
the initial estimated maximum cross-section. If this is already too low to lead to observable consequences at the LHC, running the event generator is pointless, so skipping that step for some fraction of parameter points gives a further average speed increase. Event generation is also aborted if  \pythia returns an error from the \cpp{pythia.next()} call. In both cases the contribution to the log likelihood, see Sec.~\ref{sec:stats},  is set to zero.

Taken together, these routines make it computationally tractable to run a
full Monte Carlo simulation in a global fit.

The choice of the \pythia generator is an acceptable compromise between
generality and ease of use for the first \colliderbit release. It is sufficient
for many BSM models, and is easily extendable with matrix elements for new
models via the existing \MGaMCNLO interface~\cite{Alwall:2014hca}. For
an example, see Sec.~\ref{sec:standalonemodel}. \pythia will prove insufficient,
however, in cases where NLO corrections are significant---for example in the
accurate treatment of some effective field theories of dark
matter,
where top quark loops become important~\cite{Buckley:2014fba}. These
deficiencies can be addressed in the current release via a user-supplied
interface to an appropriate Monte Carlo tool, and such interfaces will be
supplied in future \colliderbit releases.

\subsubsection{Event record}
\label{event record}

\colliderbit provides a custom set of event record classes that are independent
of the particular choice of event generator or detector simulation. These are: a
\cpp{P4} momentum 4-vector; \cpp{Particle} and \cpp{Jet}, which respectively add
particle ID and flavour-tagging information to \cpp{P4}; and an \cpp{Event}
container. The latter is used to store the particles in discrete categories of
photons, electrons, muons, taus and invisibles, as well as a jet container and a
missing momentum vector.

These event objects should be populated by conversion routines attached to the
interface to each MC generator, allowing the different event structure
conventions of each generator to be treated correctly. The conversion may be
done either at parton or particle level. Parton-level conversion is primarily
intended for speed, as it allows the most CPU-intensive parts of the event
generation to be skipped, at some cost to physical accuracy. The description
below concerns the complete particle-level variant, but the parton-level version
only differs from it in a few minor details.

First, \colliderbit loops over the contents of each event, looking separately
for decayed and stable particles. The former are only used to find $b$ quarks\footnote{We also tested final $b$-hadrons during validation, and found
  that their identification with \colliderbit differed significantly from the
  known performance of ATLAS~\cite{ATLAS-CONF-2016-001}. As experimental flavour-tagging algorithms evolve,
  it will become necessary for the tagging algorithms in \colliderbit to be made
  more configurable.} and hadronic taus for later jet tagging; following the
established \rivet MC analysis system~\cite{Buckley:2010ar}, only stable
particles are used for constructing the kinematics of truth-level events, making
the detector simulation and analysis more robust.

We require identified final-state leptons and photons to be ``prompt'', i.e.
their ancestry is recursively checked to ensure that they have not been produced
(even indirectly) in hadron decays. All final-state particles other than muons
and invisibles are used as inputs to jet finding, which is performed using the
\fastjet~\cite{Cacciari:2011ma} implementation of the anti-$k_t$ jet
algorithm~\cite{Cacciari:2008gp}. We set the anti-$k_t$ $R$ parameter to 0.4 for
Run I ATLAS BSM searches, 0.5 for corresponding CMS analyses, and 0.4 for both ATLAS \& CMS Run II analyses.
We use
$\Delta{R}$ matching between jets and the unstable
tagging objects to set appropriate jet attributes.
\colliderbit computes missing momentum from the vector sum of the momenta of the
invisible final-state particles within a geometric acceptance of $|\eta| < 5$.

The resulting \cpp{Event} is then passed on down the \colliderbit chain:
first for modification by detector simulation, and then in read-only form to
the analysis routines.

\subsubsection{Detector simulation}
\label{sec:detector_sim}

\colliderbit is structured so that the detector simulation is run during the
main parallelised event loop, implicitly speeding up the simulation step. The
user has several options for this step.

\vspace{2ex}

\paragraph{No detector simulation}
The user can choose not to perform any detector simulation, in which case the
truth-level MC events described above are passed directly to the event analysis
framework without modification. Jets may be defined directly at the parton
level, or at the hadron level. The former is only really sufficient for analyses
in which leptons are the main species of interest, in which case turning off
hadronisation can lead to a large speed increase, as seen in
Table~\ref{tab:evgencpu}.

\paragraph{\delphes}
We provide an interface to the \delphes~3.1.2 detector
simulation~\cite{Ovyn:2009tx,deFavereau:2013fsa}, which provides 
simulations of the ATLAS and CMS detectors. \delphes includes a simulation of
track propagation in the magnetic field of an LHC detector, along with a
simulation of the electron and hadron calorimeters, and the muon chambers. The
user can configure the parameters of the simulation using the normal \delphes
mechanism, but it should be noted that $b$- and $\tau$-tagging, and the ATLAS
lepton ID selection efficiencies (``medium'', ``tight'', etc.), are controlled
explicitly within the \colliderbit event analysis codes, to allow different
analyses to use different calibration settings. \delphes has been interfaced
with \colliderbit such that it can be passed single events via memory, rather
than performing several passes over a large sample of MC events in pre-produced
files, as in its usual mode of operation.

\paragraph{\buckfast}
For most purposes, a more approximate approach based on four-vector smearing is
sufficient.  We supply an internal \colliderbit detector simulation, \buckfast,
which uses particle and jet resolution \& efficiency functions based on those in
\delphes, plus parametrised ATLAS electron identification efficiencies. New
parametrisations are being added as Run~2 performance data becomes
public.

\begin{figure*}[p]
\centering
\subcaptionbox{Multiplicity of prompt electrons.}{\includegraphics[width=0.49\linewidth]{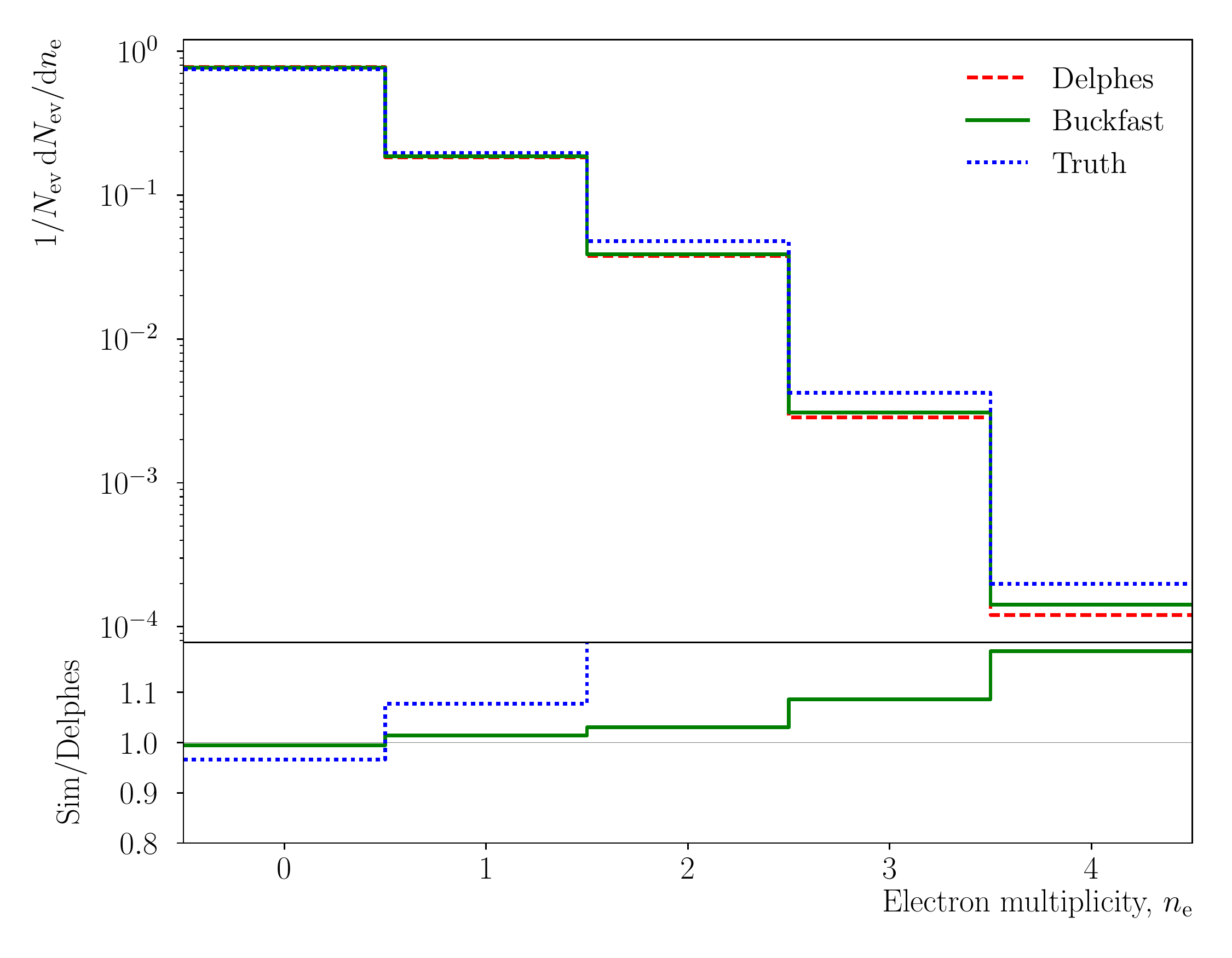}}%
\subcaptionbox{\pt distribution of leading electrons.}{\includegraphics[width=0.49\textwidth]{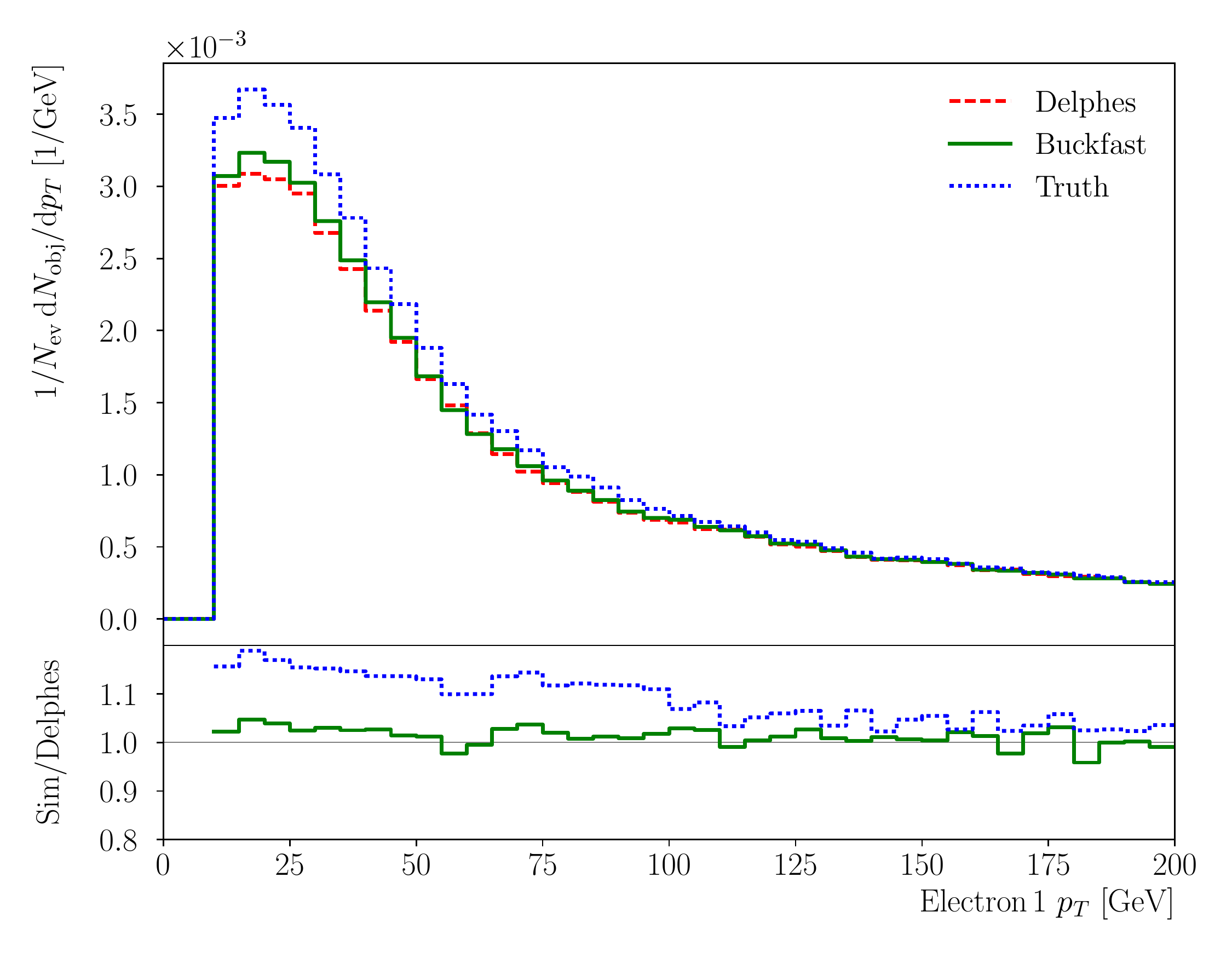}}\\[1em]
\subcaptionbox{\pt distribution of leading muons.}{\includegraphics[width=0.49\textwidth]{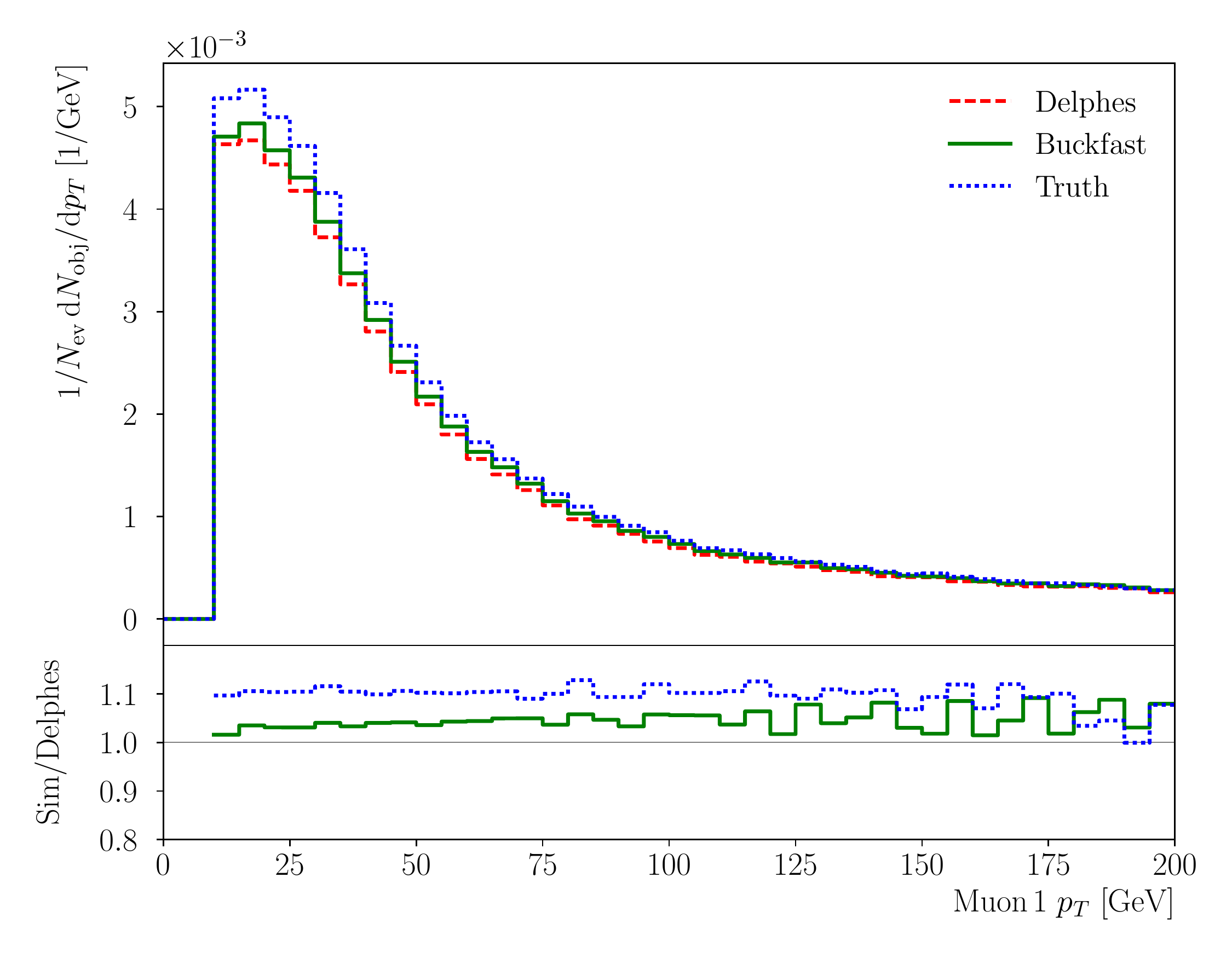}}
\subcaptionbox{\pt distribution of central ($|\eta| < 2.5$) jets.}{\includegraphics[width=0.49\textwidth]{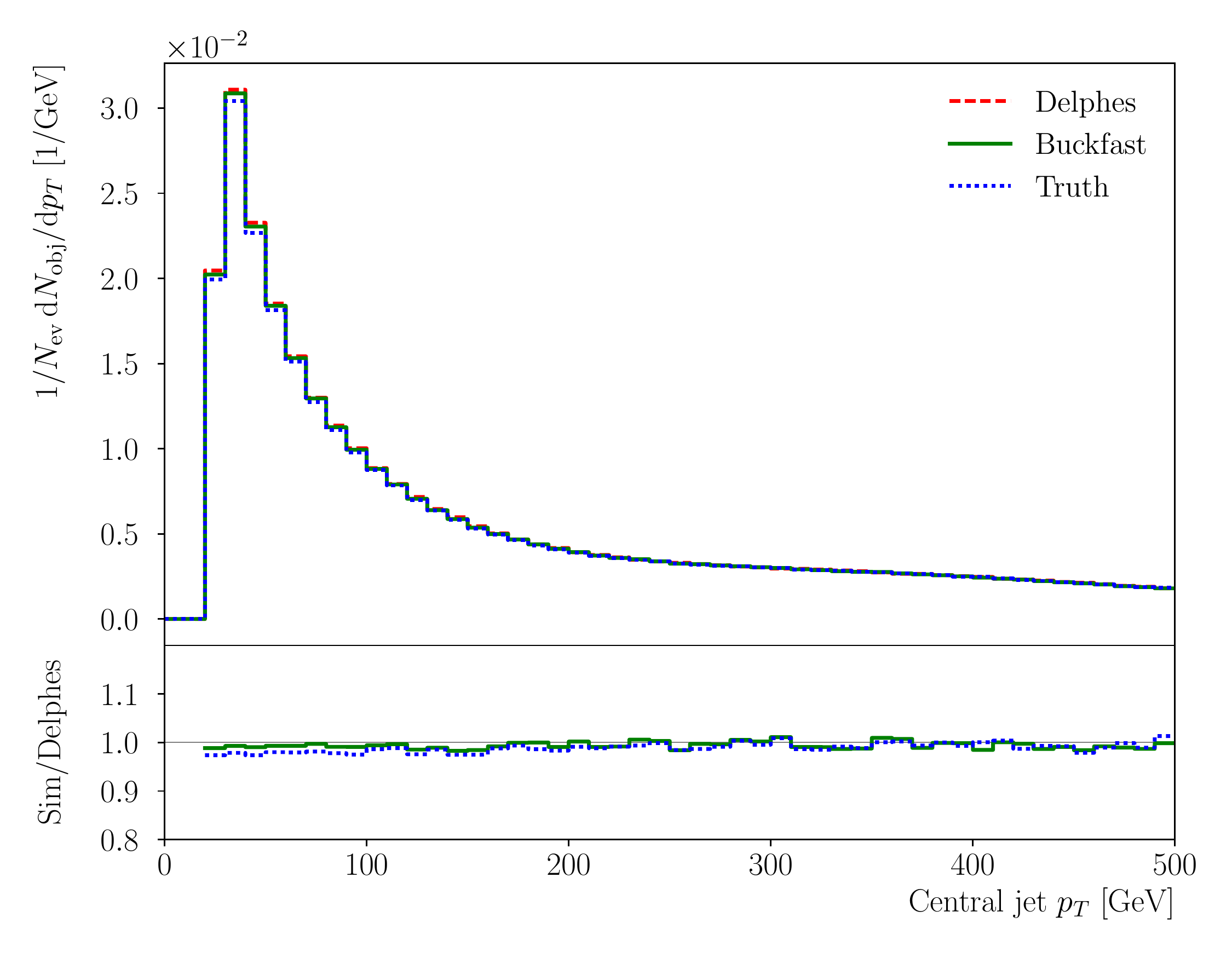}}\\[1em]
\subcaptionbox{\pt distribution of $b$-jets.}{\includegraphics[width=0.49\textwidth]{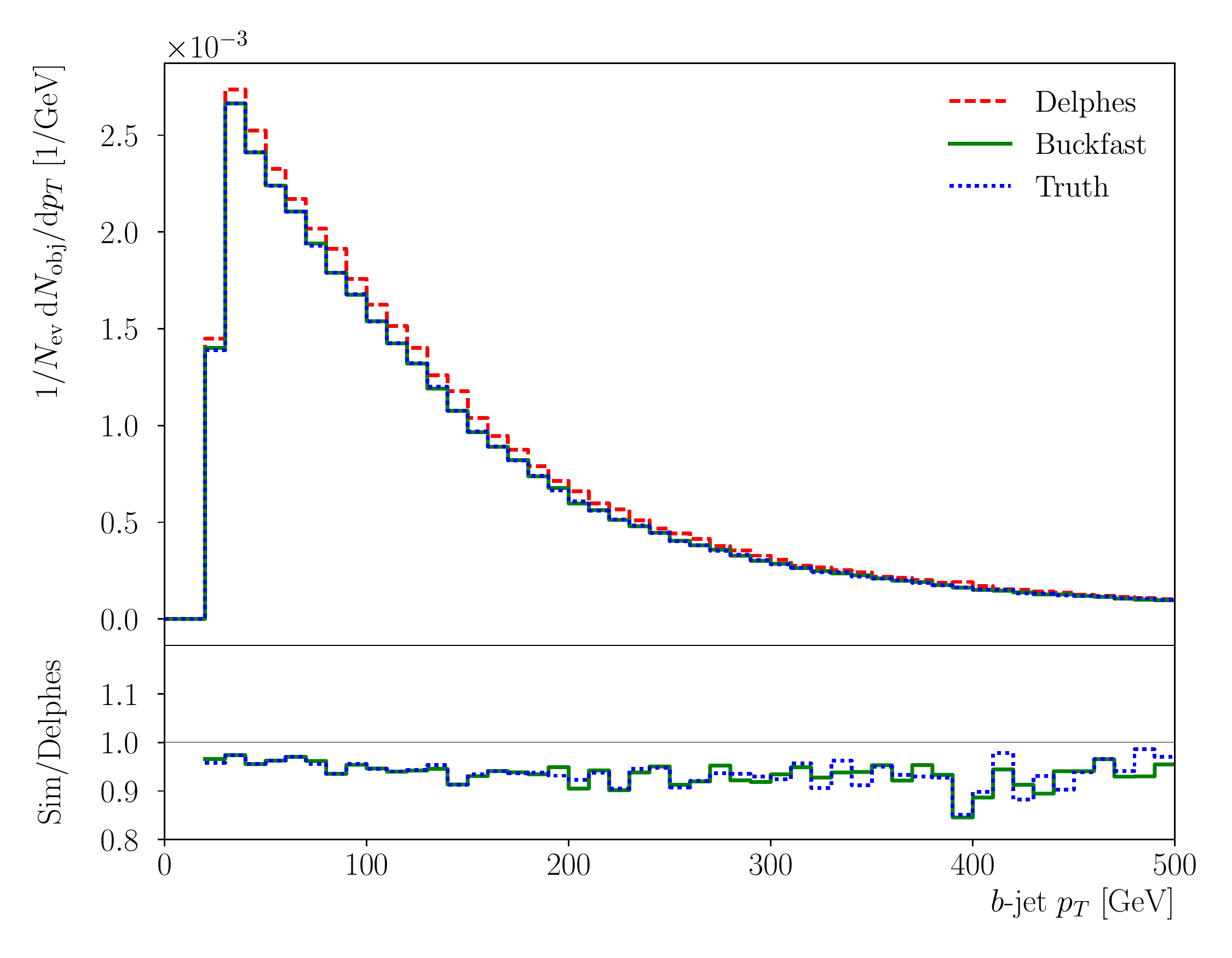}}
\subcaptionbox{Distribution of missing \et.}{\includegraphics[width=0.49\textwidth]{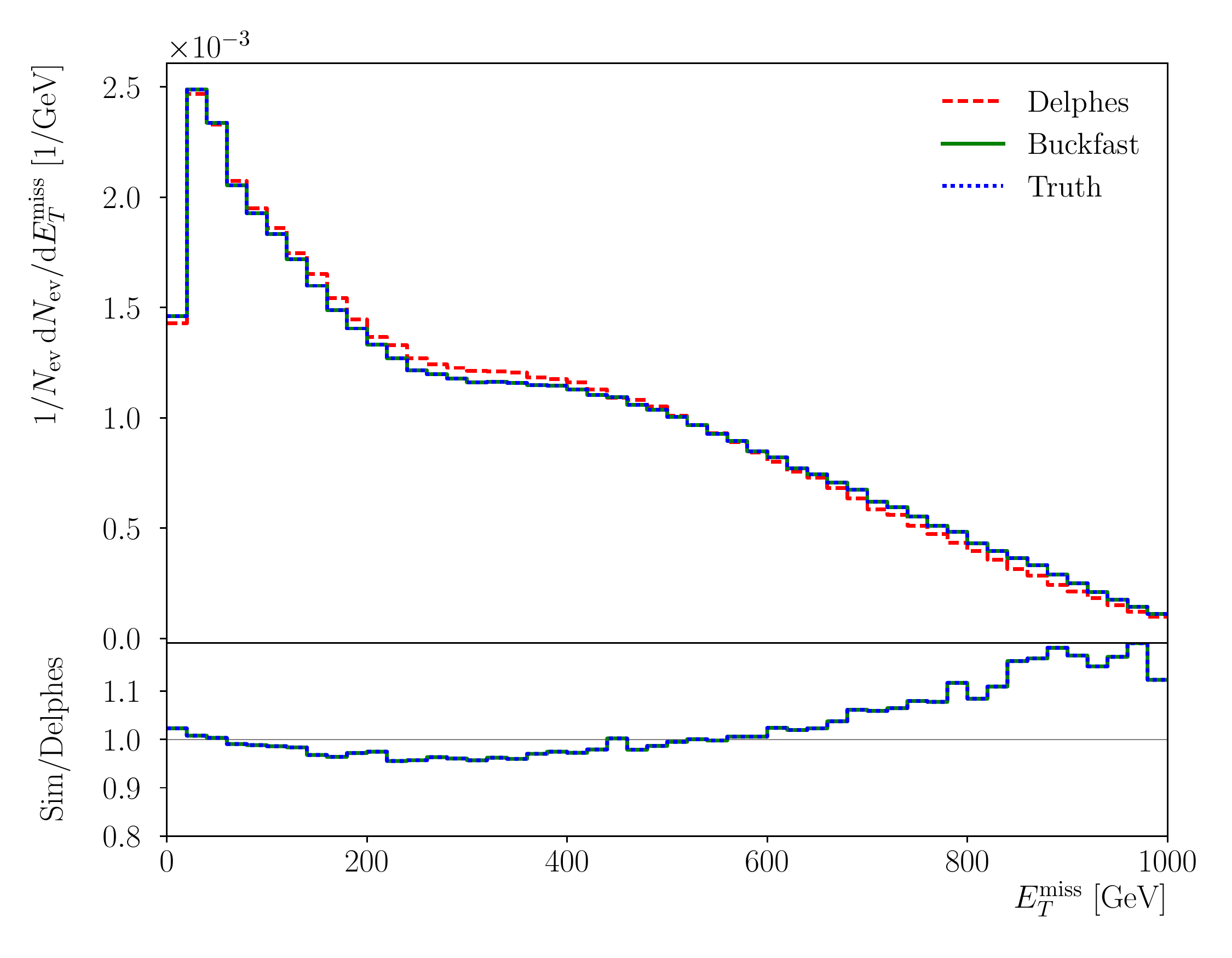}}\\[1em]
\caption{Comparisons of ATLAS event observables between the no-detector
  ``truth'' configuration, \colliderbit's \buckfast 4-vector smearing
  simulation, and the \delphes fast simulation code, for a CMSSM point near the
  current ATLAS search limit (see main text). The ratio plots are computed
  relative to \delphes, to best evaluate the performance of \buckfast.
  }
\label{fig:detsimcmp}
\end{figure*}

The components of the \buckfast simulation are:
\begin{description}
\item\textbf{Electrons: }We apply the \delphes functions for electron tracking
  efficiency, electron energy resolution and electron reconstruction efficiency
  (in that order) to the truth-level electron four-vectors. In the analysis
  step, we apply parametrisations of the ATLAS electron identification
  efficiencies as appropriate, taken from
  Refs.~\cite{ATLAS-CONF-2014-032,ATL-PHYS-PUB-2015-041}.
\item \textbf{Muons: }We apply the \delphes functions for the muon tracking
  efficiency, the muon momentum resolution and the muon reconstruction
  efficiency (in that order) to the truth-level muon four-vectors.
\item \textbf{Taus: }Hadronic taus are identified at truth level. Leptonic taus
  are discarded. For both ATLAS and CMS the hadronic tau momentum is smeared by
  a 3\% Gaussian resolution. Tau tags are applied to jets found within
  $\Delta{R} < 0.5$ of the true hadronic taus. Flat efficiencies are applied to
  tau tagging \emph{in the analysis code} to allow use of different tagging
  configurations within the analyses of each experiment.
\item \textbf{Jets: }Jets are reconstructed at hadron level using the anti-$k_t$
  algorithm, implemented in the \fastjet package.
  All fiducial final-state particles other than invisibles and muons
  are used in jet finding, mimicking typical LHC jet calibration. For both ATLAS
  and CMS the jet momentum is smeared by a 3\% Gaussian resolution chosen for
  compatibility with \delphes' constituent-level smearing.
\item \textbf{$b$-jets: }Truth-level jet tags are obtained by matching jets to
  final $b$-partons for $\Delta{R} < 0.4$; a more robust approach using final
  $b$-hadrons is also available, but by construction agrees less well with the
  parton-based \delphes and LHC Run\,1 tagging calibrations.  As for taus,
  tagging efficiencies and mistag rates parametrised in $\eta$--$\pt$ are
  applied in each analysis code to allow the use of different tagger
  configurations in different analyses.
\item \textbf{Missing energy (MET): }MET is constructed at generator level by
  summing the transverse momenta of invisible particles within the acceptance of
  the detector, and all particles outside the acceptance. No ``soft-term'' MET
  smearing is currently applied, since for events with real hard-process
  invisible particles the ATLAS reconstruction of \etmiss is within a few
  percent of the true value at all scales, and within 1\% above
  70~\GeV~\cite{Aad:2012re}.  The same approach is taken to define the ``truth
  MET'' in the ``no simulation'' mode.
\end{description}
Fig.~\ref{fig:detsimcmp} shows example performance of \buckfast, with
comparisons to \delphes and no-simulation processing.  For this example, we
choose a CMSSM point close to the current ATLAS 95\% exclusion contour,
consistent with the measured Higgs boson mass: $m_0=2000~\GeV, m_{1/2}=600~\GeV, A_0=-4000~\GeV, \tan\beta=30, \mu>0$.

The major effects of detector simulation are seen to be due to lepton efficiencies, with
explicit resolution modelling producing relatively minor effects. \buckfast
and \delphes typically agree to within a few percent for leptons, but some
larger differences remain for $b$-jets (due to truth-tag definition) and
missing \et. The latter is currently unsmeared in \buckfast, but the origin of
the deviation at high-\etmiss is unclear since the reconstructed ATLAS \etmiss
closely matches the truth value in the BSM search region above
70~\GeV~\cite{Aad:2012re}.  The impact of these discrepancies on an ATLAS
analysis dominated by $b$ jets and \etmiss is shown in Table~\ref{tab:cutflow_atlas2b}.

\buckfast is significantly faster than \delphes.  One reason for this is that
the operations it performs are computationally simpler, and should complete in
fractions of a second.  The other, more signficant, reason is that the
\textsf{ROOT} framework on which \delphes is based is not thread-safe, so must
be run serially within an \omp~\cpppragma{critical} block. In contrast,
\buckfast can be run in parallel along with our parallelised version of
\pythiaeight (cf.~Sec.~\ref{MC}), as it has no dependence on \textsf{ROOT}.

\subsubsection{LHC event analysis framework}
\label{sec:event_analysis}

\colliderbit provides a simple analysis framework, built on the event record
classes described in Sec.\ \ref{event record}. Each analysis routine is a \Cpp
class derived from the \cpp{BaseAnalysis} class, which provides the usual
interface of a pre-run \cpp{init} method and an in-run \cpp{analyze} method to
be called on each event. The user can choose which analyses to run in a given
scan directly from the \gambit configuration file. Using the generic
\colliderbit event record classes means that the analyses can be automatically
run on either unsmeared truth records or ones to which detector effects (other
than jet tagging rates) have been applied.

The result of an analysis is a set of \cpp{SignalRegionData} objects.  Each of
these encodes the predicted event counts in a particular signal region of the
analysis, from both signal and background processes.  The signal numbers are
obtained by normalising the yields of simulated events to the integrated
luminosity of the original experimental data analysis. The \cpp{BaseAnalysis}
class provides additional methods for statistically combining analyses (either
equivalent or orthogonal), and for specifying the effective luminosity simulated
in the Monte Carlo step.

\subsubsection{LHC statistics calculations}
\label{sec:stats}

To determine the basic likelihood of observing $n$ events in a certain signal
region, given a signal prediction $s$, we use the marginalised form of
Eq.~\ref{poisson_likelihood} \cite{Conrad03,Scott09c,IC22Methods}.  This allows
us to incorporate systematic uncertainties on the signal prediction ($\sigma_s$)\footnote{We choose to set this term to zero in all analyses in \colliderbit \textsf{1.0.0}, owing to our already conservative use of LO cross-sections, the error from which dwarfs uncertainties arising from finite statistics in event generation; see Sec.\ \ref{xsecs} for more details.  Future versions, employing alternative cross-section calculations, will make more extended use of $\sigma_s$.}
and background estimate ($\sigma_b$) into the calculation, by marginalising over
the probability distribution of a rescaling parameter $\xi$:
\begin{equation}
\label{likelihood}
\mathcal{L}(n|s,b) = \int_0^\infty \frac{\left[\xi(s+b)\right]^{n}e^{-\xi(b+s)}}{n!}P(\xi)\mathrm{d}\xi\,.
\end{equation}

\noindent
Note that the use of a single rescaling parameter is an approximation to avoid the need for a time-consuming 2D integration. The probability distribution for $\xi$ is peaked at $\xi=1$, and has a width characterised by $\sigma_\xi^2 = (\sigma_s^2 + \sigma_b^2) / (s+b)^2$.  The user can choose whether to assume a Gaussian form for this function,
\begin{equation}
\label{gaussian}
P(\xi|\sigma_\xi) = \frac{1}{\sqrt{2\pi}\sigma_\xi} \exp\left[-\frac{1}{2}\left(\frac{1-\xi}{\sigma_\xi}\right)^2\right] \, ,
\end{equation}
or a log-normal form,
\begin{equation}
\label{lognormal}
P(\xi|\sigma_\xi) = \frac{1}{\sqrt{2\pi}\sigma_\xi}\frac{1}{\xi}\exp\left[-\frac{1}{2}\left(\frac{\ln\xi}{\sigma_\xi}\right)^2\right] .
\end{equation}
The \colliderbit default is to use the log-normal version.  This is slower but
more correct, as it does not permit a finite probability for $\xi=0$.  In the
limit of small $\sigma_\xi$, both likelihoods give extremely similar results.
We use the highly optimised implementations of these functions contained in
\nulike \cite{IC22Methods, IC79_SUSY}.

The steps we have described so far allow \colliderbit to calculate the predicted
number of events in any given signal region, defined by a specific set of
observables and kinematic cuts, and to compute the likelihood for that region.
However, certain ATLAS and CMS analyses make use of multiple signal regions,
allowing analysis cuts to be optimised according to the specific characteristics
of each model being tested.  These signal regions may overlap, and so contain
events in common.  The likelihood functions from overlapping signal regions are
therefore not independent.  Ideally, information would be available from the
experiments about the degree to which this overlap occurs, which would allow
\gambit analyses to include all signal regions and their correlations in the
final likelihood for a given analysis.

As this information is not presently available for most analyses, \colliderbit computes the
likelihood for a given analysis on the basis of the signal region
\textit{expected} to give the strongest limit.  It does this individually for
each model, by calculating the expected number of events for every possible
signal region considered in the the original ATLAS or CMS analysis.  It then
chooses the signal region with the maximally negative log-likelihood difference
\begin{equation}
  \label{delta_pred}
  \Delta\ln\mathcal{L}_\mathrm{pred} = \ln \mathcal{L}(n=b|s,b) - \ln \mathcal{L}(n=b|s=0,b).
\end{equation}
This difference is the log of the likelihood ratio between the signal plus
background and background-only predictions, assuming that the observed counts
match the background expectation.  To calculate the likelihood for the analysis
in question, \colliderbit then computes the likelihood of the actual data in the
chosen signal region, and takes the difference with respect to the
background-only expectation in that region, giving an effective log-likelihood
\begin{equation}
  \label{lnl_eff}
  \ln\mathcal{L}_\mathrm{eff} \equiv \Delta\ln\mathcal{L}_\mathrm{true} = \ln \mathcal{L}(n|s,b) - \ln \mathcal{L}(n|s=0,b)\,.
\end{equation}
It is necessary to define the effective log-likelihood in this way because of
the selection step between different signal regions.  Signal regions can in
principle differ markedly in their number of analysis bins and expected numbers
of events, leading to very different effective likelihood normalisations.
Because of this, choosing the signal region on a per-model basis and
then adopting the raw log-likelihood from the selected signal region would
introduce erroneous model-to-model likelihood fluctuations.  Taking the
difference with respect to the background prediction not only removes the
differing (but model-independent) offsets to the log-likelihood from the
different signal regions' typical count rates, but also reduces the effective
degrees of freedom of the resulting likelihood, from $N$ (the number of analysis
bins) to just one.  This puts effective likelihoods from all signal regions on
the same footing, and allows them to be compared correctly across different
points in parameter space.

\begin{table*}
\begin{center}
\begin{tabularx}{\linewidth}{l@{}RRRR}
\toprule
Cut & ATLAS & \gambit & Ratio & \checkmate \\
\midrule
\etmiss + jet \pt cuts & 89.6\% & 91.0\% & 1.02 & 90.8\% \\[0.3ex]
$\dphi_\mathrm{min} > 0.4$ & 81.0\% & 82.5\% & 1.02 & 82.1\% \\[0.1ex]
$\etmiss/\sqrt{\mathstrut \hT} > \!15\,\GeV^{-\nicefrac{1}{2}}$ & 56.0\% & 56.8\% & 1.01 & 54.2\% \\[0.8ex]
$m_\mathrm{eff}^\mathrm{incl} > 1600\,\GeV$ & 31.6\% & 33.4\% & 1.06 & 31.9\% \\
\bottomrule
\end{tabularx}
\caption{The published ATLAS cutflow for the $2jt$ signal region taken
  from Ref.~\cite{ATLAS:0LEP_20invfb}, which searched for squarks and gluinos in
  events with jets and missing transverse momentum. The cutflow is generated for a squark pair-production simplified model (in which a pair of squarks is produced with direct decay to a quark and a lightest neutralino, all other sparticles being decoupled), with $m_{\tilde{q}}=1000$~GeV and $m_{\tilde{\chi}^0_1}=100$~GeV. This is compared with the
  \gambit cutflow obtained using \pythiaeight and \buckfast. Shown are the
  efficiencies for passing each cut (second and third columns), and the ratio of
  efficiencies (fourth column). For reference, we also show the cutflow results obtained using the \checkmate package (taken from their public validation results)~\cite{Drees:2013wra}. }
\label{tab:cutflow_2jt}
\end{center}
\end{table*}

\begin{table*}
\begin{center}
\begin{tabularx}{\linewidth}{LRRRR}
\toprule
Cut & ATLAS & \gambit & Ratio & \checkmate\\
\midrule
$\etmiss > 80$~GeV & 92.4\% & 92.6\% & 1.00 & 93.1\% \\
Lepton veto & 86.6\% & 92.6\% & 1.07 & 90.7\% \\
$\etmiss > 150$~GeV & 76.1\% & 79.1\% & 1.04 & 76.8\% \\
Jet selection & 6.85\% & 8.79\% & 1.28 & 6.46\% \\
$m_{bb}>200$~GeV & 5.52\% & 7.40\% & 1.34 & 4.60\% \\
$M_{CT}>150$~GeV & 4.72\% & 5.93\% & 1.26 & 4.01\% \\
$M_{CT}>200$~GeV & 3.86\% & 4.76\% & 1.24 & 3.32\% \\
$M_{CT}>250$~GeV & 2.93\% & 3.46\% & 1.18 & 2.50\% \\
$M_{CT}>300$~GeV & 2.01\% & 2.34\% & 1.16 & 1.69\% \\
\bottomrule
\end{tabularx}
\caption{The published ATLAS cutflow for SR\,A in
  Ref.~\cite{ATLAS:2bStop_20invfb}, a search for new physics in events with two
  $b$ jets and missing transverse momentum. The cutflow is generated for a bottom squark pair-production simplified model (in which a pair of bottom squarks is produced with direct decay to a $b$-quark and a lightest neutralino, all other sparticles being decoupled), with $m_{\tilde{b}}=500$~GeV and $m_{\tilde{\chi}^0_1}=1$~GeV. This is compared with the \gambit
  cutflow obtained using \pythiaeight and \buckfast. Shown are the percentages of the initial event sample after each cut, and the ratio of the \gambit and \atlas numbers. We also provide the \checkmate cutflow in the final column, taken from their public validation results.}
\label{tab:cutflow_atlas2b}
\end{center}
\end{table*}

This is a conservative approach, but it is the best possible treatment when one
lacks sufficient information to handle correlated data and
systematic uncertainties. When such information is made available (ideally in
a standardised format), \colliderbit will use it for a more complete likelihood
calculation.  Refs.\ \cite{Collaboration:2242860,CMS:2017kmd} are indeed very welcome recent steps in this direction.

To construct a compound likelihood from different \textit{analyses}, we assume
that all analyses have been chosen to be orthogonal, in the sense that they have
disjoint selection criteria and no single event could contribute to the signal
region counts of multiple analyses. This means that their effective
log-likelihoods can be straightforwardly summed; \colliderbit does this for all
analyses selected by the user, and returns the result to the \GB Core as a
final, combined LHC log-likelihood. It is the responsibility of the user to
ensure that they only select mutually orthogonal analyses for combination in a
\colliderbit run.

\subsubsection{Validation of \colliderbit LHC constraints}
We verified the \colliderbit LHC simulation and analysis chain by comparing
cutflows for representative model parameter points against those published by
the LHC experiments. Note that we use \pythiaeight and \buckfast for these
comparisons, so we expect to see the agreement degrade in cases where effects
not included in this chain become important, e.g.~for compressed spectra, where
a more appropriate treatment of initial state radiation is important.

Three sample cutflows are presented in Tables~\ref{tab:cutflow_2jt},~\ref{tab:cutflow_atlas2b} and
\ref{tab:cutflow_model1}, for a jets+MET search, 2$b$+MET search and a dilepton+MET search,
respectively. These show close agreement for most signal regions, rising to no
more than $\sim 50\%$ discrepancy in the worst case. These are a representative
choice of sample cutflows for all signal regions considered. For reference, we also show the publically available \checkmate cutflows where these are available. These confirm the expectation that \buckfast gives respectable performance, but does not match the ATLAS cutflows as closely as the \checkmate package, which runs a heavily tuned version of the \delphes detector simulation. The compromise in performance in \buckfast is of course compensated for by the two-fold increase in speed, resulting from the quicker simulation step, plus the fact that it can be parallelised since it does not rely on the \textsf{ROOT} framework.


\begin{table}
\begin{center}
\begin{tabularx}{\linewidth}{LRRR}
\toprule
Cut & ATLAS & \gambit & Ratio\\
\midrule
$e+e-$ & & & \\
Two leptons & 52.0 & 48.2 & 0.93 \\
Jet veto & 22.4 & 23.2 & 1.04 \\
$Z$ veto & 21.2 & 21.6 & 1.02 \\
SR MT2 90 & 12.7 & 12.6 & 0.99 \\
SR MT2 120 & 9.4 & 9.5 & 1.01 \\
SR MT2 150 & 6.2 & 6.3 & 1.02 \\
\midrule
$\mu^+\mu^-$ & & & \\
Two leptons & 47.8 & 51.2 & 1.07 \\
Jet veto & 20.7 & 25.5 & 1.23 \\
$Z$ veto & 19.3 & 23.8 & 1.23 \\
SR MT2 90 & 11.5 & 13.8 & 1.20 \\
SR MT2 120 & 8.7 & 9.8 & 1.12 \\
SR MT2 150 & 5.7 & 6.6 & 1.16\\
\midrule
$e^{\pm}\mu^{\mp}$ & & & \\
Two leptons & 77.7 & 102.7 & 1.32 \\
Jet veto & 32.4 & 50.8 & 1.6 \\
$Z$ veto & 32.4 & 42.1 & 1.49 \\
SR MT2 90 & 19.1 & 27.2 & 1.42 \\
SR MT2 120 & 14.7 & 20.1 & 1.37 \\
SR MT2 150 & 10.1 & 13.6 & 1.34 \\
\bottomrule
\end{tabularx}
\caption{The published ATLAS cutflow for Model~1 in
  Ref.~\cite{ATLAS:2LEPEW_20invfb}, a search for new physics in events with two
  leptons and missing transverse momentum. This is compared with the \gambit
  cutflow obtained using \pythiaeight and \buckfast. Shown are the numbers of
  events expected in 20.1~\invfb of 8~\TeV ATLAS data, and the ratio of the
  \gambit and \atlas numbers. Note that for the \gambit numbers, we used the same value of the SUSY production cross-section as that assumed in the ATLAS cutflow (and thus our cutflow does not include the effect of the LO cross-section that we use in our SUSY scans). }
\label{tab:cutflow_model1}
\end{center}
\end{table}

To illustrate the effect of changing the \pythiaeight settings on the physics performance of \colliderbit, we show the ATLAS 0 lepton cutflow for four different \pythiaeight configurations in Table~\ref{tab:cutflow_2jt_pythiachecks}. This should be the most strongly affected cutflow, since the settings are all relevant for jet physics. In fact, we observe only a slight degradation of the cutflow performance as various approximations (e.g. removing hadronisation and FSR) are made, which validates the removal of certain \pythiaeight features in the interests of speed. We caution that for models with compressed particle decays where the effects of final state radiation may become more important, this conclusion is not expected to hold, but a thorough investigation is clearly physics-dependent and beyond the scope of this paper.

\begin{table*}
\begin{center}
\begin{tabularx}{\linewidth}{l@{}RRRR}
\toprule
Cut & MPI+FSR+HAD & FSR+HAD & HAD & None \\
\midrule
\etmiss + jet \pt cuts & 91.0\% & 90.7\% & 91.4\% & 91.0\% \\[0.3ex]
$\dphi_\mathrm{min} > 0.4$ & 82.4\% & 82.4\% & 82.7\% & 81.7\% \\[0.1ex]
$\etmiss/\sqrt{\mathstrut \hT} > \!15\,\GeV^{-\nicefrac{1}{2}}$ & 56.8\% & 57.1\% & 57.7\% & 57.0\% \\[0.8ex]
$m_\mathrm{eff}^\mathrm{incl} > 1600\,\GeV$ & 33.0\% & 32.7\% & 33.7\% & 34.2\% \\
\bottomrule
\end{tabularx}
\caption{Reproduction of the same ATLAS 0 lepton cutflow as Table~\ref{tab:cutflow_2jt}, with each column representing different \pythiaeight settings. The baseline in the final ``None'' column has tau spin correlations turned off (since they have no effect for SUSY models in any case), ISR turned on, and hadronisation (HAD), FSR and multiple parton interactions (MPI) turned off. \pythiaeight is configured to produce light squark pairs only, and the parton-level events are reconstructed with the parton \buckfast settings. The first three columns add hadronization, FSR and multiple parton interactions. It is worth noting that none of these configurations match the cutflow configuration in Table~\ref{tab:cutflow_2jt}, which includes a tuning of the minimum $p_T$ threshold for the \pythiaeight~\cpp{TimeShower}.}
\label{tab:cutflow_2jt_pythiachecks}
\end{center}
\end{table*}

\begin{figure}[tp]
\centering
\includegraphics[width=\columnwidth]{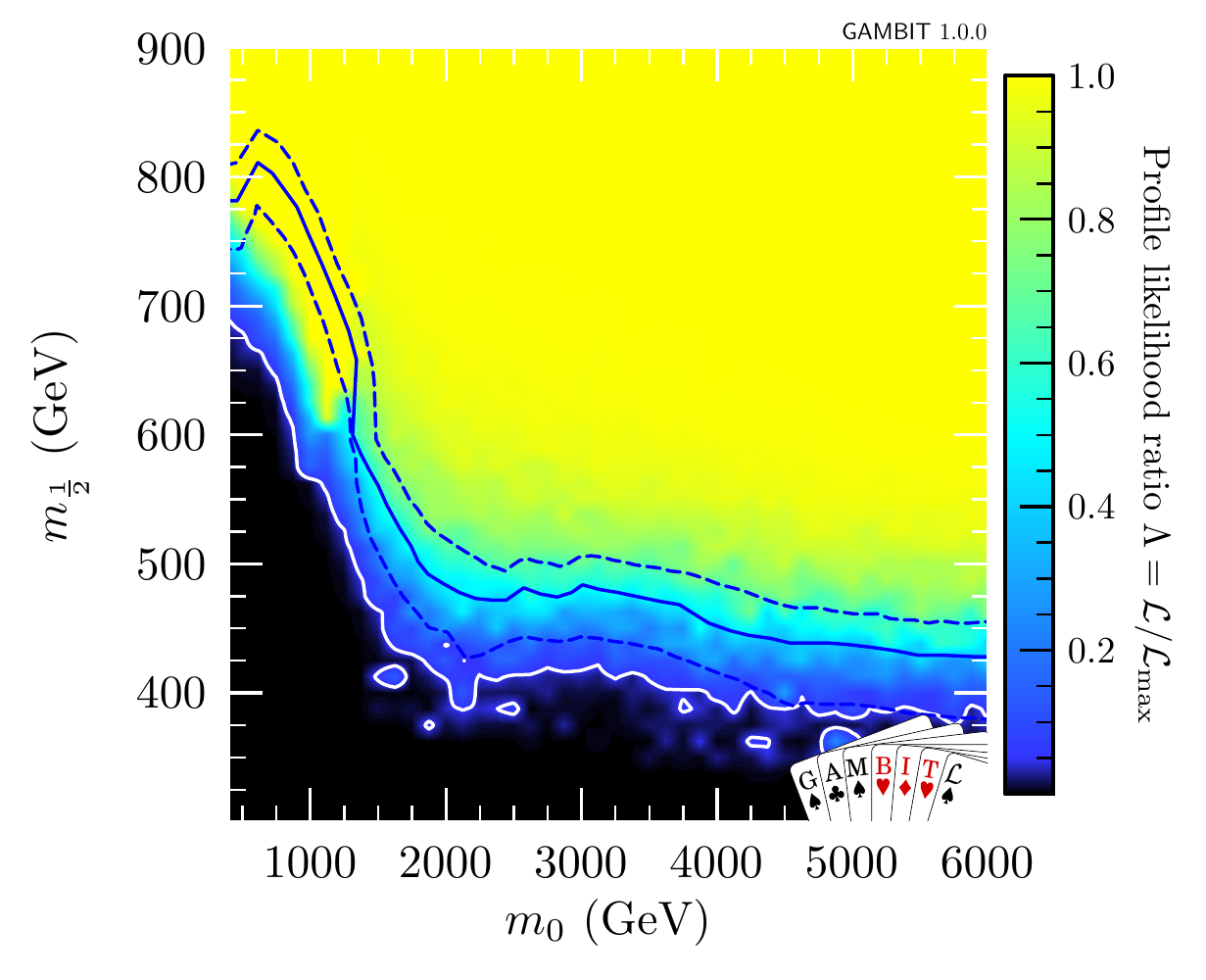}
\caption{Output from a \colliderbit CMSSM grid scan over $m_0$ and $m_{1/2}$ with
  $\tan\beta=30$, $A_0=-2m_0$ and $\mu>0$, using 50 grid points in each direction. The likelihood
  only includes the ATLAS zero lepton SUSY search, with 20,000 MC events
  generated per point. The colour map shows the profile likelihood ratio
  $\mathcal{L}/\mathcal{L}_{max}$ and the solid white line indicates the \gambit
  95\%~\CL exclusion contour, defined by $\mathcal{L}/\mathcal{L}_{max} = 0.05$. The blue solid line shows the ATLAS 95\%~\CL observed
  exclusion limit, taken from Ref.~\cite{ATLAS:0LEP_20invfb}, with the blue
  dashed lines showing the reported $\pm 1 \sigma$ theoretical (cross section)
  uncertainty. \label{fig:atlaslimitdebug}}
\end{figure}

In Figure~\ref{fig:atlaslimitdebug}, we compare the observed ATLAS Run~1 zero lepton
CMSSM 95\% \CL exclusion limit in the $m_0$--$m_{1/2}$ plane
from~\cite{ATLAS:0LEP_20invfb} with a \gambit \colliderbit scan performed with
the same model.  Here $m_0$ and $m_{1/2}$ are free parameters, $\tan\beta=30$,
$A_0=-2m_0$ and $\mu>0$. Since there are only two free parameters we perform a simple grid scan with $50\times50$ grid points. The \colliderbit likelihood includes only the LHC likelihood contribution, which in turn uses only the ATLAS zero lepton analysis, with 20,000 MC events generated per parameter point.
The white solid line show the 95\%~\CL exclusion contour, defined by the likelihood ratio $\mathcal{L}/\mathcal{L}_{max} = 0.05$. For comparison, the observed limit from the ATLAS analysis is plotted as a solid blue line, with dashed blue lines showing the reported $\pm 1 \sigma$ theoretical uncertainty on this limit.

The \colliderbit exclusion limit is more conservative than the ATLAS result, as expected from the different cross-sections used (LO for \colliderbit, NLO+NLL for the ATLAS result). We have checked how our limit would change with NLO+NLL cross-sections from \nllfast~\textsf{2.1} for a number of points close to the observed ATLAS limit. In the region where $m_0 \gg m_{1/2}$ we find close agreement with the ATLAS limit; the rescaled \colliderbit limit ends within the uncertainty band of the ATLAS limit.
In the low-$m_0$ part of the plane, where $m_0 \sim m_{1/2}$, we see a somewhat larger discrepancy with the observed ATLAS limit also after rescaling our results with NLO+NLL cross-sections.  The reason for this discrepancy is that \colliderbit here differ from the ATLAS analysis in what signal region is predicted to have the best expected sensitivity. ATLAS uses the 4-jet region \textsf{4jt} while \colliderbit chooses the 3-jet region \textsf{3j}.
Coincidentally, the \textsf{4jt} region observes a small downwards fluctuation relative to the background expectation, leading ATLAS to set stronger limits than expected, while at the same time there is a small upwards fluctuation in the event count for the region \textsf{3j}, giving a weaker \colliderbit limit than expected. In this part of parameter space the squarks and gluinos are rather close in mass, implying that some of the jets will be soft. The choice between a 3-jet and a 4-jet signal region is therefore likely to be sensitive to the details of jet handling in the different event generators used (\textsf{HERWIG++\,2.5.2} for the ATLAS result and \textsf{Pythia\,8.212} for \colliderbit). We note that if we by hand force \colliderbit to use the \textsf{4jt} region, our limit again agrees nicely with the ATLAS limit after NLO+NLL scaling. For instance, our rescaled limit at $m_0 = 800$~\GeV moves up to $m_{1/2} = 780$~\GeV.

\begin{figure}[tp]
\centering
\includegraphics[width=\columnwidth]{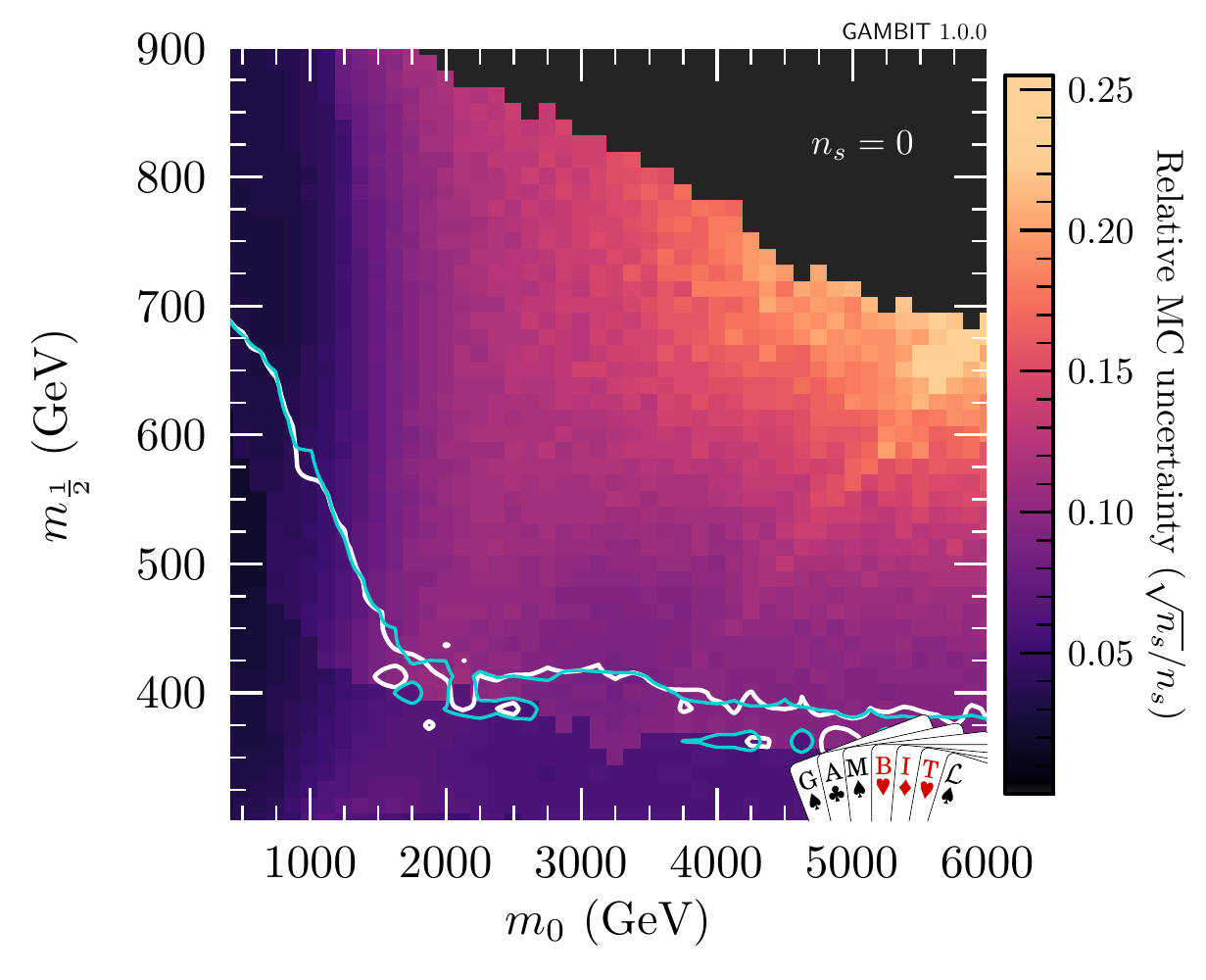}
\caption{The relative MC uncertainty $\sqrt{n_s}/n_s$ for \colliderbit simulations of the ATLAS zero lepton SUSY search across the plane of $m_0$ and $m_{1/2}$, using 20,000 MC events per parameter point. Here $n_s$ is the number of accepted MC signal events. As in Fig.\ \ref{fig:atlaslimitdebug}, the remaining CMSSM parameters are given by $\tan\beta=30$, $A_0=-2m_0$ and $\mu>0$. The solid lines show the \gambit 95\%~\CL exclusion contours obtained using 20,000 (white) and 100,000 (cyan) MC events. \label{fig:atlaslimitMCerr}}
\end{figure}

In the above $50\times50$ grid scan of the $m_0$--$m_{1/2}$ plane we only generated 20,000 MC events per parameter point. This scan completed in less than 80 minutes using 48 CPUs. Clearly, for a low-dimensional scan like this one can afford a much higher number of events per point to reduce the MC uncertainty. But for large global fits in many-dimensional parameter spaces 20,000 events may be a realistic trade-off between speed and accuracy. In Figure \ref{fig:atlaslimitMCerr} we show a colour map of the relative MC uncertainty, $\sqrt{n_s}/n_s$, across the $m_0$--$m_{1/2}$ plane for the \colliderbit simulation of the ATLAS zero lepton search using 20,000 events per point.  Here $n_s$ is the number of accepted MC events for the signal region chosen by \colliderbit for the given parameter point. As in Fig.\ \ref{fig:atlaslimitdebug}, the white line depicts the 95\%~\CL exclusion contour obtained with 20,000 generated events. For comparison the cyan line shows the limit obtained when generating 100,000 events per point.  We see that for the signal regions used to calculate the likelihood for this analysis, the relative MC uncertainty stays below 13\% in the parameter regions around the exclusion limit. At large $m_0$ and $m_{1/2}$ the uncertainty increases as we approach the production threshold. (The apparent cut-off at $\sqrt{n_s}/n_s \sim 0.25$ is due to the grid step size.)

In contrast to the simple grid scan presented in Figures~\ref{fig:atlaslimitdebug} and~\ref{fig:atlaslimitMCerr}, a large global fit will typically involve sampling millions of parameter points. Limited MC event statistics then increases the chance of a few points ending up with a spurious good likelihood due to MC fluctuations. This may in particular affect the result of a frequentist analysis where the preferred parameter regions are determined relative to the best-fit point, as with the likelihood ratio $\mathcal{L}/\mathcal{L}_{max}$ used above. A spurious good likelihood $\mathcal{L}_{max}$ for the best-fit point will result in a falsly strong constraint on the preferred parameter regions. One simple way to ensure conservative and more stable limits in a large scan is to cap the \colliderbit effective log-likehood in Eq.~\ref{lnl_eff} at the value given by the background-only expectation ($s=0$), \ie to force $\ln\mathcal{L}_\mathrm{eff} \leq 0$. This then becomes an ``exclusion only'' likelihood, as all points where $s > 0$ gives an improved fit to the data are assigned the likelihood corresponding to $s=0$. Of course, this method is not appropriate if the aim of the parameter scan is to fit the model to a potential new signal in the data.

\begin{figure}[tp]
\centering
\includegraphics[width=\columnwidth]{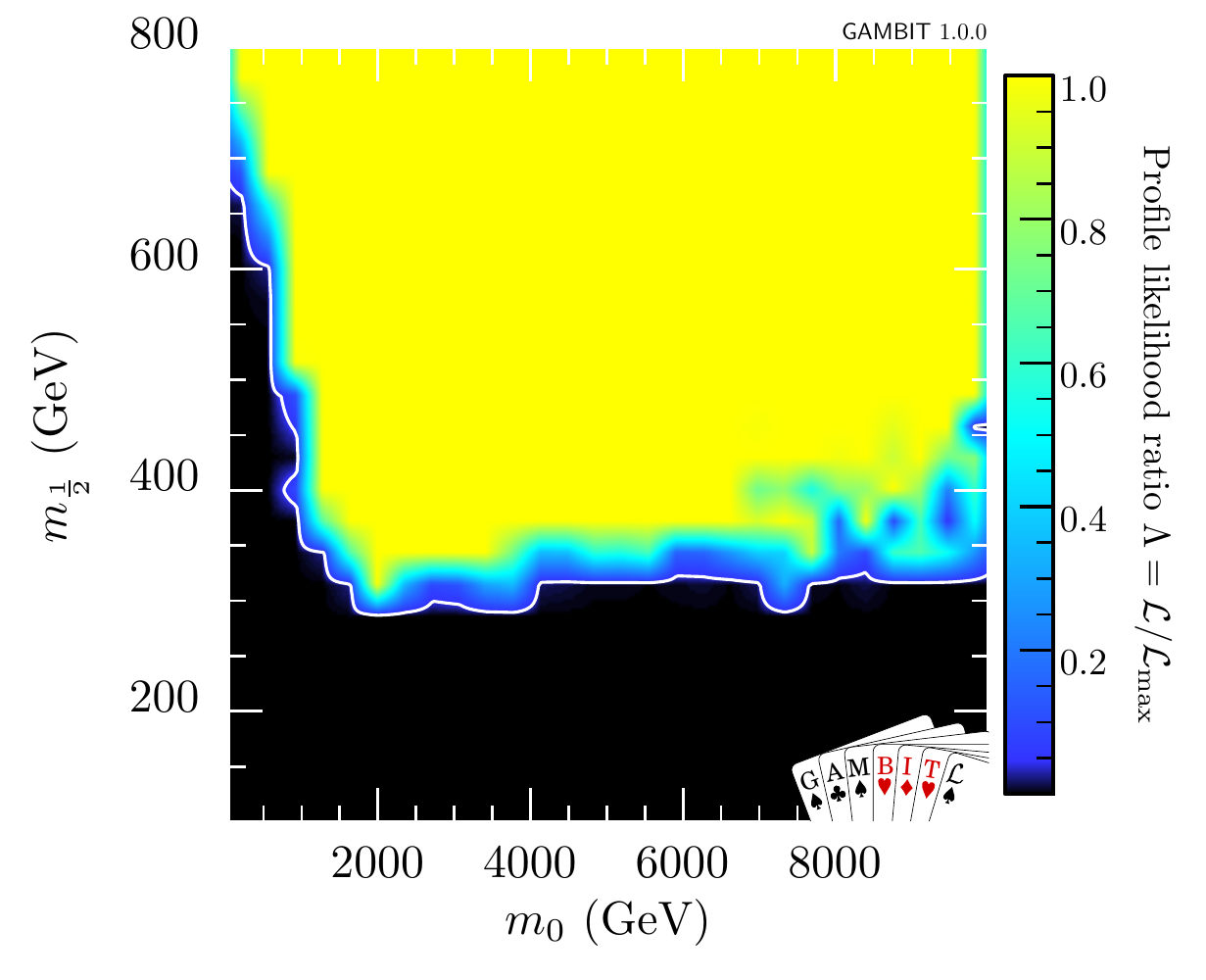}
\caption{Output from a \colliderbit CMSSM scan over $m_0$, $m_{1/2}$,
  $\tan\beta$, and $A_0$, with SM parameters set to default values, using \diver with \GB production settings (\protect\fortran{convthresh} = $10^{-5}$ and \protect\fortran{NP} = 19200). The likelihood includes all of the ATLAS and CMS Run I analyses summarised in the text, but no other contributions. The colour map shows the profile likelihood ratio
  $\mathcal{L}/\mathcal{L}_{max}$ and the solid white line indicate the \gambit
  95\%~\CL exclusion contours at $\mathcal{L}/\mathcal{L}_{max} = 0.05$. \label{fig:lhclimitdebug}}
\end{figure}

The approach of capping the effective likelihood was used for the result shown in Fig.\ \ref{fig:atlaslimitdebug}. We also apply it in the analysis shown in Figure~\ref{fig:lhclimitdebug}. This gives an example of a typical use-case for \colliderbit, in which multiple LHC searches are included in the combined LHC likelihood and a larger parameter space is scanned. Here we show the CMSSM 95\%~\CL exclusion limit in the $m_0$--$m_{1/2}$ plane, following a scan of $m_0$, $m_{1/2}$, $\tan\beta$, and $A_0$ using \diver \cite{ScannerBit} with \GB production settings (\fortran{convthresh} = $10^{-5}$ and \fortran{NP} = 19200), with the SM parameters held to their default values. All of the LHC Run I analyses listed above are included in the LHC combined likelihood, and no other likelihoods are used. One obtains an exclusion contour of similar shape to the ATLAS zero lepton limit for fixed $A_0$ and $\tan\beta$, but it is shifted to lower values of $m_{1/2}$.

\subsection{LEP likelihood calculation}
\label{sec:leplike}
Despite the huge improvement in lower limits provided by high-energy LHC data,
limits from direct searches at the LEP experiments are still important for some
BSM models. This is true in particular for SUSY models that only have
significant production of slepton or neutralino/chargino pairs with masses below
half the maximum LEP centre-of-mass energy. In most available codes, LEP limits
from direct searches take the form of hard lower limits on sparticle masses, at
e.g.\ 95\%~\CL.  This is how such limits are implemented in
\ds~\cite{Gondolo:2004sc} and \micromegas~\cite{micromegas}, for
example. Such limits generally rely on model-dependent assumptions, which are
not always clearly stated.

As an example, in \ds~\textsf{5.1.3} the selectron mass is bounded by
$m_{{\tilde e^{\phantom{*}}}_R}>95$ GeV if
$m_{{\tilde e^{\phantom{*}}}_R}-m_{{\tilde\chi}_1^0}>15$ \GeV,
based on a search by the ALEPH experiment~\cite{Heister:2001nk}, and
$m_{{\tilde e^{\phantom{*}}}_R}>87.1$ \GeV if $m_{{\tilde e^{\phantom{*}}}_R}-m_{{\tilde\chi}_1^0}>5$ GeV,
based on results from the OPAL experiment~\cite{Abbiendi:1999as}. However, if
one looks closely at the details of these limits there are indeed remaining
model assumptions, e.g.\ the ALEPH experiment assumes $\mu = -200$~GeV and
$\tan\beta=2$ for the production cross-section, and a branching ratio
$\BR({\tilde e^{\phantom{*}}}_R\to e {\tilde\chi}_1^0)=1$. In contrast, in an analysis of the
same data using a more general MSSM parameter space (but still assuming gaugino
mass unification, scalar mass unification, no slepton mixing, and negligible
squark mixing), the selectron mass limit becomes 73~GeV~\cite{Heister:2002jca}.
The weakening of this limit is due to possibile cascade decays of the selectron.

In \colliderbit we take a different approach, which is free from model-dependent assumptions, using the direct cross-section limits for sparticle pair production
of sleptons, neutralinos and charginos at LEP. Our approach includes not only model-dependent effects in the cross section, but also in the decay rates, where we make no assumptions on the branching ratios, relying instead on an explicit calculation.

\begin{figure}[tp]
\centering
\includegraphics[width=\columnwidth]{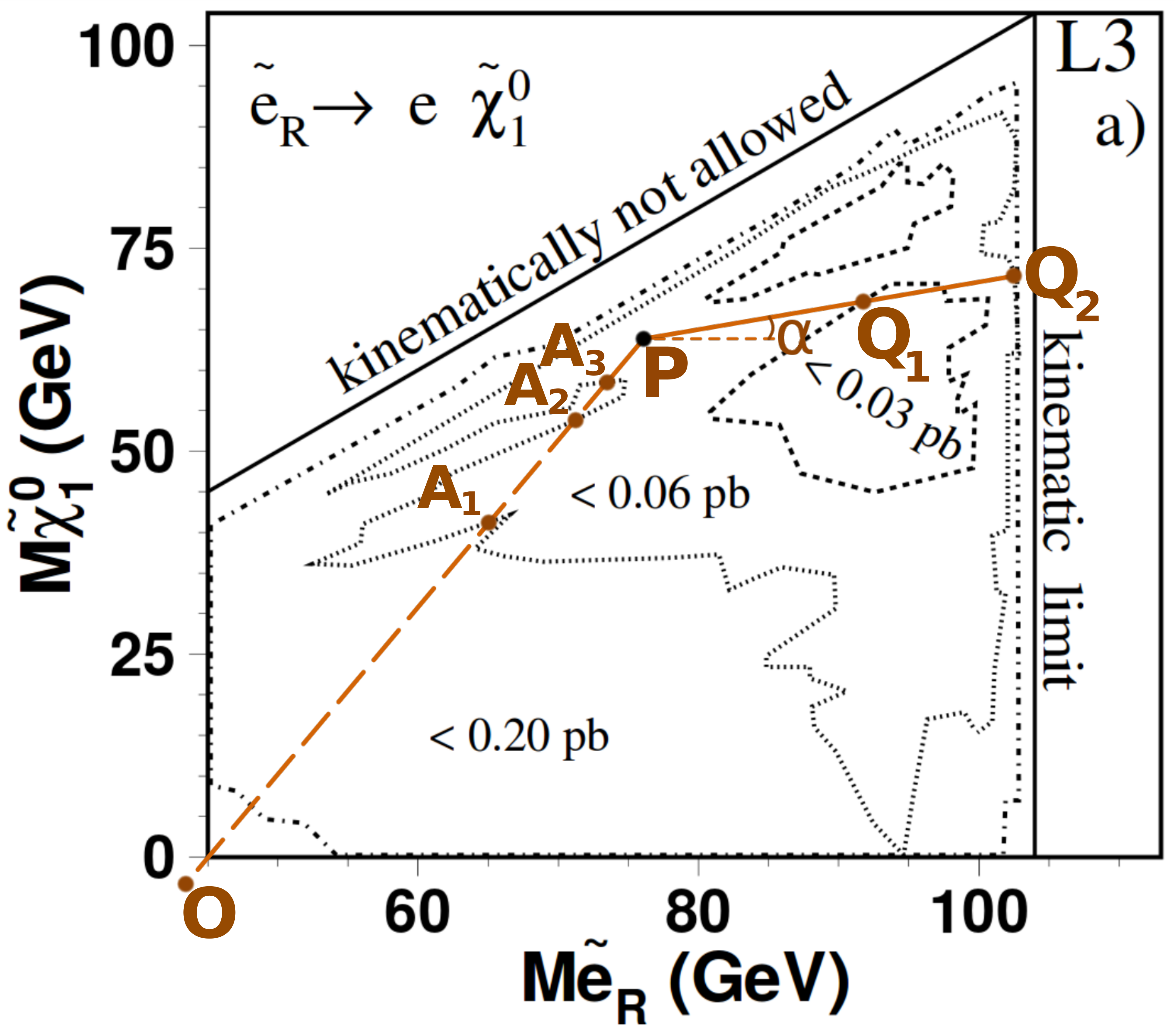}
\caption{Example of the limit interpolation process, based upon Fig.~2a of
  Ref.~\cite{L3:sleptons_squarks}. The line segment $\mathrm{OP}$ is used to
  find the intersection points $\mathrm{A_1}$, $\mathrm{A_2}$, and
  $\mathrm{A_3}$, which determine that $\mathrm{P}$ is within the 0.06~\pb
  limit. Then, for each angle $\alpha \in [0, 2\pi]$, the line segments
  $\mathrm{PQ_1}$ and $\mathrm{PQ_2}$ contribute to the weighted average limit
  at the point $\mathrm{P}$. More details of this procedure are described in the
  main body of text. \label{fig:limitinterp}}
\end{figure}

Continuing the example of the selectron case, we now discuss how we model the cross-section limit for slepton pair production
and decay into the lightest neutralino from the L3 experiment. This has been given as
a function of the selectron and neutralino mass in Fig.~2a of
Ref.~\cite{L3:sleptons_squarks}, which we reproduce here in
Fig.~\ref{fig:limitinterp} for demonstration. Corresponding results for smuons
and staus are used in the same manner. These results cover slepton masses from
45~GeV up to the kinematic limit of 104~GeV, with neutralino masses from zero up
to the slepton mass.  For a particular model point the theoretical slepton pair
production cross-section is calculated in a separate
routine. This uses leading order results on the
cross-section taken from Ref.~\cite{Dawson:1983fw,Bartl:1987zg}, which includes
t-channel contributions from neutralinos. We treat contributions to a possible
signal cross-section from ${\tilde e^*}_L{\tilde e^{\phantom{*}}}_L$ and ${\tilde e^*}_R{\tilde e^{\phantom{*}}}_R$ pair production separately, taking into account the relevant branching ratios for the decay to the lightest neutralino using \decaybit~\cite{SDPBit}, which can be interfaced to, e.g., \SUSYHIT~\cite{Djouadi:2006bz}. Hereafter, we refer to this cross-section times branching ratio as $\sigma\times\BR$.

We estimate the dominant theoretical uncertainty on $\sigma\times\BR$ using the
mass uncertainties of the sleptons, as reported by \specbit \cite{SDPBit}. For slepton mass
values of $m_1 \pm \delta_1$ and $m_2 \pm \delta_2$, we calculate the central
value of $\sigma\times\BR$ for $m_1$ and $m_2$. Then, we recalculate
$\sigma\times\BR$ with the upper and lower mass values and use the
maximum and minimum of these as estimates for the overall
$\sigma\times\BR$ uncertainty.

Once the $\sigma\times\BR$ has been calculated this way, we can look up
the appropriate limit with which to compare from Fig.~\ref{fig:limitinterp}. We do this by digitising each cross-section limit contour, and using inverse distance-weighted interpolation \cite{Shepard} to estimate the cross-section limits in regions between contours.  The weighted averaging prevents the noisiness of the LEP limit curves from strongly influencing the interpolant (an advantage over e.g.\ spline or bilinear interpolation), whilst at the same time forcing the interpolant to exactly reproduce the published cross-section contours (an advantage over e.g.\ data smoothing algorithms).

\begin{figure*}[tp]
\centering
\includegraphics[width=\linewidth]{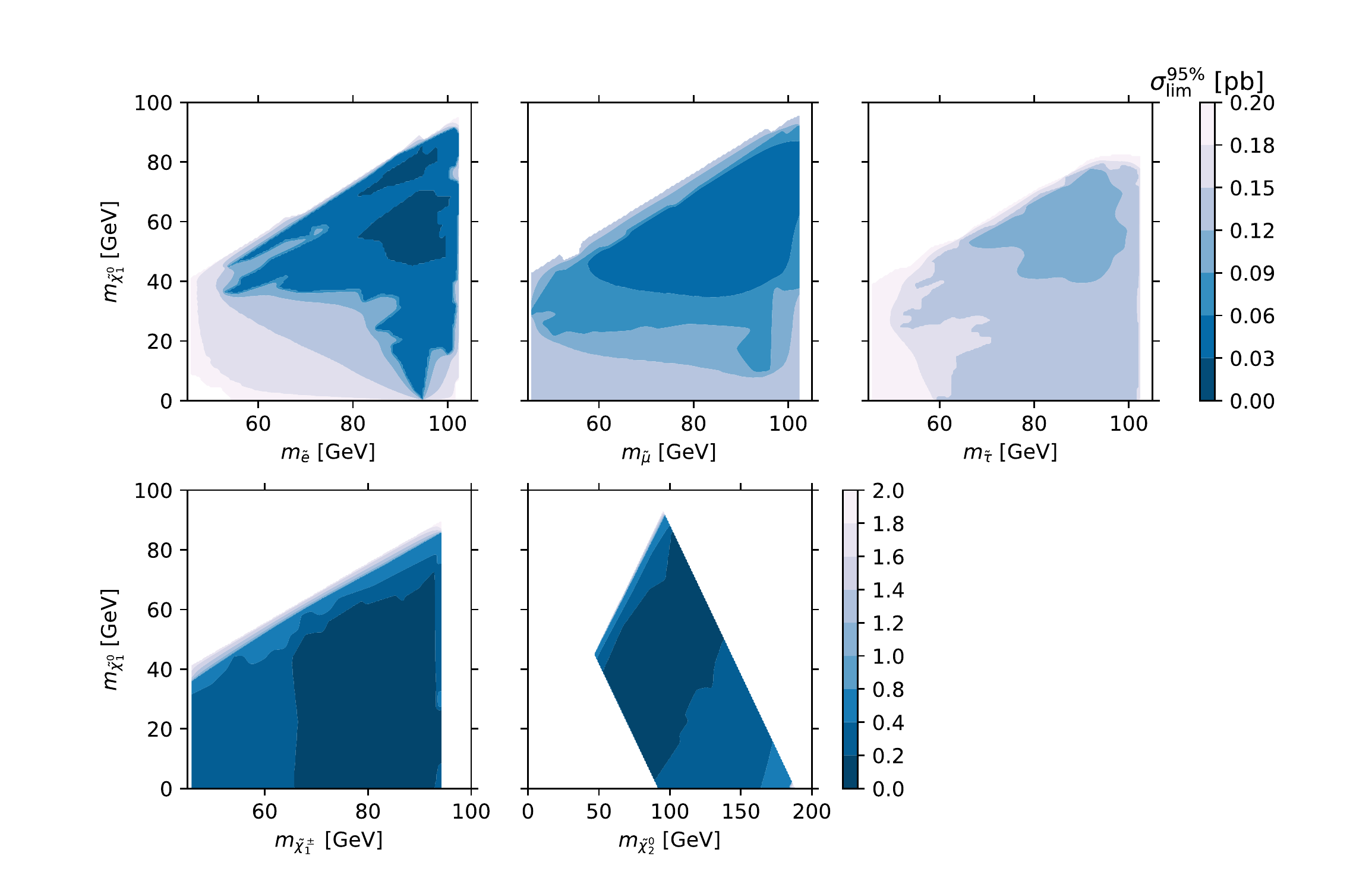}
\caption{Interpolated 95\% CL on the cross section for pair production of selectrons (upper left), smuons (upper centre), staus (upper right), charginos (lower left) and the next-to-lightest neutralino (lower right), as a function of the produced sparticle mass and the mass of the lightest neutralino. This interpolation is based on results by the L3 experiment at LEP~\cite{L3:sleptons_squarks,L3:gauginos}.}
\label{fig:lep_limits_interpolation_L3}
\end{figure*}

Our algorithm works as follows. Given a point on the $m_{\tilde e}$--$m_{{\tilde\chi}_1^0}$ plane, such as point $\mathrm{P}$ in Fig.~\ref{fig:limitinterp}, we first determine which contours contain this point. This can be achieved by drawing a line segment from this point to any point $\mathrm{O}$ outside of the plot. Then, for each contour, if this line segment $\mathrm{OP}$ intersects the contour an odd number of times, say at $\mathrm{A_1}$, $\mathrm{A_2}$, and $\mathrm{A_3}$, then the point lies within the contour. Using this method, we find the two limit contours, 0.06~\pb and 0.03~\pb, between which the point $\mathrm{P}$ lies. Next, for a large number of angles $\alpha \in [0, 2\pi]$, we draw a line segment $\mathrm{PQ_2}$ from $\mathrm{P}$ to where it intersects the outer limit of 0.06~\pb. We also note if this line segment intersects the 0.03~\pb contour, such as at $\mathrm{Q_1}$.  If the point lies directly on top of one of the contours, we simply take that contour as the correct limit. If not, we calculate the limit as a weighted average of all bounding cross-section limits over all angles. We weight each cross-section limit sample by $\overline{\mathrm{PQ}}^{-p}$, where $\overline{\mathrm{PQ}}$ is the length of the line segment $\mathrm{PQ}$, and $p$ is the so-called `power' parameter of the inverse distance-weighted interpolation algorithm. We choose $p=0.5$, to avoid artificially endowing the interpolating functions with local minima and maxima around the sample points, a known shortcoming of the algorithm for choices of $p$ greater than or equal to 1. The results of this interpolation proceedure can be seen in Fig.~\ref{fig:lep_limits_interpolation_L3}, which shows the interpolated 95\% confidence limits for all of the results from the L3 experiment that we use here. In particular the top left plot can be compared directly to Fig.~\ref{fig:limitinterp}.

Comparing the values of $\sigma\times\BR$ to the 95\% confidence interpolated limit drawn from the digitised limit plot,
we can now calculate the likelihood using the error function and the estimated theoretical uncertainty on $\sigma\times\BR$.

To increase the constraining power of the direct LEP searches, we also use the corresponding
cross-section limits set by the ALEPH experiment, calculating a second
likelihood in the same manner. For this, we consider the searches for scalar
leptons in the same mass range, using the model-independent results of Fig.~3 in
Ref.~\cite{Heister:2001nk}. We treat the data of the two experiments as
independent.\footnote{This should be a good approximation; common uncertainties across the experiments, such as luminosity, are subdominant to the systematic
uncertainty from Monte Carlo statistics in the experimental result.}

\begin{figure*}[tp]
\centering
\includegraphics[width=0.49\linewidth]{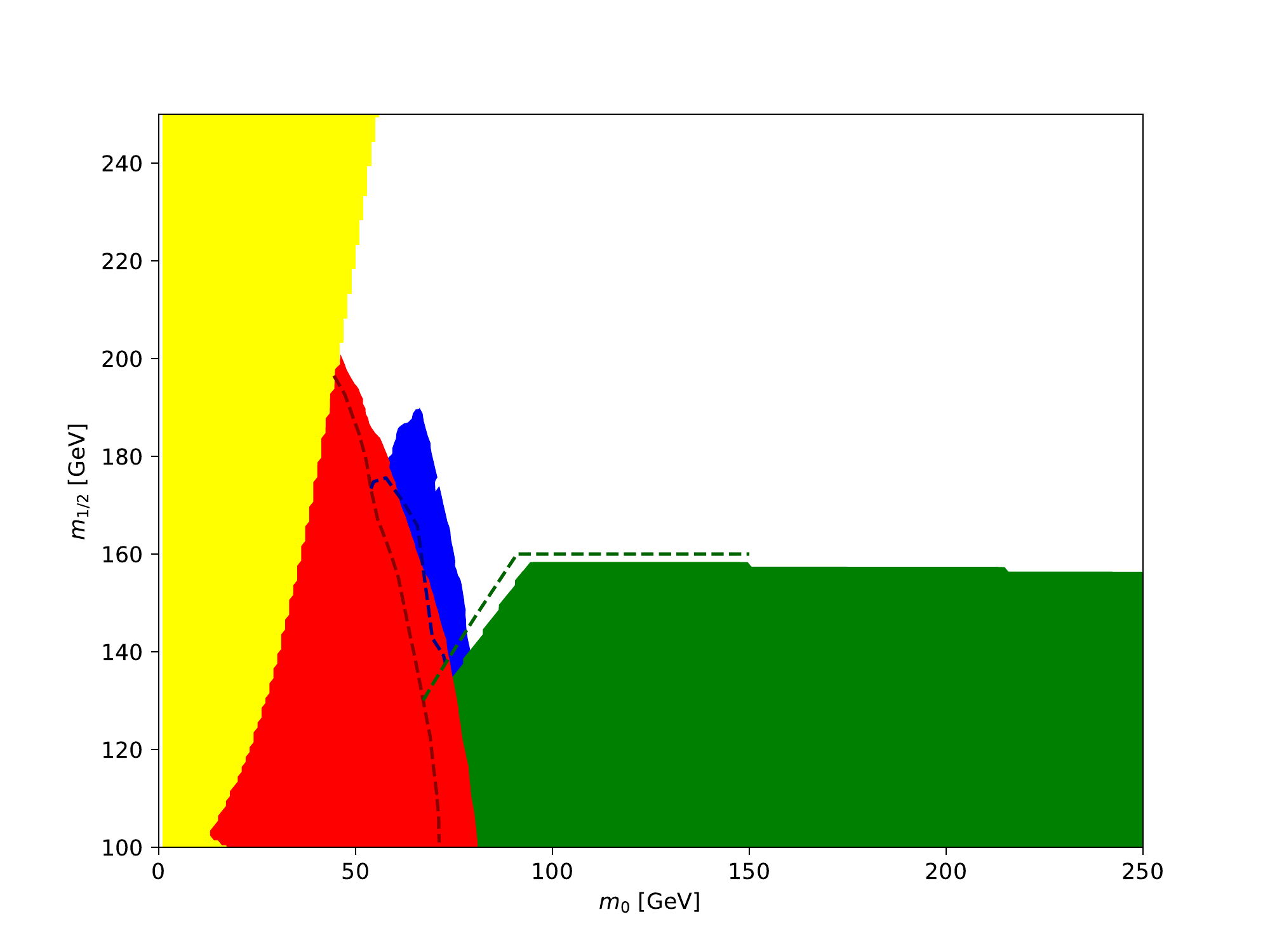}
\includegraphics[width=0.49\linewidth]{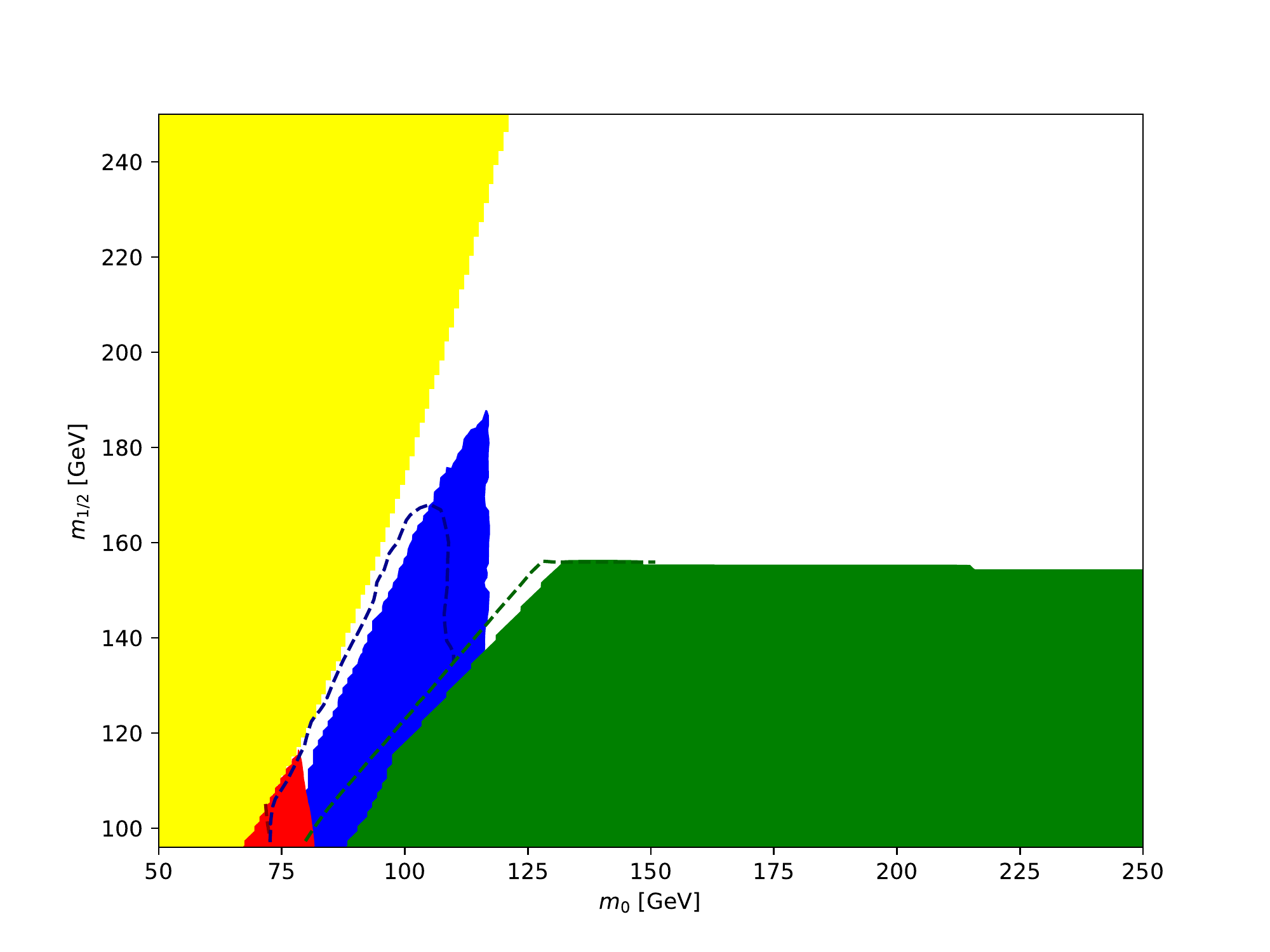}
\caption{Limits from direct sparticle pair production searches at LEP shown in the CMSSM  $(m_0,m_{1/2})$-mass plane with fixed $\tan\beta=15$ (left) and $\tan\beta=30$ (right), $A_0=0$ GeV and $\mu>0$. The 95\%~\CL excluded areas from chargino searches (green), stau searches (blue), and selectron searches (red)  are shown separately and overlaid in the sequence listed here. Theoretically forbidden regions are shown in yellow.
Included for comparison are the corresponding results from the ALEPH experiment alone taken from Fig.~6 of~\cite{Heister:2002jca}, indicated by the dashed lines.
\label{fig:LEP_limits}}
\end{figure*}

In Fig.~\ref{fig:LEP_limits} we show the \colliderbit exclusion limits from the combination of ALEPH and L3 searches for slepton pair production, in the CMSSM $(m_0,m_{1/2})$-mass plane for two different values of $\tan\beta$. The results in Fig.~\ref{fig:LEP_limits} can be compared to the corresponding CMSSM exclusion limits from ALEPH alone given in~\cite{Heister:2002jca} (dashed lines). We have checked that the observed differences are mainly due to the higher constraining power of the two experiments combined, with some remaining
unavoidable differences caused by the RGE codes used for the spectrum generation.

We take similar limits for the neutralino and chargino pair production cross-sections,
with decays into the lightest neutralino, from searches by the OPAL
and L3 experiments. The corresponding theoretical leading-order cross-sections
are from Refs.~\cite{Bartl:1986hp} and~\cite{Bartl:1985fk}, again taking
into account the relevant branching ratio for each model point. For neutralino
pairs, the limits are set on $\tilde\chi_2^0\tilde\chi_1^0$ production with
subsequent decay of the $\tilde\chi_2^0$. We take OPAL results from
Fig.~9 in Ref.~\cite{OPAL:gauginos}, which applies to hadronic decays, giving
bounds for $m_{\tilde\chi_2^0}$ from 100 GeV to the kinematic limit of 204 GeV,
while $m_{\tilde\chi_1^0}$ ranges from zero to $m_{\tilde\chi_2^0}$. The region
$m_{\tilde\chi_1^0}+m_{\tilde\chi_2^0}<$ 100~GeV is not bounded. From L3 we have
limits on leptonic decays $\tilde\chi_2^0\to ll \tilde\chi_1^0$ from Fig. 3b of
Ref.~\cite{L3:gauginos} for $m_{\tilde\chi_2^0}$ from 91 GeV to the kinematic
limit of 189 GeV. Again, no limit applies in the low mass region
$m_{\tilde\chi_1^0}+m_{\tilde\chi_2^0}<$ 91~GeV. Our interpolation from the L3 results is shown in Fig.~\ref{fig:lep_limits_interpolation_L3} (bottom right).\footnote{In this region limits
  from the decay of the $Z$-boson would apply unless the neutralinos are purely
  bino/wino combinations. In that case there must also be a light chargino
  (wino), for which the limits below apply.}

For chargino pair production, the OPAL experiment~\cite{OPAL:gauginos} sets
limits on hadronic, semi-leptonic and leptonic decays separately in Figs.~5--7
of that article. The limits are set from a chargino mass of 75~GeV up to the
kinematical limit of 104~GeV, and for neutralino masses from zero up to the
chargino mass. For each channel we take into account the branching ratios
$\BR(\tilde\chi_1^\pm\to q\bar q' \tilde\chi_1^0)$ and
$\BR(\tilde\chi_1^\pm\to l\nu \tilde\chi_1^0)$ of the model point. We use
an older, compatible, limit from the L3 experiment on
 fully leptonic decays taken from Fig. 2b of Ref.~\cite{L3:gauginos}. This extends from 45~GeV chargino
masses up to a kinematic limit of 94.5~GeV. Unfortunately, the L3 experiment
does not give separate model-independent cross-section limits for the other two
channels. Our interpolation from the L3 results is again shown in Fig.~\ref{fig:lep_limits_interpolation_L3} (bottom left).

We also include results on chargino and neutralino pair production from both OPAL (Fig.~8 in Ref.~\cite{OPAL:gauginos}) and L3 (Figs.~2a and~3a in Ref.~\cite{L3:gauginos}), where the limits assume that the fermions in $\tilde\chi_1^\pm\to ff' \tilde\chi_1^0$ and $\tilde\chi_2^0\to f \bar f \tilde\chi_1^0$ are represented as per the normal $W$ and $Z$ branching ratios into two fermions. This must be used with some care, as light sfermions may affect the assumption.

We note that while the experimental limits have been set on  $\tilde\chi_2^0\tilde\chi_1^0$ and $\tilde\chi_1^+\tilde\chi_1^-$ production, the \colliderbit likelihood can be calculated for production of any $\tilde\chi_i^0\tilde\chi_1^0$ and $\tilde\chi_i^+\tilde\chi_i^-$, as long as we consider the same experimental signature in the decay. Again, we show the resulting exclusion limits in the CMSSM in Fig.~\ref{fig:LEP_limits}. Here we observe very good agreement with earlier ALEPH results on chargino pair production~\cite{Heister:2002jca}, and we have checked that the difference is dominated by differences in the RGE codes used. \gambit relies on \flexiblesusy\cite{Athron:2014yba}, while the ALEPH analysis was carried out with {\textsf ISASUSY~7.51}~\cite{Baer:1999sp}.

\subsection{Higgs likelihood calculation}
\colliderbit includes likelihoods relating to constraints on
extended Higgs sectors from collider experiments, and to measurements of the SM-like
Higgs mass and production cross-sections at the LHC. These likelihoods are
provided through an interface to
\higgsbounds~\cite{Bechtle:2008jh,Bechtle:2011sb} and
\higgssignals~\cite{Bechtle:2013wla}.

Although constructing a likelihood from null search results at colliders generally
requires event simulation, the information provided by the combined LEP Higgs
search results \cite{Schael:2006cr} allows for the construction of an
approximate likelihood for neutral Higgs bosons. \higgsbounds interpolates the full \CLsb distribution
from the combined model-independent LEP searches, for all Higgs mass
combinations, over varying production cross sections. Using a
Gaussian approximation valid in the asymptotic limit, it employs the \CLsb distribution to calculate an approximate likelihood.

With the direct observation of an SM-like Higgs
boson~\cite{Chatrchyan2012,Aad:2012tfa}, measurements of the new particle's
mass, production cross section, and branching ratios can be used to constrain
the neutral Higgs sector of BSM models.  In channels where measurements of the
neutral boson's mass are available, \higgssignals calculates contributions to the mass likelihood as a $\chi^2$, taking into account both experimental and theoretical
uncertainties. For each channel, it minimises the $\chi^2$ independently over
the possibility of each neutral state in the Higgs sector being responsible for
the signal, including the simultaneous appearance of multiple resonances if they
are nearly degenerate in mass.  For signal strengths, it uses measurements over all available channels
to construct a single $\chi^2$, using the associated
$N_{\rm meas}-$dimensional covariance matrix to account for reported
experimental uncertainties, including correlations due to common
channels between experiments and the uncertainty in the integrated luminosity.
It then combines this signal-strength likelihood with the mass likelihood to form a combined LHC neutral Higgs sector likelihood.

For both the LEP and LHC likelihoods implemented in \colliderbit, the theoretical masses (with
uncertainties, when available), couplings and branching ratios come from other
\gambit modules, namely \decaybit and \specbit~\cite{SDPBit}.  In particular, we use the Higgs couplings provided
via \cpp{HiggsCouplingsTable} objects from \specbit to estimate the neutral Higgs boson
production cross sections.  We calculate the ratios of the production cross-sections for each Higgs in a given BSM theory to an SM Higgs of the same mass, assuming them to be given by
the ratio of squared couplings for the relevant processes.

\section{User interface}
\label{sec:interface}
\begin{table*}[tp]
\centering
\scriptsize
\begin{tabular}{l|p{3.7cm}|l|l|l}
  \textbf{Capability}
      & \multirow{2}{*}{\parbox{3.7cm}{\textbf{Function} (\textbf{Return type}):
             \\  \textbf{Brief description}}}
          & \textbf{Dependencies}
          & \multirow{2}{*}{\parbox{2cm}{\textbf{Backend} \\ \textbf{requirements}}}
          & \textbf{Options} (\textbf{Type})
          \\ & & & &
  \\ \hline
  \cpp{ColliderOperator}
      & \multirow{5}{*}{\parbox{3.7cm}{ \cpp{operateLHCLoop} (\cpp{void}):
              \\ Controls the parallelisation and execution
              of the entire event loop of the collider simulation.}}
          & & \pythiaeight
          & \cpp{pythiaNames}
          \\ & & & & \ (\cpp{vector<string>})
          \\ & & & & \cpp{nEvents}
          \\ & & & & \ (\cpp{vector<int>})
          \\ & & & & \cpp{silenceLoop} (\cpp{bool})
  \\ \hline
  \cpp{LHC\_Combined\_LogLike}
      & \multirow{6}{*}{\parbox{3.7cm}{\cpp{calc\_LHC\_LogLike}
              (\cpp{double}):
              \\ Combines the results from different analyses
              together into a single delta-log-likelihood value.}}
          & \cpp{DetAnalysisNumbers} & \nulike &
          \\ & & \cpp{ATLASAnalysisNumbers} & &
          \\ & & \cpp{CMSAnalysisNumbers} & &
          \\ & & \cpp{IdentityAnalysisNumbers} & &
          \\ & & \ (Table~\ref{tab:analysis}) & &
          \\ & & & &
  \\ \hline
\end{tabular}
\caption{The capabilities provided by \colliderbit that control the simulation
event loop and calculate the likelihood. The \cpp{operateLHCLoop} function
requires classes from \pythiaeight, which is connected to \gambit via
\BOSS~\cite{gambit}.
The options are read at runtime from the \gambit\ \YAML file (or configured in the
\colliderbit standalone code).  For readability, here and in the following tables we suppress the namespace \cpp{std} for standard \Cpp types such as \cpp{std::vector} and \cpp{std::string}.  The \cpp{pythiaNames} option tells the \cpp{operateLHCLoop} function the names of the \pythiaeight configurations for
which it should run simulation loops (one loop per configuration). The
\cpp{nEvents} option tells it how many events to generate per loop, while the \cpp{silenceLoop} option (default \cpp{true}) is used to suppress output to \protect\term{stdout} during the simulation loops.
\label{tab:loopexterior}}
\end{table*}

\begin{table*}[tp]
\centering
\scriptsize
\begin{tabular}{l|p{5.4cm}|l|l|l}
  \textbf{Capability}
      & \multirow{2}{*}{\parbox{5.4cm}{\textbf{Function} (\textbf{Return type}):
             \\ \textbf{Brief description}}}
          & \textbf{Dependencies}
          & \multirow{2}{*}{\parbox{1.2cm}{\textbf{Backend} \\ \textbf{req.}}}
          & \textbf{Options} (\textbf{Type})
          \\ & & & &
  \\ \hline
  \cpp{HardScatteringSim}
      & \multirow{5}{*}{\parbox{5.4cm}{\cpp{getPythia}
              \\ (\cpp{ColliderBit::SpecializablePythia}):
              \\ Provides a \pythiaeight instance within a
              container that is ready to simulate collision events for a
              model chosen by \scannerbit.}}
          & \cpp{decay\_rates} & \pythiaeight
          & \cpp{Pythia\_doc\_path}
          \\ & & A relevant \cpp{Spectrum} & & \ (\cpp{string})
          \\ & & object & & \metavar{Pythia\_config}
          \\ & & & & \ (\cpp{vector<string>})
          \\ & & & & \cpp{xsec_vetos}
          \\ & & & &\ (\cpp{vector<double>})
          \\ \cmidrule{2-5}
      & \multirow{5}{*}{\parbox{5.4cm}{\cpp{getPythiaFileReader}
              \\  (\cpp{ColliderBit::SpecializablePythia}):
              \\  Provides a \pythiaeight instance within a
              container that is ready to simulate collision events based upon
              some SLHA files.}}
          & & \pythiaeight
          & \cpp{Pythia\_doc\_path}
          \\ & & & & \ (\cpp{string})
          \\ & & & & \metavar{Pythia\_config}
          \\ & & & & \ (\cpp{vector<string>})
          \\ & & & & \cpp{SLHA\_filenames}
          \\ & & & & \ (\cpp{vector<string>})
          \\ & & & & \cpp{xsec_vetos}
          \\ & & & &\ (\cpp{vector<double>})
  \\ \hline
  \cpp{HardScatteringEvent}
      & \multirow{3}{*}{\parbox{5.4cm}{\cpp{generatePythia8Event}
              (\cpp{Pythia8::Event}):
              \\ Uses the given \cpp{HardScatteringSim} to
              generate the next event of the collider simulation chain.}}
          & \cpp{HardScatteringSim} & \pythiaeight
          \\ & & & &
          \\ & & & &
          \\ & & & &
          \\ & & & &
  \\ \hline
\end{tabular}
\caption{The collider capabilities provided by \colliderbit. In addition to the
dependencies shown above, all of these
functions also depend on the \cpp{ColliderOperator} capability in
Table~\ref{tab:loopexterior}, because they all execute within the event loop.
These functions need classes from \pythiaeight, which is connected to \gambit
via \BOSS~\cite{gambit}. The \cpp{decay\_rates} and \cpp{MSSM\_spectrum}
dependencies can be fulfilled by \decaybit and \specbit~\cite{SDPBit},
respectively. The options are read at runtime from the \gambit\ \YAML file (or
configured in the \colliderbit standalone code). The \cpp{Pythia\_doc\_path}
option points to the \cpp{xmldoc} directory of \pythia. The \metavar{Pythia\_config} option
is a list of \pythia settings. One such \metavar{Pythia\_config} list is required per \pythia configuration name, as given in \cpp{pythiaNames} (Table~\ref{tab:loopexterior}). The \cpp{SLHA\_filenames} option is a list of the SLHA files that the user wants to run using \cpp{getPythiaFileReader}. Finally, the \cpp{xsec_vetos} option specifies limits on the
maximum total cross-section (in \fb), as estimated by \pythia at the beginning of a run, below which the simulation should be skipped. One cross-section limit can be set per \pythia configuration (default \cpp{0}).
\label{tab:collider}}
\end{table*}

The \gambit code consists of a series of separate code modules that calculate likelihoods for new physics models using data from flavour physics~\cite{FlavBit}, astrophysics~\cite{DarkBit}, electroweak precision physics~\cite{SDPBit} and collider physics (the present paper). These modules can be used as standalone tools (using a custom C++ driving code), or they can be used via the \gambit core framework that resolves dependencies between calculations, and steers scans with the aid of a dedicated scanning and statistics module~\cite{ScannerBit}. The advantage of using the latter is that it is by far the easiest way to define models, calculate spectra and perform decay width calculations.

There are thus two ways to take advantage of the high-energy collider likelihoods
provided by \colliderbit: either via the \gambit
framework or by interfacing to \colliderbit as a standalone tool. Here we describe each in turn.

\subsection{\gambit interface}

The \gambit framework~\cite{gambit} defines two sorts of function that can be
used by each \cross{module} within the framework:

\begin{itemize}
\item \doublecross{Module functions}{module function}: \Cpp functions within the \gambit code itself.
\item \doublecross{Backend functions}{backend function}: functions that live within an external code, such as \pythiaeight.
\end{itemize}

In \gambit, each module function is given a tag, called a \cross{capability}, that
describes what it can calculate, be it an observable, e.g.\ the number of
events expected at the LHC, or a likelihood, e.g.\ the combined likelihood of a
set of LHC searches. Module functions may also have \doublecross{dependencies}{dependency} on
other module functions --- which may live either in the same module or in another
\gambit module --- or \doublecross{backend requirements}{backend requirement} that are satisfied by \doublecross{backend functions}{backend function} or \doublecross{backend variables}{backend variable}. A concrete example from \colliderbit is the capability of the combined LHC likelihood calculation (Table~\ref{tab:loopexterior}),
which has dependencies on the numbers of events expected in CMS and ATLAS
searches, and has a backend requirement relating to the functional form of the likelihood
required.

\colliderbit interfaces with the \gambit Core to communicate its capabilities,
dependencies, and backend requirements. The Core then runs its dependency
resolution routine to connect and execute the module functions in the
order that fulfils all the dependencies. As most of this machinery is
described in the main \gambit paper~\cite{gambit}, in this section we shall
simply describe each of the \colliderbit capabilities and as their uses.

\subsubsection{LHC simulation capabilities}
\label{sec:LHC_simulation_cap}
These capabilities are grouped within \colliderbit into three categories,
which correspond to the three main steps of simulation: collider, detector,
and analysis. There are also two additional capabilities needed to complete a
collider simulation. One is a capability meant simply to control the
parallelisation and execution of the event generation loop (\cpp{ColliderOperator}), and the other
calculates the likelihood as a final result (\cpp{LHC_Combined_LogLike}). These two capabilities
are shown in Table~\ref{tab:loopexterior}.  Tables~\ref{tab:collider},
\ref{tab:detector}, and \ref{tab:analysis} show the collider, detector, and
analysis capabilities, respectively.

\begin{table*}[tp]
\centering
\scriptsize
\begin{tabular}{l|p{5.1cm}|l|l|l}
  \textbf{Capability}
      & \multirow{2}{*}{\parbox{5.1cm}{\textbf{Function} (\textbf{Return type}):
             \\  \ \textbf{Brief description}}}
          & \textbf{Dependencies}
          & \multirow{2}{*}{\parbox{1.1cm}{\textbf{Backend} \\ \textbf{req.}}}
          & \textbf{Options} (\textbf{Type})
          \\ & & & &
  \\ \hline
  \cpp{DetectorSim}
      & \multirow{4}{*}{\parbox{5.1cm}{\cpp{getDelphes}
              \\ (\cpp{ColliderBit::DelphesVanilla}):
              \\ Provides a \delphes instance within a
              container that is ready to perform detector simulation.}}
          & & \pythiaeight & \cpp{delphesConfigFiles}
          \\ & & & & \ (\cpp{vector<string>})
          \\ & & & & \cpp{useDetector}
          \\ & & & & \ (\cpp{vector<bool>})
          \\ & & & &
  \\ \hline
  \cpp{SimpleSmearingSim}
      & \multirow{6}{*}{\parbox{5.1cm}{\cpp{getBuckFastATLAS}
              \\ (\cpp{ColliderBit::BuckFastSmearATLAS}):
              \\ Provides a set of \buckfast functions within a
              container that is ready to apply ATLAS smearing and reconstruction
              efficiencies to an event.}}
          & & & \cpp{antiktR}
          \\ & & & & \ (\cpp{vector<double>})
          \\ & & & &\cpp{partonOnly}
          \\ & & & & \ (\cpp{vector<bool>})
          \\ & & & &\cpp{useDetector}
          \\ & & & & \ (\cpp{vector<bool>})
          \\ \cmidrule{2-5}
      & \multirow{6}{*}{\parbox{5.1cm}{\cpp{getBuckFastCMS}
              \\ (\cpp{ColliderBit::BuckFastSmearCMS}):
              \\ Provides a set of \buckfast functions within a
              container that is ready to apply CMS smearing and reconstruction
              efficiencies to an event.}}
          & & & \cpp{antiktR}
          \\ & & & & \ (\cpp{vector<double>})
          \\ & & & &\cpp{partonOnly}
          \\ & & & & \ (\cpp{vector<bool>})
          \\ & & & &\cpp{useDetector}
          \\ & & & & \ (\cpp{vector<bool>})
          \\ \cmidrule{2-5}
      & \multirow{6}{*}{\parbox{5.1cm}{\cpp{getBuckFastIdentity}
              \\ (\cpp{ColliderBit::BuckFastIdentity}):
              \\ Provides a function that does absolutely
              nothing to a given event within a container similar to those
              returned by other \cpp{SimpleSmearingSim} capabilities.}}
          & & & \cpp{antiktR}
          \\ & & & & \ (\cpp{vector<double>})
          \\ & & & &\cpp{partonOnly}
          \\ & & & & \ (\cpp{vector<bool>})
          \\ & & & &\cpp{useDetector}
          \\ & & & & \ (\cpp{vector<bool>})
  \\ \hline
  \cpp{ReconstructedEvent}
      & \multirow{5}{*}{\parbox{5.1cm}{\cpp{reconstructDelphesEvent}
              \\ (\cpp{HEPUtils::Event}):
              \\ Uses the given \cpp{DetectorSim} to perform
              detector simulation upon the given \cpp{HardScatteringEvent}.}}
          & \cpp{HardScatteringEvent}& \pythiaeight &
          \\ & & \ (Table~\ref{tab:collider}) & &
          \\ & &  \cpp{DetectorSim} & &
          \\ & & & &
          \\ & & & &
  \\ \hline
  \cpp{ATLASSmearedEvent}
      & \multirow{4}{*}{\parbox{5.1cm}{\cpp{smearEventATLAS}
              (\cpp{HEPUtils::Event}):
              \\ Uses the given \cpp{SimpleSmearingSim} to
              apply smearing and reconstruction efficiencies upon the given
              \cpp{HardScatteringEvent}.}}
          & \cpp{HardScatteringEvent} &
          \\ & & \ (Table~\ref{tab:collider}) &
          \\ & & \cpp{SimpleSmearingSim} of  &
          \\ & & type \cpp{BuckFastSmear}- &
          \\ & & \cpp{ATLAS} & &
  \\ \hline
  \cpp{CMSSmearedEvent}
      & \multirow{4}{*}{\parbox{5.1cm}{\cpp{smearEventCMS}
              (\cpp{HEPUtils::Event}):
              \\ Uses the given \cpp{SimpleSmearingSim} to
              apply smearing and reconstruction efficiencies upon the given
              \cpp{HardScatteringEvent}.}}
          & \cpp{HardScatteringEvent} &
          \\ & & \ (Table~\ref{tab:collider}) &
          \\ & & \cpp{SimpleSmearingSim} of &
          \\ & & type \cpp{BuckFastSmearCMS} &
          \\ & & & &
  \\ \hline
  \cpp{CopiedEvent}
      & \multirow{4}{*}{\parbox{5.1cm}{\cpp{copyEvent}
              (\cpp{HEPUtils::Event}):
              \\ Uses the given \cpp{SimpleSmearingSim} to
              do absolutely nothing to the given \cpp{HardScatteringEvent}.}}
          & \cpp{HardScatteringEvent} &
          \\ & & \ (Table~\ref{tab:collider}) &
          \\ & & \cpp{SimpleSmearingSim} of &
          \\ & & type  \cpp{BuckFastIdentity} &
  \\ \hline
\end{tabular}
\caption{The detector capabilities provided by \colliderbit. In addition to the
dependencies shown above, all of these functions also depend on the
\cpp{ColliderOperator} capability in Table~\ref{tab:loopexterior}, since
they all execute within the event loop. Some of these functions need classes
from \pythiaeight, which is connected to \gambit via \BOSS~\cite{gambit}.
The options are read at runtime from the \gambit\ \YAML file (or
configured in the \colliderbit standalone code). The \cpp{delphesConfigFiles}
option specifies the TCL files used by \delphes for its configuration. The
\cpp{antiktR} options (default \cpp{0.4}) control the $R$ value used by \fastjet's
anti-$k_T$ jet algorithm. The \cpp{partonOnly} options (default
\cpp{false}) tell the smearing sims to consider only the partonic states of
the event. Finally, the \cpp{useDetector} option switches a given detector simulation on or off, along with all analyses relying on that detector (default \cpp{true} for \cpp{getBuckFastATLAS}
and \cpp{getBuckFastCMS} and \cpp{false} for \cpp{getDelphes} and \cpp{getBuckFastIdentity}).
All the options in this table are vectors that require one entry per
\pythia configuration in \cpp{pythiaNames} (Table~\ref{tab:loopexterior}).
\label{tab:detector}}
\end{table*}

\begin{table*}[t]
\centering
\scriptsize
\begin{tabular}{l|p{6.2cm}|l|l}
  \textbf{Capability}
      & \textbf{Function} (\textbf{Return type}): \textbf{Brief description}
          & \textbf{Dependencies}
          & \textbf{Options} (\textbf{Type})
  \\ \hline
  \cpp{DetAnalysis}
      & \multirow{4}{*}{\parbox{6.2cm}{\cpp{getDetAnalysisContainer}
              \\ (\cpp{ColliderBit::HEPUtilsAnalysisContainer}):
              \\ Provides a list of analyses within a container
              that is ready to apply them to an event.}}
          & \cpp{HardScatteringSim}
          & \cpp{analyses}
          \\ \ \cpp{Container} & & \ (Table~\ref{tab:collider}) & \ (\cpp{vector<vector<string>>})
          \\ & & &
          \\ & & &
  \\ \hline
  \cpp{ATLASAnalysis}
      & \multirow{4}{*}{\parbox{6.2cm}{\cpp{getATLASAnalysisContainer}
              \\ (\cpp{ColliderBit::HEPUtilsAnalysisContainer}):
              \\ Provides a list of ATLAS analyses within a container
              that is ready to apply them to an event.}}
          & \cpp{HardScatteringSim}
          & \cpp{analyses}
          \\ \ \cpp{Container} & & \ (Table~\ref{tab:collider}) & \ (\cpp{vector<vector<string>>})
          \\ & & &
          \\ & & &
  \\ \hline
  \cpp{CMSAnalysis}
      & \multirow{4}{*}{\parbox{6.2cm}{\cpp{getCMSAnalysisContainer}
              \\ (\cpp{ColliderBit::HEPUtilsAnalysisContainer}):
              \\ Provides a list of CMS analyses within a container
              that is ready to apply them to an event.}}
          & \cpp{HardScatteringSim}
          & \cpp{analyses}
          \\ \ \cpp{Container} & & \ (Table~\ref{tab:collider}) & \ (\cpp{vector<vector<string>>})
          \\ & & &
          \\ & & &
  \\ \hline
  \cpp{IdentityAnalysis}
      & \multirow{5}{*}{\parbox{6.2cm}{\cpp{getIdentityAnalysisContainer}
              \\ (\cpp{ColliderBit::HEPUtilsAnalysisContainer}):
              \\ Provides a list of ``identity'' analyses (no detector
              \\ smearing) within a container
              that is ready to apply them to an event.}}
          & \cpp{HardScatteringSim}
          & \cpp{analyses}
          \\ \ \cpp{Container} & & \ (Table~\ref{tab:collider}) & \ (\cpp{vector<vector<string>>})
          \\ & & &
          \\ & & &
          \\ & & &
  \\ \hline
  \cpp{DetAnalysis}
      & \multirow{4}{*}{\parbox{6.2cm}{\cpp{runDetAnalyses}
              \\ (\cpp{ColliderBit::AnalysisNumbers}):
              \\ Uses the given \cpp{DetAnalysisContainer} to
              perform all its analyses upon the given \cpp{ReconstructedEvent}.}}
          & \cpp{ReconstructedEvent} &
          \\ \ \cpp{Numbers} & &  \ (Table~\ref{tab:detector}) &
          \\ & & \cpp{HardScatteringSim}&
          \\ & &  \ (Table~\ref{tab:collider})  &
           \\ & &\cpp{DetAnalysisContainer}  &
  \\ \hline
  \cpp{ATLASAnalysis}
      & \multirow{4}{*}{\parbox{6.2cm}{\cpp{runATLASAnalyses}
              \\ (\cpp{ColliderBit::AnalysisNumbers}):
              \\ Uses the given \cpp{ATLASAnalysisContainer} to
              perform all its analyses upon the given \cpp{ATLASSmearedEvent}.}}
          & \cpp{ATLASSmearedEvent} &
          \\ \ \cpp{Numbers}& &  \ (Table~\ref{tab:detector}) &
          \\ & & \cpp{HardScatteringSim}&
          \\ & &  \ (Table~\ref{tab:collider})  &
          \\ & &\cpp{ATLASAnalysis} &
          \\ & &  \ \cpp{Container} &
  \\ \hline
  \cpp{CMSAnalysis}
      & \multirow{4}{*}{\parbox{6.2cm}{\cpp{runCMSAnalyses}
              \\ (\cpp{ColliderBit::AnalysisNumbers}):
              \\ Uses the given \cpp{CMSAnalysisContainer} to
              perform all its analyses upon the given \cpp{CMSSmearedEvent}.}}
          & \cpp{CMSSmearedEvent} &
          \\ \ \cpp{Numbers} & &  \ (Table~\ref{tab:detector}) &
          \\ & &  \cpp{HardScatteringSim}&
          \\ & &  \ (Table~\ref{tab:collider}) &
          \\ & & \cpp{CMSAnalysisContainer} &
  \\ \hline
  \cpp{IdentityAnalysis}
      & \multirow{5}{*}{\parbox{6.2cm}{\cpp{runIdentityAnalyses}
              \\ (\cpp{ColliderBit::AnalysisNumbers}):
              \\ Uses the given \cpp{IdentityAnalysisContainer} to
              perform all its analyses upon the given \cpp{CopiedEvent}.}}
          & \cpp{CopiedEvent} &
          \\ \ \cpp{Numbers} & &  \ (Table~\ref{tab:detector}) &
          \\ & &  \cpp{HardScatteringSim}&
          \\ & &  \ (Table~\ref{tab:collider}) &
          \\ & & \cpp{IdentityAnalysis}- &
          \\ & & \cpp{Container} &
  \\ \hline
\end{tabular}
\caption{The analysis capabilities provided by \colliderbit. In addition to the
dependencies shown above, all of these functions depend on the
\cpp{ColliderOperator} capability in Table~\ref{tab:loopexterior}, since
they all execute within the event loop. The options are read at runtime from
the \gambit\ \YAML file (or configured in the \colliderbit standalone code). The
\cpp{analyses} options tell the \cpp{getDetAnalysisContainer},
\cpp{getATLASAnalysisContainer}, \cpp{getCMSAnalysisContainer} and \cpp{getIdentityAnalysisContainer} functions the names of all the analyses the user wishes to include for the collider simulations. One vector of analysis names (possibly empty) is required per \pythia configuration in \cpp{pythiaNames} (Table~\ref{tab:loopexterior}).
\label{tab:analysis}}
\end{table*}

Since there can be a variety of configurations for collider simulation and
a variety of experimental analyses, we designed these components so new
configurations and analyses could be easily added.

For instance, a user who wishes to add a new \pythiaeight configuration must
only complete the following steps:
\begin{enumerate}
\item \emph{Create a} \cpp{SpecializablePythia} \emph{initialisation function.} These
functions are defined in the file \term{colliders/SpecializablePythia.cpp}.\footnote{Within this
section, all header files (\term{*.hpp}) mentioned are found in the
 \term{ColliderBit/include/gambit/ColliderBit} directory, while source files
(\term{*.cpp}) are found in the \term{ColliderBit/src} directory.}
Each such function must have its own namespace and a call signature of
\cpp{void init(SpecializablePythia*} \cpp{specializeMe)}. Within the init function,
settings can be sent to \pythiaeight as strings using the
\cpp{SpecializablePythia::addToSettings} function. For example, the following
would be a valid init function:
\begin{minipage}{\columnwidth - 5mm}
\begin{lstcpp}
namespace Pythia_ttbar_LHC_13TeV
{
  void init(SpecializablePythia* specializeMe)
  {
    specializeMe->addToSettings(
        "Beams:eCM = 13000");
    specializeMe->addToSettings(
        "Top:qqbar2ttbar = on");
    specializeMe->addToSettings(
       "Top:gg2ttbar = on");
  }
}
\end{lstcpp}
\end{minipage}
This would initialize a \pythiaeight configuration named \cpp{Pythia_ttbar_LHC_13TeV} to simulate $t\bar t$ production through $q\bar q\to t\bar t$ and $gg\to t\bar t$ at 13~\TeV. Init functions may also ``inherit'' from other existing init functions by explicitly calling them. For instance:
\begin{minipage}{\columnwidth - 5mm}
\begin{lstcpp}
namespace Pythia_alltop_LHC_13TeV
{
  void init(SpecializablePythia* specializeMe)
  {
    Pythia_ttbar_LHC_13TeV::init(specializeMe);
    specializeMe->addToSettings(
        "Top:qq2tq(t:W) = on");
    specializeMe->addToSettings(
        "Top:ffbar2ttbar(s:gmZ) = on");
    specializeMe->addToSettings(
        "Top:ffbar2tqbar(s:W) = on");
    specializeMe->addToSettings(
        "Top:gmgm2ttbar = on");
  }
}
\end{lstcpp}
\end{minipage}
This creates a second \pythiaeight configuration named \cpp{Pythia_alltop_LHC_13TeV} for simulating the full set of top quark production processes. The first line of the init function calls the init function of \cpp{Pythia_ttbar_LHC_13TeV}, which sets the energy to 13~\TeV and turns on the $q\bar q\to t\bar t$ and $gg\to t\bar t$ processes. Then follows four calls to \cpp{addToSettings} that switch on the simulation of the additional production processes $qq'\to tq''$ (t-channel $W$ exchange), $f\bar f\to t\bar t$ (s-channel $\gamma$/$Z$ exchange), $f\bar f'\to tq''$ (s-channel $W$ exchange) and $\gamma\gamma\to t\bar t$.

\item \emph{Add the initialisation function namespace within}
\cpp{SpecializablePythia::resetSpecialization}. This function is located at
the end of \term{colliders/SpecializablePythia.cpp}, and it allows for runtime
selection of the \pythiaeight specialisation via a \cpp{std::string}. The namespaces
for the new init functions must be included here using the \cpp{IF_X_SPECIALIZEX}
macro. Thus, for our example top production init functions above, we would add:

\begin{minipage}{\columnwidth - 5mm}
\begin{lstcpp}
IF_X_SPECIALIZEX(Pythia_ttbar_LHC_13TeV)
IF_X_SPECIALIZEX(Pythia_alltop_LHC_13TeV)
\end{lstcpp}
\end{minipage}

\item \emph{Recompile \gambit.} (See Appendix~\ref{sec:quickstart}.)

\item \emph{Activate the new init function.} The choice of init function is
specified within the \gambit\ \YAML file in the ``Rules'' section for the
\cpp{operateLHCLoop} module function. For example, to activate our ``alltop''
\pythiaeight configuration, this would be the \YAML entry:

\begin{minipage}{\columnwidth - 5mm}
\begin{lstyaml}
  - capability:  ColliderOperator
    function: operateLHCLoop
    options:
      nEvents: [1000]
      pythiaNames: ["Pythia_alltop_LHC_13TeV"]
      silenceLoop: true
\end{lstyaml}
\end{minipage}

The last of these options removes much of the output from the Pythia event generator. We may also supply our chosen configuration with additional options right in the
\YAML file. We do this in the rules for the \cpp{getPythia} module function:
\begin{minipage}{\columnwidth - 5mm}
\begin{lstyaml}
  - capability:  HardScatteringSim
    function:  getPythia
    options:
      Pythia_alltop_LHC_13TeV: ["Print:quiet = on",
            "PartonLevel:MPI = off",
            "PartonLevel:ISR = on",
            "PartonLevel:FSR = on",
            "HadronLevel:all = on"]
\end{lstyaml}
\end{minipage}
\end{enumerate}

We recommend registering commonly-used \pythiaeight configurations in \term{colliders/SpecializablePythia.cpp} as described above. However, it is also possible to set up custom  \pythia configurations directly in the \YAML file by specifying all relevant \pythia options there. In this case the configuration name given in \cpp{pythiaNames} must not match any of the registered init functions in \term{colliders/SpecializablePythia.cpp}.

For each \pythia configuration, the choice of analyses and detector simulations can be varied. Many of the options detailed in Tables~\ref{tab:collider}, \ref{tab:detector} and \ref{tab:analysis} are therefore vectors expecting one element per \pythia configuration, in the same order as the configurations in \cpp{pythiaNames}.

Adding a new analysis is nearly as simple as adding a \pythiaeight
configuration. An annotated minimal example is given in \term{analyses/Analysis\_Minimum.cpp}.
This contains the minimum required to print out the number of jets,
$b$-jets and leptons, and the missing energy in each event, plus pass an arbitrary set of signal region cuts. To add a new analysis:
\begin{enumerate}
\item \emph{Copy the template example,} \term{Analysis\_Minimum.cpp},
to a new location in the \term{analyses} folder, for example \term{analyses/Analysis\_ATLAS\_TYPE\_20invfb.cpp}.
Within the new file, replace every instance of \cpp{Minimum} by \cpp{ATLAS\_TYPE\_20invfb}.

\item \emph{Edit the new analysis file to include the required cuts.} This
includes the option to add extra signal regions. The existing repository of
analyses provides examples of how to apply complex cuts.

\item \emph{Add an analysis factory declaration to
\term{analyses/} \term{HEPUtilsAnalysisContainer.cpp},} by adding the line:

\begin{minipage}{\columnwidth - 5mm}
\begin{lstcpp}
DECLARE_ANALYSIS_FACTORY(ATLAS_TYPE_20invfb);
\end{lstcpp}
\end{minipage}

\item \emph{Add a factory definition to \term{analyses/} \term{HEPUtilsAnalysisContainer.cpp},} by adding a line:

\begin{minipage}{\columnwidth - 5mm}
\begin{lstcpp}
IF_X_RTN_CREATEX(ATLAS_TYPE_20invfb);
\end{lstcpp}
\end{minipage}

\item \emph{Recompile \gambit.} (See Appendix~\ref{sec:quickstart}.)

\item \emph{Activate the new analysis.} The user may now run the analysis by
adding it to the list of analyses in the \YAML file, within the rules for the
relevant \cpp{AnalysisContainer}. For our \cpp{ATLAS\_TYPE\_20invfb}
example, we would add it here:

\begin{minipage}{\columnwidth - 5mm}
\begin{lstyaml}
  - capability: ATLASAnalysisContainer
    function: getATLASAnalysisContainer
    options:
      analyses: [["ATLAS_0LEP_20invfb",
            "ATLAS_TYPE_20invfb"]]
\end{lstyaml}
\end{minipage}
\end{enumerate}

Although the current framework only supports cut-and-count analyses, the
user could easily add more complicated likelihoods by adding new module functions.

\subsubsection{LEP supersymmetry limit capabilities}
\colliderbit contains functions that calculate the cross-section for various
SUSY particle productions within the context of the LEP collider. The capabilities
for these functions are described in Table~\ref{tab:lepxsec}. Using these functions
along with SUSY particle decay information, we calculate the cross-section times
branching ratio for each production mechanism associated with LEP model-independent
limits. The capabilities and functions that compare this calculation with each LEP
limit are described in Tables~\ref{tab:leplimitslepton} and~\ref{tab:leplimitgaugino}.

\begin{table*}[tp]
\centering
\scriptsize
\begin{tabular}{l|p{9.2cm}|l|l}
  \textbf{Capability}
      & \textbf{Function} (\textbf{Return type}): \textbf{Brief description}
          & \multirow{2}{*}{\parbox{1.3cm}{\textbf{Energies} \\ \textbf{for \metavar{E}}}}
          & \multirow{2}{*}{\parbox{1.8cm}{\textbf{Eigenstates} \\ \textbf{for \metavar{X} and \metavar{Y}}}}
          \\ & & &
  \\ \hline
  \cpp{LEP}\metavar{E}\cpp{\_xsec\_se}\metavar{X}\cpp{se}\metavar{Y}\cpp{bar}
      & \multirow{3}{*}{\parbox{9.2cm}{
              \cpp{LEP}\metavar{E}\cpp{\_SLHA1\_convention\_xsec\_se}\metavar{X}\cpp{se}\metavar{Y}\cpp{bar} (\cpp{triplet<double>}):
              \\ Calculates the LEP selectron pair production
              cross-section for centre of mass energy \metavar{E}, with selectron
              eigenstates \metavar{X} and \metavar{Y}.}}
          & \cpp{208}
          & \cpp{l}, \cpp{r} (helicity)
          \\ & & \cpp{205} & \cpp{1}, \cpp{2} (mass)
          \\ & & \cpp{188} &
  \\ \hline
  \cpp{LEP}\metavar{E}\cpp{\_xsec\_smu}\metavar{X}\cpp{smu}\metavar{Y}\cpp{bar}
      & \multirow{3}{*}{\parbox{9.2cm}{
              \cpp{LEP}\metavar{E}\cpp{\_SLHA1\_convention\_xsec\_smu}\metavar{X}\cpp{smu}\metavar{Y}\cpp{bar} (\cpp{triplet<double>}):
              \\ Calculates the LEP smuon pair production
              cross-section for centre of mass energy \metavar{E}, with smuon
              eigenstates \metavar{X} and \metavar{Y}.}}
          & \cpp{208}
          & \cpp{l}, \cpp{r} (helicity)
          \\ & & \cpp{205} & \cpp{1}, \cpp{2} (mass)
          \\ & & \cpp{188} &
  \\ \hline
  \cpp{LEP}\metavar{E}\cpp{\_xsec\_stau}\metavar{X}\cpp{stau}\metavar{Y}\cpp{bar}
      & \multirow{3}{*}{\parbox{9.2cm}{
              \cpp{LEP}\metavar{E}\cpp{\_SLHA1\_convention\_xsec\_stau}\metavar{X}\cpp{stau}\metavar{Y}\cpp{bar} (\cpp{triplet<double>}):
              \\ Calculates the LEP stau pair production
              cross-section for centre of mass energy \metavar{E}, with stau
              eigenstates \metavar{X} and \metavar{Y}.}}
          & \cpp{208}
          & \cpp{l}, \cpp{r} (helicity)
          \\ & & \cpp{205} & \cpp{1}, \cpp{2} (mass)
          \\ & & \cpp{188} &
  \\ \hline
  \cpp{LEP}\metavar{E}\cpp{\_xsec\_chi00\_}\metavar{XY}
      & \multirow{3}{*}{\parbox{9.2cm}{
              \cpp{LEP}\metavar{E}\cpp{\_SLHA1\_convention\_xsec\_chi00\_}\metavar{XY} (\cpp{triplet<double>}):
              \\ Calculates the LEP neutralino pair production
              cross-section for centre of mass energy \metavar{E}, with neutralino
              mass eigenstates \metavar{X} and \metavar{Y}.}}
          & \cpp{208} & \cpp{1}, \cpp{2}, \cpp{3}, \cpp{4}
          \\ & & \cpp{205} &
          \\ & & \cpp{188} &
  \\ \hline
  \cpp{LEP}\metavar{E}\cpp{\_xsec\_chipm\_}\metavar{XY}
      & \multirow{3}{*}{\parbox{9.2cm}{
              \cpp{LEP}\metavar{E}\cpp{\_SLHA1\_convention\_xsec\_chipm\_}\metavar{XY} (\cpp{triplet<double>}):
              \\ Calculates the LEP chargino pair production
              cross-section for centre of mass energy \metavar{E}, with chargino
              mass eigenstates \metavar{X} and \metavar{Y}.}}
          & \cpp{208}
          & \cpp{1}, \cpp{2}
          \\ & & \cpp{205} &
          \\ & & \cpp{188} &
  \\ \hline
\end{tabular}
\caption{The capabilities provided by \colliderbit that calculate SUSY particle
production cross-sections within the context of the LEP collider. All of these
functions return a triplet of \cpp{double}s.  These correspond to the maximum,
central, and minimum cross-sections calculated while varying the SUSY particle
masses according to their estimated uncertainties. All of these functions
depend on \gambit's \cpp{MSSM30atMGUT} model parameters~\cite{gambit}, \specbit's
\cpp{MSSM\_spectrum}, and \decaybit's \cpp{Z\_decay\_rates}~\cite{SDPBit}.
Versions of these functions exist with many different \metavar{E}, \metavar{X} and \metavar{Y} values, corresponding to the energy (in \GeV) and particle
eigenstates used in the calculation.
\label{tab:lepxsec}}
\end{table*}

\begin{table*}[tp]
\centering
\scriptsize
\begin{tabular}{l|p{9cm}|l}
  \textbf{Capability}
      & \textbf{Function} (\textbf{Return type}): \textbf{Brief description} & \textbf{Dependencies}
  \\ \hline
  \cpp{ALEPH\_Selectron\_LLike}
      & \multirow{4}{*}{\parbox{9cm}{\cpp{ALEPH\_Selectron\_Conservative\_LLike} (\cpp{double}):
              \\ Compares the cross section times branching ratio
              for selectron pair production to the model-independent limit according
              to the ALEPH collaboration. Returns a log likelihood value.}}
          & \cpp{LEP208\_xsec\_selselbar}
          \\ & & \cpp{LEP208\_xsec\_serserbar}
          \\ & & \cpp{selectron\_l\_decay\_rates}
          \\ & & \cpp{selectron\_r\_decay\_rates}
  \\ \hline
 \cpp{ALEPH\_Smuon\_LLike}
      & \multirow{4}{*}{\parbox{9cm}{\cpp{ALEPH\_Smuon\_Conservative\_LLike} (\cpp{double}):
              \\ Compares the cross section times branching ratio
              for smuon pair production to the model-independent limit according
              to the ALEPH collaboration. Returns a log likelihood value.}}
          & \cpp{LEP208\_xsec\_smulsmulbar}
          \\ & & \cpp{LEP208\_xsec\_smursmurbar}
          \\ & & \cpp{smuon\_l\_decay\_rates}
          \\ & & \cpp{smuon\_r\_decay\_rates}
  \\ \hline
  \cpp{ALEPH\_Stau\_LLike}
      & \multirow{4}{*}{\parbox{9cm}{\cpp{ALEPH\_Stau\_Conservative\_LLike} (\cpp{double}):
              \\ Compares the cross section times branching ratio
              for stau pair production to the model-independent limit according
              to the ALEPH collaboration. Returns a log likelihood value.}}
          & \cpp{LEP208\_xsec\_stau1stau1bar}
          \\ & & \cpp{LEP208\_xsec\_stau2stau2bar}
          \\ & & \cpp{stau\_1\_decay\_rates}
          \\ & & \cpp{stau\_2\_decay\_rates}
  \\ \hline
  \cpp{L3\_Selectron\_LLike}
      & \multirow{4}{*}{\parbox{9cm}{\cpp{L3\_Selectron\_Conservative\_LLike} (\cpp{double}):
              \\ Compares the cross section times branching ratio
              for selectron pair production to the model-independent limit according
              to the L3 collaboration. Returns a log likelihood value.}}
          & \cpp{LEP205\_xsec\_selselbar}
          \\ & & \cpp{LEP205\_xsec\_serserbar}
          \\ & & \cpp{selectron\_l\_decay\_rates}
          \\ & & \cpp{selectron\_r\_decay\_rates}
  \\ \hline
  \cpp{L3\_Smuon\_LLike}
      & \multirow{4}{*}{\parbox{9cm}{\cpp{L3\_Smuon\_Conservative\_LLike} (\cpp{double}):
              \\  Compares the cross section times branching ratio
              for smuon pair production to the model-independent limit according
              to the L3 collaboration. Returns a log likelihood value.}}
          & \cpp{LEP205\_xsec\_smulsmulbar}
          \\ & & \cpp{LEP205\_xsec\_smursmurbar}
          \\ & & \cpp{smuon\_l\_decay\_rates}
          \\ & & \cpp{smuon\_r\_decay\_rates}
  \\ \hline
  \cpp{L3\_Stau\_LLike}
      & \multirow{4}{*}{\parbox{9cm}{\cpp{L3\_Stau\_Conservative\_LLike} (\cpp{double}):
              \\  Compares the cross section times branching ratio
              for stau pair production to the model-independent limit according
              to the L3 collaboration. Returns a log likelihood value.}}
          & \cpp{LEP205\_xsec\_stau1stau1bar}
          \\ & & \cpp{LEP205\_xsec\_stau2stau2bar}
          \\ & & \cpp{stau\_1\_decay\_rates}
          \\ & & \cpp{stau\_2\_decay\_rates}
  \\ \hline
\end{tabular}
\caption{The slepton LEP limit capabilities provided by \colliderbit. In
addition to the dependencies shown above, all of these functions also depend
on \gambit's \cpp{MSSM30atMGUT}~\cite{gambit} model parameters and \specbit's
\cpp{MSSM\_spectrum}~\cite{SDPBit} capability. Each of the
\cpp{decay\_rates} can be provided by \decaybit~\cite{SDPBit}. These
functions have no options to be specified in the \YAML file.
\label{tab:leplimitslepton}}
\end{table*}

\begin{table*}[tp]
\centering
\scriptsize
\begin{tabular}{l|p{9.2cm}|l}
  \textbf{Capability}
      & \textbf{Function} (\textbf{Return type}): \textbf{Brief description} & \textbf{Dependencies}
  \\ \hline
  \cpp{L3\_Neutralino\_}
      & \multirow{4}{*}{\parbox{9.2cm}{\cpp{L3\_Neutralino\_All\_Channels\_Conservative\_LLike} (\cpp{double}):
              \\ Compares the cross section times branching ratio
              for neutralino pair production to the model-independent limit according
              to the L3 collaboration. Returns a log likelihood value.}}
          & \cpp{LEP188\_xsec\_chi00\_12}
          \\ \ \cpp{All\_Channels\_LLike} & & \cpp{LEP188\_xsec\_chi00\_13}
          \\ & & \cpp{LEP188\_xsec\_chi00\_14}
          \\ & & \cpp{decay\_rates}
  \\ \hline
  \cpp{L3\_Neutralino\_}
      & \multirow{4}{*}{\parbox{9.2cm}{\cpp{L3\_Neutralino\_Leptonic\_Conservative\_LLike} (\cpp{double}):
              \\ Compares the cross section times branching ratio
              for neutralino pair production (with leptonically decaying Z bosons)
              to the model-independent limit according to the L3 collaboration.
              Returns a log likelihood value.}}
          & \cpp{LEP188\_xsec\_chi00\_12}
          \\ \ \cpp{Leptonic\_LLike} & & \cpp{LEP188\_xsec\_chi00\_13}
          \\ & & \cpp{LEP188\_xsec\_chi00\_14}
          \\ & & \cpp{decay\_rates}
  \\ \hline
  \cpp{L3\_Chargino\_}
      & \multirow{4}{*}{\parbox{9.2cm}{\cpp{L3\_Chargino\_All\_Channels\_Conservative\_LLike} (\cpp{double}):
              \\ Compares the cross section times branching ratio
              for chargino pair production to the model-independent limit
              according to the L3 collaboration. Returns a log likelihood value.}}
          & \cpp{LEP188\_xsec\_chipm\_11}
          \\ \ \cpp{All\_Channels\_LLike} & & \cpp{LEP188\_xsec\_chipm\_22}
          \\ & & \cpp{decay\_rates}
          \\ & &
  \\ \hline
  \cpp{L3\_Chargino\_}
      & \multirow{4}{*}{\parbox{9.2cm}{\cpp{L3\_Chargino\_Leptonic\_Conservative\_LLike} (\cpp{double}):
              \\ Compares the cross section times branching ratio
              for chargino pair production (with leptonically decaying W bosons)
              to the model-independent limit according to the L3 collaboration.
              Returns a log likelihood value.}}
          & \cpp{LEP188\_xsec\_chipm\_11}
          \\ \ \cpp{Leptonic\_LLike} & & \cpp{LEP188\_xsec\_chipm\_22}
          \\ & & \cpp{decay\_rates}
          \\ & &
  \\ \hline
  \cpp{OPAL\_Neutralino\_}
      & \multirow{4}{*}{\parbox{9.2cm}{\cpp{OPAL\_Neutralino\_Hadronic\_Conservative\_LLike} (\cpp{double}):
              \\  Compares the cross section times branching ratio
              for neutralino pair production (with hadronically decaying Z bosons)
              to the model-independent limit according to the OPAL collaboration.
              Returns a log likelihood value.}}
          & \cpp{LEP208\_xsec\_chi00\_12}
          \\ \ \cpp{Hadronic\_LLike} & & \cpp{LEP208\_xsec\_chi00\_13}
          \\ & & \cpp{LEP208\_xsec\_chi00\_14}
          \\ & & \cpp{decay\_rates}
  \\ \hline
  \cpp{OPAL\_Chargino\_}
      & \multirow{4}{*}{\parbox{9.2cm}{\cpp{OPAL\_Chargino\_All\_Channels\_Conservative\_LLike} (\cpp{double}):
              \\ Compares the cross section times branching ratio
              for chargino pair production to the model-independent limit
              according to the OPAL collaboration. Returns a log likelihood value.}}
          & \cpp{LEP208\_xsec\_chipm\_11}
          \\ \ \cpp{All\_Channels\_LLike} & & \cpp{LEP208\_xsec\_chipm\_22}
          \\ & & \cpp{decay\_rates}
          \\ & &
  \\ \hline
  \cpp{OPAL\_Chargino\_}
      & \multirow{4}{*}{\parbox{9.2cm}{\cpp{OPAL\_Chargino\_Hadronic\_Conservative\_LLike} (\cpp{double}):
              \\ Compares the cross section times branching ratio
              for chargino pair production (with hadronically decaying W bosons)
              to the model-independent limit according to the OPAL collaboration.
              Returns a log likelihood value.}}
          & \cpp{LEP208\_xsec\_chipm\_11}
          \\ \ \cpp{Hadronic\_LLike} & & \cpp{LEP208\_xsec\_chipm\_22}
          \\ & & \cpp{decay\_rates}
          \\ & &
  \\ \hline
  \cpp{OPAL\_Chargino\_}
      & \multirow{4}{*}{\parbox{9.2cm}{\cpp{OPAL\_Chargino\_Leptonic\_Conservative\_LLike} (\cpp{double}):
              \\ Compares the cross section times branching ratio
              for chargino pair production (with leptonically decaying W bosons)
              to the model-independent limit according to the OPAL collaboration.
              Returns a log likelihood value.}}
          & \cpp{LEP208\_xsec\_chipm\_11}
          \\ \ \cpp{Leptonic\_LLike} & & \cpp{LEP208\_xsec\_chipm\_22}
          \\ & & \cpp{decay\_rates}
          \\ & &
  \\ \hline
  \cpp{OPAL\_Chargino\_}
      & \multirow{5}{*}{\parbox{9.2cm}{\cpp{OPAL\_Chargino\_SemiLeptonic\_Conservative\_LLike} (\cpp{double}):
              \\ Compares the cross section times branching ratio
              for chargino pair production (with one leptonic and one hadronic decaying W boson)
              to the model-independent limit according to the OPAL collaboration.
              Returns a log likelihood value.}}
          & \cpp{LEP208\_xsec\_chipm\_11}
          \\ \ \cpp{SemiLeptonic\_LLike} & & \cpp{LEP208\_xsec\_chipm\_22}
          \\ & & \cpp{decay\_rates}
          \\ & &
          \\ & &
  \\ \hline
\end{tabular}
\caption{The gaugino LEP limit capabilities provided by \colliderbit. In
addition to the dependencies shown above, all of these functions also depend
on \gambit's \cpp{MSSM30atMGUT}~\cite{gambit} model parameters and \specbit's
\cpp{MSSM\_spectrum}~\cite{SDPBit} capability. Each of the
\cpp{decay\_rates} can be provided by \decaybit~\cite{SDPBit}. These
functions have no options to be specified in the \YAML file. Note that the \cpp{All_Channels} likelihoods assume that the neutralino or chargino decay to fermions follows the same branching pattern as the corresponding on-shell gauge boson, and should be used with care.\label{tab:leplimitgaugino}}
\end{table*}

\subsubsection{Higgs likelihood capabilities}
\colliderbit provides likelihoods from experimental searches for Higgs bosons at LEP and the LHC, through interfaces
to \higgsbounds and
\higgssignals. The capability
\cpp{LEP\_Higgs\_LogLike} is provided by the function \cpp{calc_HB_LEP_LogLike}, which uses \higgsbounds to calculate an approximate likelihood constructed from the results from searches for neutral and
charged Higgs bosons at LEP. Similarly, capability \cpp{LHC\_Higgs\_LogLike} is provided by function \cpp{calc_HS_LHC_LogLike}, which employs \higgssignals to compute a likelihood including
constraints from measurements of the Higgs boson production rates and
mass at the LHC. These functions are detailed in Table~\ref{tab:higgscapabilities},
along with their dependencies.

Both functions depend on being provided with a \higgsbounds/\textsf{Signals}-specific data object containing all the input parameters needed to run either of these two external codes.  \colliderbit constructs one of these objects from the \cpp{Higgs_Couplings} provided by \specbit.  There are separate functions to do this for a pure SM Higgs, an MSSM Higgs sector with three neutral and one charged Higgs, and a Higgs sector containing just one SM-like Higgs and possible invisible states for it to decay to, as in the scalar singlet and other such singlet Higgs portal models (e.g.\ \cite{Cline13b,Beniwal,SSDM}).

\begin{table*}[tp]
\centering
\scriptsize
\begin{tabular}{l|p{7cm}|l|l}
  \textbf{Capability}
      & \multirow{2}{*}{\parbox{7cm}{\textbf{Function} (\textbf{Return type}):
             \\  \ \textbf{Brief description}}}
          & \textbf{Dependencies}
          & \multirow{2}{*}{\parbox{1.7cm}{\textbf{Backend} \\ \textbf{requirements}}}
          \\ & & &
  \\ \hline
  \cpp{LEP\_Higgs\_LogLike}
      & \multirow{3}{*}{\parbox{7cm}{\cpp{calc\_HB\_LEP\_LogLike}(\cpp{double}):
              \\  Provides log-likelihood for combined
              model-independent LEP neutral Higgs searches.}}
          & \cpp{HB\_ModelParameters} & \higgsbounds
          \\ & &
          \\ & &
  \\ \hline
  \cpp{LHC\_Higgs\_LogLike}
      & \multirow{3}{*}{\parbox{7cm}{\cpp{calc\_HS\_LHC\_LogLike}(\cpp{double}):
              \\ Provides log-likelihood for LHC Higgs
              mass and signal strength measurements.}}
          & \cpp{HB\_ModelParameters} & \higgssignals
          \\ & &
          \\ & &
  \\ \hline
  \cpp{HB\_ModelParameters}
      & \multirow{4}{*}{\parbox{7cm}{\cpp{SMHiggs\_ModelParameters}(\cpp{hb\_ModelParameters}):
              \\ Provides inputs for LEP and LHC Higgs likelihood calculations with \higgsbounds and \higgssignals, for a Higgs sector consisting only of an SM Higgs.}}
          & \cpp{Higgs_Couplings}
          \\ & & \cpp{SM\_spectrum}
          \\ & &
          \\ & &
          \\ \cmidrule{2-4}
      & \multirow{5}{*}{\parbox{7cm}{\cpp{SMHiggs\_ModelParameters}(\cpp{hb\_ModelParameters}):
              \\ Provides inputs for LEP and LHC Higgs likelihood calculations  with \higgsbounds and \higgssignals, for a Higgs sector consisting only of a single neutral Higgs, with possible decays to additional invisible particles.}}
          & \cpp{Higgs_Couplings}
          \\ & & A relevant \cpp{Spectrum} object
          \\ & &
          \\ & &
          \\ & &
          \\ \cmidrule{2-4}
      & \multirow{4}{*}{\parbox{7cm}{\cpp{MSSMHiggs\_ModelParameters}(\cpp{hb\_ModelParameters}):
              \\ Provides inputs for LEP and LHC Higgs likelihood calculations with \higgsbounds and \higgssignals, for an MSSM Higgs sector.}}
          & \cpp{Higgs_Couplings}
          \\ & & \cpp{MSSM\_spectrum}
          \\ & &
          \\ & &
  \\ \hline
  \cpp{FH\_HiggsProd}
      & \multirow{3}{*}{\parbox{7cm}{\cpp{FH\_HiggsProd}(\cpp{fh\_HiggsProd}):
              \\ Provides estimated MSSM Higgs
              production cross sections through an interface to \feynhiggs.}}
          & & \feynhiggs
          \\ & &
          \\ & &
  \\ \hline
\end{tabular}
\caption{The capabilities provided by \colliderbit for calculating LEP
  and LHC likelihoods from Higgs-sector-related experimental
  constraints. Final likelihood calculations are performed by the external code
  packages \higgsbounds and \higgssignals, using
  interfaces for input and output of model parameters incorporated
  into the \gambit framework. \cpp{Higgs_Couplings} are typically
  provided by \specbit~\cite{SDPBit}.
\label{tab:higgscapabilities}}
\end{table*}

\subsection{Standalone interface}
As described in~\cite{gambit}, \gambit routines can be called in a standalone
code provided that the code specifies the module functions and backend functions
that the user requires, along with any necessary options.
In addition, the user must resolve the dependencies of
each module function ``by hand''.

An annotated example program for running
\colliderbit independently of the \gambit framework can be found in
\term{ColliderBit/examples/} \term{ColliderBit\_standalone\_example.cpp}.
This example uses \colliderbit with a custom version of \pythia (\textsf{8.212.EM})
to calculate the LHC likelihood for a simple BSM model,
with the required couplings, masses and branching ratios input
via an SLHA file. The details of how to connect the custom \pythia
version to \colliderbit and run the standalone are given in Sec.~\ref{sec:standalonemodel}. Here we go through the structure of the code in \term{ColliderBit\_standalone\_example.cpp}.

The program consists of three main parts: dependency resolution, configuration of \colliderbit and \pythia, and execution of the simulation loop plus calculation of the LHC log-likelihood. It is the second part that the user typically will want to edit, as this is where the settings for event generation and detector simulation are specified, along with which LHC analyses to include.

To simplify the syntax a bit we use the following typedefs in \term{ColliderBit\_standalone\_example.cpp}:
\begin{minipage}{\columnwidth - 2mm}
\begin{lstcpp}
using namespace std;
typedef vector<int> vint;
typedef vector<double> vdouble;
typedef vector<bool> vbool;
typedef vector<string> vstr;
typedef vector<vector<string> > vvstr;
\end{lstcpp}
\end{minipage}
The configuration section begins by setting up the function \cpp{operateLHCLoop} with settings for the LHC simulation loop. In this example we set up two \pythia configurations, \cpp{"Pythia_EM_8Tev"} and \cpp{"Pythia_EM_13Tev"}, which will generate 20,000 events each. We will also allow detailed output to \term{stdout} during the event loops:
\begin{minipage}{\columnwidth - 2mm}
\begin{lstcpp}
operateLHCLoop.setOption<vstr>("pythiaNames",
    vstr {"Pythia_EM_8Tev", "Pythia_EM_13TeV"});
operateLHCLoop.setOption<vint>("nEvents",
    vint {20000, 20000});
operateLHCLoop.setOption<bool>("silenceLoop",
    false);
\end{lstcpp}
\end{minipage}

Then the \cpp{getPythiaFileReader} function can be configured with vectors of settings for the two \pythia configurations. For \cpp{"Pythia_EM_8Tev"} we have:
\begin{minipage}{\columnwidth - 2mm}
\begin{lstcpp}
getPythiaFileReader.setOption<vstr>(
    "Pythia_EM_8Tev", vstr {
        "UserModel:all = on",
        "Beams:eCM = 8000",
        "PartonLevel:MPI = off",
        "PartonLevel:ISR = on",
        "PartonLevel:FSR = off",
        "HadronLevel:all = off",
        "TauDecays:mode = 0",
        "Random:setSeed = on"});
\end{lstcpp}
\end{minipage}
Here the \cpp{"UserModel:all = on"} setting turns on the processes in the new BSM model, as detailed in Sec.~\ref{sec:standalonemodel}. The \cpp{"Pythia_EM_13Tev"} configuration is set up in a similar way, this time using \cpp{"Beams:eCM = 13000"}. The \cpp{getPythiaFileReader} function must also be given the path to the \textsf{XML} directory of \pythia \textsf{8.212.EM}:
\begin{minipage}{\columnwidth - 2mm}
\begin{lstcpp}
getPythiaFileReader.setOption<string>(
    "Pythia_doc_path", "Backends/installed/
        Pythia/8.212/share/Pythia8/xmldoc/");
\end{lstcpp}
\end{minipage}
In \term{ColliderBit\_standalone\_example.cpp} the path to a single input SLHA file is taken as a command line argument and stored in a variable \cpp{inputFileName}, which can then be passed to \cpp{getPythiaFileReader}:
\begin{minipage}{\columnwidth - 2mm}
\begin{lstcpp}
getPythiaFileReader.setOption<vstr>(
    "SLHA_filenames", vstr {inputFileName});
\end{lstcpp}
\end{minipage}

Finally, we choose which detector simulators and LHC analyses to include. To use the ATLAS configuration of \buckfast with both \cpp{"Pythia_EM_8Tev"} and \cpp{"Pythia_EM_13Tev"} we configure the \colliderbit function \cpp{getBuckFastATLAS} as follows:
\begin{minipage}{\columnwidth - 2mm}
\begin{lstcpp}
getBuckFastATLAS.setOption<vbool>(
    "useDetector", vbool {true, true});
getBuckFastATLAS.setOption<vdouble>(
    "antiktR", vdouble {0.4, 0.4});
getBuckFastATLAS.setOption<vbool>(
    "partonOnly", vbool {false, false});
\end{lstcpp}
\end{minipage}
The names of the ATLAS analyses to include are then passed to the function \cpp{getATLASAnalysisContainer}. Here we include the 0-lepton searches at 8 and 13~\TeV:
\begin{minipage}{\columnwidth - 2mm}
\begin{lstcpp}
getATLASAnalysisContainer.setOption<vvstr>(
    "analyses", vvstr {{"ATLAS_0LEP_20invfb"},
        {"ATLAS_13TeV_0LEP_13invfb"}});
\end{lstcpp}
\end{minipage}
Note that the two analyses are given in separate subvectors, one for each \pythia configuration (see Table~\ref{tab:analysis}). CMS analyses are similarly included by configuring \cpp{getBuckFastCMS} and \cpp{getCMSAnalysisContainer}. In our example, we only use a Run I CMS analysis, which is therefore only applied to the 8 TeV Pythia configuration.

The full LHC simulation loop and likelihood calculation is run in the third part of the main program, by executing the \colliderbit functions \cpp{operateLHCLoop} and \cpp{calc_LHC_LogLike}:
\begin{minipage}{\columnwidth - 2mm}
\begin{lstcpp}
operateLHCLoop.reset_and_calculate();
calc_LHC_LogLike.reset_and_calculate();
\end{lstcpp}
\end{minipage}

\section{Examples}
\label{sec:examples}
\subsection{CMSSM example}

An annotated example of a \YAML file for scanning the CMSSM with \gambit using only functions from \colliderbit is provided in \term{yaml_files/ColliderBit\_CMSSM.yaml}. The file demonstrates how to specify the model parameters (and priors), choose
and configure a sampler, choose a printer (either \textsf{hdf5} or
\textsf{ascii}), run the LHC and LEP collider likelihoods, run the \higgsbounds
and \higgssignals Higgs likelihoods, and configure details of the detector
simulation and Monte Carlo event generator.

\subsection{Generic \pythia model example}
\label{sec:standalonemodel}
The recommended method of using \colliderbit with a new model is to define and run the model
within the full \gambit framework, allowing access to the model declaration and
scanning routines, in addition to non-collider likelihood functions should these
be of interest. However, if the user only wants to check single parameter points with \colliderbit, the standalone interface described in the previous section presents a more minimal alternative. Regardless of which interface is used, \colliderbit must be set up to work with a version of \pythia that can generate events for the new model. Here we go through an example of how to achieve this.

Our physics model example consists of the SM augmented by a new scalar
singlet field $\phi_1$ and a new, coloured Dirac fermion $U$. The model is a
stripped down version of that featured in~\cite{Ask:2012sm}, which contains a
complete tutorial for how to implement the model in Monte Carlo generators. The
new particles have the following mass terms:
\begin{equation}
\mathcal{L}_\textrm{mass}=-\frac{m_1^2}{2} \phi_1^2 + M_U\bar{U}U.
\end{equation}

The new fermion interacts with the new scalar via the Lagrangian term
\begin{equation}
\mathcal{L}_\textrm{yuk}=\lambda_1\phi_1\bar{U}P_R u + h.c.,
\end{equation}
where $u$ is the SM up-quark field. We will simulate the process
$pp\rightarrow \bar{U}U$ where the $U$ subsequently decays via
$U\rightarrow u\phi_1$.

To use this model with \colliderbit, we make use of the \MGaMCNLO--\pythiaeight
interface to generate matrix element code that can be used to supplement the
internal processes in \pythia. Sample \Mathematica notebook and \feynrules model files for generating UFO
output are provided in \term{ColliderBit/data/ExternalModel}. The
\MGaMCNLO commands for generating matrix element code for coloured fermion
production in proton collisions are as follows, assuming that the UFO model has
been placed in the \MGaMCNLO~ \term{models} directory:
\begin{lstpy}
import model GambitDemo_UFO
generate p p > uv uv~
output pythia8
\end{lstpy}

The resulting \Cpp code can be found in the \term{src} and \term{include} subdirectories of \term{Backends/patches/} \term{pythia/8.212.EM/ExternalModel}. This directory also contains two \pythia \textsf{XML} files that declare a new \pythia setting \cpp{UserModel:all}, and a version of the \pythia file \term{ProcessContainer.cc} that connects this setting to the generated matrix element code.

The \gambit build system can be used to make a new version of the \pythia backend (\textsf{8.212.EM}) with
\begin{lstterm}
make pythia_8.212.EM
\end{lstterm}
This command performs the following tasks:
\begin{itemize}
\item downloads \pythia in the usual way, but into a new location;
\item copies the new matrix element code to the new location;
\item updates the \pythia\ \textsf{XML} configuration files to define the new
  \cpp{UserModel} setting;
\item updates the \pythia file \term{ProcessContainer.cc} to allow the user to
  run the new matrix elements using the setting \cpp{UserModel:all = on};
\item runs \pythia through \BOSS \cite{gambit} to construct the interface to \gambit;
\item builds the new \pythia version.
\end{itemize}

To implement a different BSM physics model, the existing \MGaMCNLO-generated files in \term{Backends/patches/pythia/8.212.EM/ExternalModel} must be replaced with the files generated for the new model, and \term{ProcessContainer.cc} must be updated accordingly. The \gambit build system will then take care of updating the \pythia backend to use the new code.

It remains to tell \colliderbit to use the new \pythia \textsf{8.212.EM} backend rather than the old
one. Since this contains all previous \pythia functionality in addition to the
new matrix elements, it can be used in all places where \pythia\ \textsf{8.212} was previously used. To change the version of \pythia used, the user
must change the default version in \term{Backends/include/gambit/Backends/} \term{default\_bossed\_versions.hpp}, using:
\begin{lstcpp}
#define Default_Pythia 8_212_EM
\end{lstcpp}
Note that \gambit must be rebuilt after this change. Also, the \colliderbit option \cpp{Pythia\_doc\_path} (Table~\ref{tab:collider}) must be set to \term{Backends/installed/Pythia/8.212.EM/} \term{share/Pythia8/xmldoc} in the input \YAML file when \colliderbit is used as part of a \gambit run, or directly in the standalone code as shown in \term{ColliderBit\_standalone\_example.cpp}.

An example \YAML file showing how to run \GB with the new \pythia \textsf{8.212.EM} backend can be found in \term{yaml_files/ColliderBit_ExternalModel.yaml}.

After compilation (see Appendix~\ref{sec:quickstart}), the standalone example that makes use of the new \pythia \textsf{8.212.EM} backend can be run as
\begin{lstterm}
./ColliderBit_standalone
    ColliderBit/data/ExternalModel_point.slha
\end{lstterm}
This instructs \pythia to produce $\bar{U}U$ pairs in proton collisions,
but they will not decay unless instructed to do so via the input SLHA file.
An example SLHA file generated with \MGaMCNLO is provided in \term{ColliderBit/data/ExternalModel\_point.slha}.
This file contains a decay table for the $U$ particle with a 100\% branching ratio to an
up quark and a $\phi_1$.

We remind the reader that the standalone example is only intended as a minimal way of running
single points of a new model through \colliderbit. For a comprehensive study, including scanning over model parameters,
the user should add the model in the \gambit model database and implement spectrum and decay calculations through the \gambit modules \specbit and \decaybit as required.

Finally, there is an important subtlety regarding invisible particles. At the
time of writing, the default PDG ID codes of new particles in \feynrules do not
always correspond to those of invisible, uncharged particles. In the
\colliderbit simulation chain, this means that the particles will not appear as
missing energy in the detector simulation. According to the PDG ID code standard,
invisible particles may have a PID of 12, 14 or 16 (SM neutrinos), 1000022
(lightest neutralino in a superysmmetric model), or 50--60 (for generic new BSM
particles). The user can thus obtain correct behaviour for an invisible species
by including the PDG code definition in the \feynrules field definition as in
the following example:
\begin{lstterm}
 S[10] == {
    ClassName       -> p1,
    SelfConjugate   -> True,
    Indices         -> {},
    Mass            -> {Mp1, 10},
    PDG             -> {51},
    Width           -> {Wp1, 0}
  }
\end{lstterm}

A less satisfactory option is to change the following code in \term{contrib/heputils/include/HEPUtils/Event.h} that implements the PDG ID
standard for invisibles:
\begin{lstcpp}
if (p->abspid() == 12 || p->abspid() == 14 ||
    p->abspid() == 16 || p->pid() == 1000022 ||
    in_range(p->pid(), 50, 60))
  _invisibles.push_back(p);
\end{lstcpp}

\section{Conclusions}
\label{sec:conclusions}
\colliderbit is a new modular software code for the application of high-energy collider constraints to generic BSM physics models, written in the \gambit framework. This paper serves as an introduction to the code, and as a reference manual for users wishing to add new analyses or features.

The code provides a rigorous and fast implementation of LHC constraints through a parallelised Monte Carlo simulation interfaced with several detector simulation options, including a new simulation based on four-vector smearing. A custom event analysis class allows the user to apply the same LHC analysis code to any level of detector simulation,  and we supply likelihood routines capable of reproducing LHC cut and count searches, or binned shape fits. An interface to the \pythiaeight event generator allows the user to add matrix elements for new models.

LEP constraints are handled via a new code based on a sophisticated interpolation of the cross-section limits on slepton, neutralino and chargino pair production. Higgs limits, for both LEP and the LHC, are currently handled via an interface to the \higgssignals and \higgsbounds packages, but there exists scope to provide and interface new likelihood calculations in future \colliderbit releases.

The code can function either as a standalone tool for quick checks of specific model points, or it can be run within the \gambit framework to provide a complete tool for BSM inference from high energy collider data.

\begin{acknowledgements}
We thank the other members of the \GB Collaboration for helpful discussions, comments and support.  We are very grateful to Torbj\"orn Sj\"ostrand and Peter Skands for helpful discussions on the use of the \pythia event generator, and for code modifications to improve the efficiency and flexibility of its process selection and settings database.  \gambitacknos
\end{acknowledgements}

\appendix

\section{Quick start guide}
\label{sec:quickstart}
Instructions for how to get \colliderbit and \gambit can be found at
\href{http://gambit.hepforge.org}{gambit.hepforge.org}. Here, we give a list
of steps to follow in order to build and run \colliderbit, either in its
standalone version or linked with \gambit. Additional details about configuring
and building \gambit can be found in~\cite{gambit}.

\subsection{Building and running the standalone example}

The basic commands to build the standalone example are:
\begin{lstterm}
cd gambit
mkdir build
cd build
cmake ..
make -j@\metavar{n}@ ColliderBit_standalone
\end{lstterm}
Here, $n$ is the number of logical cores the user wishes to use during the
compilation.

The backends used by the standalone example must also be built:
\begin{lstterm}
make nulike
make pythia_8.212.EM
\end{lstterm}

The user can set the number of OpenMP threads to use during \colliderbit's
parallelisation step with a system variable:
\begin{lstterm}
export OMP_NUM_THREADS=@\metavar{m}@
\end{lstterm}
Here, $m$ is the number of threads to use during runtime.

Finally, the standalone example can be run from the main \gambit directory:
\begin{lstterm}
cd ..
./ColliderBit_standalone
    ColliderBit/data/ExternalModel_point.slha
\end{lstterm}

\subsection{Running the \colliderbit example in \GB}

The basic commands to build \gambit and run a minimal \colliderbit example
are very similar to those shown above, except that we now also need the backend \susyhit, and we use the default version \pythia (\textsf{8.212}):
\begin{lstterm}
cd gambit
mkdir build
cd build
cmake ..
make -j@\metavar{n}@ gambit
make nulike
make pythia
make susyhit
export OMP_NUM_THREADS=@\metavar{m}@
cd ..
./gambit -f yaml_files/ColliderBit_CMSSM.yaml
\end{lstterm}

\section{\colliderbit classes}
\label{sec:colliderbitclasses}
Users who wish to add their own custom functions to \colliderbit may find it
useful to use our inheritance scheme. For such users, we here describe the
main base classes and inheritance scheme of \colliderbit.
We expect such users to be familiar with adding capabilities, module functions,
and (possibly) backend functions, as described in the main \gambit paper~\cite{gambit}.

There are four categories of functions within \colliderbit associated with abstract base classes: Collider simulation is associated with
the \cpp{BaseCollider} class, detector simulation with \cpp{BaseDetector},
analysis with \cpp{BaseAnalysis}, and the limit-setting application with
\cpp{BaseLimitContainer}.

The \cpp{BaseCollider} class is templated on the type of collider event
(\cpp{EventT}) that it can provide. Each subclass of \cpp{BaseCollider<EventT>} will
inherit the virtual functions described in Table~\ref{tab:basecollider}.
Thus, creating a subclass of \cpp{BaseCollider} will force the user to
define these functions, which are the usual things to be expected of collider
simulation tools. A very simple example of this can be found in the header
file\footnote{Within this Appendix, the headers paths (\term{*.hpp}) are realtive to \term{ColliderBit/} \term{include/gambit/ColliderBit}
, while source files (\term{*.cpp}) are in
\term{ColliderBit/src}.} \term{colliders/SimplePythia.hpp}.

\begin{table*}[t]
\centering
\scriptsize
\begin{tabular}{lp{8.5cm}}
  \toprule
  \textbf{\Cpp function signature} & \textbf{Intended purpose}
  \\ \midrule
  \cpp{virtual void clear()}$^{\rm a}$
  & Clear the internal memory of this instance so that it may be reused.
  \\ \addlinespace
  \cpp{virtual void nextEvent(EventT\& event)} \cpp{const}$^{\rm b}$
  & Generate the next collider event, storing the result into the given
          \cpp{event}.
  \\ \addlinespace
  \cpp{virtual double xsec\_pb()} \cpp{const}$^{\rm b}$
  & Return the total cross section (in \pb) of generated events.
  \\ \addlinespace
  \cpp{virtual double xsecErr\_pb()} \cpp{const}$^{\rm b}$
  & Return the absolute error estimate of the cross section (in \pb) of
          generated events.
  \\ \addlinespace
  \cpp{virtual void init(const std::vector<std::string>\&)}$^{\rm a}$
  & Initialise the collider simulator with a set of options given as a vector
          of strings.
  \\ \addlinespace
  \cpp{virtual void init()}$^{\rm a}$
  & Initialise the collider simulator with no options.
  \\ \bottomrule
\end{tabular}
\caption{Inherited functions for subclasses of \cpp{BaseCollider<EventT>}.
Functions marked with $^{\rm a}$ do nothing unless overridden by the subclass
author. Functions marked with $^{\rm b}$ \emph{must} be overridden by the
subclass author. \label{tab:basecollider}}
\end{table*}

Within this file, we see the definition of the \cpp{SimplePythia} class, which
inherits from \cpp{BaseDetector<Pythia8::Event>}. The class defines overrides for
each of the virtual functions shown in Table~\ref{tab:basecollider}. A more
complicated example of this can be seen for the \cpp{SpecializablePythia} class,
which also inherits from \cpp{BaseDetector<Pythia8::Event>}. It is declared and
defined within the files \term{colliders/SpecializablePythia.hpp} and \term{colliders/SpecializablePythia.cpp}.

The addition of custom detectors to \colliderbit involves subclasses of the
\cpp{BaseDetector} class, which is templated on both the type of event that it
can accept for simulation (\cpp{EventIn}), and the type of event that it
will return after detector simulation (\cpp{EventOut}). In a similar
way as described above for colliders, a user may add a fully custom detector by
creating a subclass of \cpp{BaseDetector<EventIn, EventOut>} and writing
overrides for the virtual functions, as described in Table~\ref{tab:basedetector}.

\begin{table*}[t]
\centering
\scriptsize
\begin{tabular}{lp{7.2cm}}
  \toprule
  \textbf{\Cpp function signature} & \textbf{Intended purpose}
  \\ \midrule
  \cpp{virtual void clear()}$^{\rm a}$
  & Clear the internal memory of this instance so that it may be reused.
  \\ \addlinespace
  \cpp{virtual void processEvent(const EventIn\&, EventOut\&)} \cpp{const}$^{\rm b}$
  & Apply detector simulation to the given \cpp{EventIn}, storing the
          result into the \cpp{EventOut}.
  \\ \addlinespace
  \cpp{virtual void init(const std::vector<std::string>\&)}$^{\rm a}$
  & Initialise the collider simulator with a set of options given as a vector
          of strings.
  \\ \addlinespace
  \cpp{virtual void init()}$^{\rm a}$
  & Initialise the collider simulator with no options.
  \\ \bottomrule
\end{tabular}
\caption{Inherited functions for subclasses of \cpp{BaseDetector<EventIn,
EventOut>}. Functions marked with $^{\rm a}$ do nothing unless overridden by
the subclass author. Functions marked with $^{\rm b}$ \emph{must} be
overridden by the subclass author. \label{tab:basedetector}}
\end{table*}

The analysis base class, \cpp{BaseAnalysis}, is templated on the type of
event that it can analyze (\cpp{EventT}). Subclasses of
\cpp{BaseAnalysis<EventT>} inherit the functions as described in
Table~\ref{tab:baseanalysis}, some of which must be overridden by the subclass
author. Existing analyses provide examples of how to do this.

\begin{table*}[t]
\centering
\scriptsize
\begin{tabular}{lp{8.4cm}}
  \toprule
  \textbf{\Cpp function signature} & \textbf{Intended purpose}
  \\ \midrule
  \cpp{virtual void clear()}$^{\rm a}$
  & Clear the internal memory of this instance so that it may be reused.
  \\ \addlinespace
  \cpp{void analyze(const EventT\& e)}
  & This version of \cpp{analyze} simply calls the pointer version below.
  \\ \addlinespace
  \cpp{virtual void analyze(const EventT*)}$^{\rm a}$
  & Analyze the given event, storing the result internally.
  \\ \addlinespace
  \cpp{double num\_events()}
  & Return the total number of events analyzed.
  \\ \addlinespace
  \cpp{double xsec()}
  & Return the total cross section (in fb) of events analyzed.
  \\ \addlinespace
  \cpp{double xsec\_err()}
  & Return the cross section uncertainty (in fb).
  \\ \addlinespace
  \cpp{double xsec\_relerr()}
  & Return the relative cross section uncertainty.
  \\ \addlinespace
  \cpp{double xsec\_per\_event()}
  & Return the cross section per event (in fb) of events analyzed.
  \\ \addlinespace
  \cpp{double luminosity()}
  & Return the integrated luminosity (in fb$^{-1}$) of events analyzed.
  \\ \addlinespace
  \cpp{void set\_xsec(double xs, double xserr)}
  & Set the cross section, and its uncertainty (in fb).
  \\ \addlinespace
  \cpp{void set\_luminosity(double lumi)}
  & Set the luminosity (in fb$^{-1}$).
  \\ \addlinespace
  \cpp{std::vector<SignalRegionData> get\_results()}
  & Return the results as a \cpp{vector} of \cpp{SignalRegionData} objects.
  \\ \addlinespace
  \cpp{void add\_result(const SignalRegionData\& res)}
  & Add a result to the internal results list.
  \\ \addlinespace
  \cpp{virtual void collect\_results()}$^{\rm b}$
  & Collect all results of this analysis together in preparation for a
    likelihood calculation.
  \\ \addlinespace
  \cpp{virtual void init(const std::vector<std::string>\&)}$^{\rm a}$
  & Initialise the analysis with a set of options given as a vector of strings.
  \\ \addlinespace
  \cpp{virtual void init()}$^{\rm a}$
  & Initialise the analysis with no options.
  \\ \addlinespace
  \cpp{virtual void scale(double factor)}$^{\rm a}$
  & Scale the results of this analysis by the given factor, which is optional.
    If no factor is given, the scale factor is set instead by the luminosity.
  \\ \addlinespace
  \cpp{virtual void add(BaseAnalysis* other)}$^{\rm a}$
  & Adds the results of an identical analysis to this one.
  \\ \addlinespace
  \cpp{void add\_xsec(double xs, double xserr)}
  & Add the given cross section to the stored total and recompute the uncertainty.
  \\ \addlinespace
  \cpp{void improve\_xsec(double xs, double xserr)}
  & Improve the stored cross section by averaging it with the given one, and
    recompute the uncertainty.
  \\ \bottomrule
\end{tabular}
\caption{Inherited functions for subclasses of \cpp{BaseAnalysis<EventT>}.
Functions that are \cpp{virtual} and marked with $^{\rm a}$ perform only
simple operations on variables in the base class, unless they are overridden
by the subclass author. Functions that are \cpp{virtual} and marked with
$^{\rm b}$ \emph{must} be overridden by the subclass author. Non-\cpp{virtual}
functions are not intended to be overridden. Analysis results are contained
within each subclass in a \cpp{SignalRegionData} instance, which is described
in Table~\ref{tab:signalregiondata}. \label{tab:baseanalysis}}
\end{table*}

\begin{table*}[t]
\centering
\scriptsize
\begin{tabular}{lp{12.4cm}}
  \toprule
  \textbf{\Cpp Member Variable} & \textbf{Intended purpose}
  \\ \midrule
  \cpp{std::string analysis\_name}
  & The name of the analysis that contains this signal region.
  \\ \addlinespace
  \cpp{std::string sr\_label}
  & A label for this particular signal region.
  \\ \addlinespace
  \cpp{double n\_observed}
  & The number of events passing selection for this signal region, as reported
    by the experiment.
  \\ \addlinespace
  \cpp{double n\_signal}
  & The number of simulated model events passing selection for this signal region.
  \\ \addlinespace
  \cpp{double n\_signal\_at\_lumi}
  & The number of simulated model events passing selection for this signal region,
    scaled to the experimental luminosity.
  \\ \addlinespace
  \cpp{double n\_background}
  & The number of Standard Model events passing selection for this signal
    region, as reported by the experiment.
  \\ \addlinespace
  \cpp{double signal\_sys}
  & The absolute systematic error of \cpp{n\_signal}.
  \\ \addlinespace
  \cpp{double background\_sys}
  & The absolute systematic error of \cpp{n\_background}.
  \\ \bottomrule
\end{tabular}
\caption{Member variables of the \cpp{SignalRegionData} struct, which is used
in the \cpp{BaseAnalysis} class as a container for the analysis results of each
signal region. The \cpp{BaseAnalysis} class is described in
Table~\ref{tab:baseanalysis}. \label{tab:signalregiondata}}
\end{table*}

The addition of custom limits and limit curve interpolation to \colliderbit
requires that the user declare new module functions in
\term{ColliderBit\_rollcall.hpp} and define them in
\term{ColliderBit.cpp}. However, if the user wishes to use \colliderbit's
limit interpolation system (as described in Sec.~\ref{sec:leplike}), they
can create a subclass of the \cpp{BaseLimitContainer} class and override
the functions shown in Table~\ref{tab:baselimitcontainer}.

\begin{table*}[t]
\centering
\scriptsize
\begin{tabular}{p{8.9cm}p{7.5cm}}
  \toprule
  \textbf{\Cpp function signature} & \textbf{Intended purpose}
  \\ \addlinespace
  \cpp{virtual P2 convertPt()} \cpp{const}$^{\rm b}$
  & Convert a point from pixel units to axis units, creating a \cpp{P2}.
  \\ \addlinespace
  \cpp{virtual bool isWithinExclusionRegion(double x, double y,}
  \phantom{\cpp{virtual bool isWithinExclusionRegion(}}\cpp{double mZ)} \cpp{const}$^{\rm b}$
  & Check to see if the point specified by \cpp{x} and \cpp{y} is within
    the exclusion region. This may depend on the $Z$ boson mass, \cpp{mZ}.
  \\ \addlinespace
  \cpp{virtual double specialLimit(double, double)} \cpp{const}
  &  Return a special value for the limit when outside of the exclusion region.
    This function returns zero unless overridden.
  \\ \addlinespace
  \cpp{double limitAverage(double x, double y, double mZ)} \cpp{const}
  & Uses the limit interpolation described in Sec.~\ref{sec:leplike} to
    return a limit at the point specified by \cpp{x} and \cpp{y}. This
    may depend on the $Z$ boson mass, \cpp{mZ}.
  \\ \addlinespace
  \cpp{void dumpPlotData(double xlow, double xhigh, double ylow, }
  \phantom{\cpp{void dumpPlotData(}}\cpp{double yhigh, double mZ, std::string}
  \phantom{\cpp{void dumpPlotData(}}\cpp{filename, int ngrid)} \cpp{const}
  & Creates a file, \cpp{filename}, containing the results of
    \cpp{limitAverage} calls using a grid of
    \cpp{ngrid} $\times$ \cpp{ngrid} points within the rectangle defined
    by \cpp{xlow}, \cpp{xhigh}, \cpp{ylow}, and \cpp{yhigh}.
  \\ \addlinespace
  \cpp{void dumpLightPlotData(std::string filename,}
  \phantom{\cpp{void dumpLightPlotData(}}\cpp{int nperLine)} \cpp{const}
  & Creates a file, \cpp{filename}, containing the limit contour data as a
    series of points, using \cpp{nperLine} points for each line of the contour.
  \\ \bottomrule
\end{tabular}
\caption{Inherited functions for subclasses of \cpp{BaseLimitContainer}.
Functions that are \cpp{virtual} and marked with $^{\rm b}$ \emph{must} be
overridden by the subclass author. Non-\cpp{virtual} functions are not
intended to be overridden. Note that \cpp{P2} is a simple class that
represents a point found on the limit plane and is defined in the header file
\cpp{limits/PointsAndLines.hpp}. \label{tab:baselimitcontainer}}
\end{table*}

\startglossary

\gitem{backend}\input{"glossary/backend.glossentry"}
\gitem{backend function}\input{"glossary/backend_function.glossentry"}
\gitem{backend requirement}\input{"glossary/backend_requirement.glossentry"}
\gitem{backend variable}\input{"glossary/backend_variable.glossentry"}
\newcommand{\seecompdatabase}{see Sec.\ 10.7 of Ref.\ \cite{gambit}}
\gitem{capability}\input{"glossary/capability.glossentry"}
\gitem{dependency}\input{"glossary/dependency.glossentry"}
\gitem{frontend}\input{"glossary/frontend.glossentry"}
\gitem{frontend header}\input{"glossary/frontend_header.glossentry"}
\gitem{module}\input{"glossary/module.glossentry"}
\gitem{module function}\input{"glossary/module_function.glossentry"}
\gitem{physics module}\input{"glossary/physics_module.glossentry"}
\gitem{rollcall header}\input{"glossary/rollcall_header.glossentry"}
\gitem{type}\input{"glossary/type.glossentry"}

\finishglossary

\bibliography{R1}

\end{document}